\documentclass[11pt]{article}

\usepackage[T1]{fontenc}
\usepackage{microtype}
\usepackage{booktabs}
\usepackage{multirow}
\usepackage{colortbl}
\usepackage{array}
\usepackage{amsmath,amsthm}
\usepackage{aliascnt}
\usepackage{newpxtext,newpxmath}
\usepackage[margin=15pt,font=small,labelfont={bf}]{caption}
\usepackage[margin=1in]{geometry}
\usepackage[english]{babel}
\usepackage[utf8]{inputenc}
\usepackage{csquotes}

\usepackage[backend=biber,style=alphabetic,maxalphanames=3,minalphanames=3,maxbibnames=99,minbibnames=99]{biblatex}
\usepackage{graphicx}
\usepackage{algorithm}
\usepackage[noend]{algpseudocode}
\usepackage[bottom]{footmisc}

\AtBeginDocument{
  }
\usepackage{xcolor}
\usepackage[colorlinks=true,
    citecolor=darkgreen,
    linkcolor=darkred
]{hyperref}
\usepackage{aliascnt}
\definecolor{darkgreen}{rgb}{0,0.5,0.5}
\definecolor{darkred}{rgb}{0.5,0,0}
\definecolor{purple}{rgb}{0.5,0,0.5}
\definecolor{resultyellow}{HTML}{BBEEAA}
\newcolumntype{Y}{>{\columncolor{resultyellow}}c}
\newcommand\drop[1]{}

\newcommand*{\newaliastheorem}[3]{%
  \newaliascnt{#1}{#2}%
  \newtheorem{#1}[#1]{#3}%
  \aliascntresetthe{#1}%
  \expandafter\providecommand\csname #1autorefname\endcsname{#3}%
}

\newtheorem{theorem}{Theorem}[section]
\newaliastheorem{proposition}{theorem}{Proposition}
\newaliastheorem{lemma}{theorem}{Lemma}
\newaliastheorem{corollary}{theorem}{Corollary}
\newaliastheorem{definition}{theorem}{Definition}
\newaliastheorem{conjecture}{theorem}{Conjecture}
\newaliastheorem{example}{theorem}{Example}
\newaliastheorem{remark}{theorem}{Remark}

\addto\extrasenglish{

}

\title{Online Graph Coloring: Competitive Ratios Separating Randomized and Deterministic Algorithms}
\author{Ken-ichi Kawarabayashi\thanks{National Institute of Informatics \& The University of Tokyo, Tokyo, Japan, \texttt{k\_keniti@nii.ac.jp}.\\ 
    Supported by JSPS Kakenhi 22H05001, JP25K24465, and by JST ASPIRE JPMJAP2302.}
    \and 
    Hirotaka Yoneda\thanks{The University of Tokyo, Tokyo, Japan, \texttt{squar37@gmail.com}.\\ 
    Supported by JSPS JP25K24465, JST ASPIRE JPMJAP2302 and by JST ACT-X JPMJAX25CT.}
    \and 
    Masataka Yoneda\thanks{The University of Tokyo, Tokyo, Japan, \texttt{e869120@gmail.com}.\\ 
    Supported by JSPS JP25K24465 and JST ASPIRE JPMJAP2302.}}
\date{}
\begin{document}
\maketitle

\pagenumbering{gobble}

\begin{abstract}
    We study online coloring, focusing on the competitive ratio. The best previously known results were (1) a lower bound of $\Omega(n / \log^2 n)$ for both deterministic and randomized algorithms by Halld\'{o}rsson and Szegedy (1994), (2) an upper bound of $O(n / \log n)$ for randomized algorithms by Halld\'{o}rsson (1997), and (3) an upper bound of $O(n / \log \log n)$ for deterministic algorithms by Kawarabayashi, Yoneda, and Yoneda (2025). In this paper, we improve these results as follows:

    \begin{enumerate}
        \item A new lower bound of $\Omega(n / \log n)$ for deterministic algorithms.
        \item A new upper bound of $O(n (\log \log n)^2 / \log^2 n)$ for randomized algorithms.
        \item A new upper bound of $O(n / (\log \log n)^{2-\varepsilon})$ for deterministic algorithms for every fixed $\varepsilon>0$.
    \end{enumerate}

    For randomized algorithms, we nearly close the gap between the lower and upper bounds; the remaining gap is only $O((\log \log n)^2)$. 

    We establish the asymptotic separation between randomized and deterministic online graph coloring. Against an oblivious adversary, a polynomial-time randomized algorithm achieves a competitive ratio asymptotically smaller than any deterministic algorithm can guarantee, even with unlimited computation.
\end{abstract}

\newpage
{
    \setcounter{tocdepth}{2}
    \tableofcontents
    \clearpage
}
\pagenumbering{arabic}
\setcounter{page}{1}

\section{Introduction}\label{sec:1}

A central problem both in discrete math and in theoretical computer science, \emph{graph coloring}, is to color each vertex of a given graph $G = (V, E)$ with as few colors as possible so that no two adjacent vertices have the same color. We study the \emph{online coloring} problem, the online version of graph coloring, in which the vertices are given one by one (together with their incident edges). Immediately after a vertex arrives, we must assign it a color before the next vertex arrives. Once the color is assigned, it cannot be changed. Online coloring is one of the most central questions in online algorithms, alongside online matching, $k$-server, and the list update problem \cite{Alb03}.

\subsection{Historical Background}

\begin{figure}[b]
  \centering
  \includegraphics[width=\linewidth]{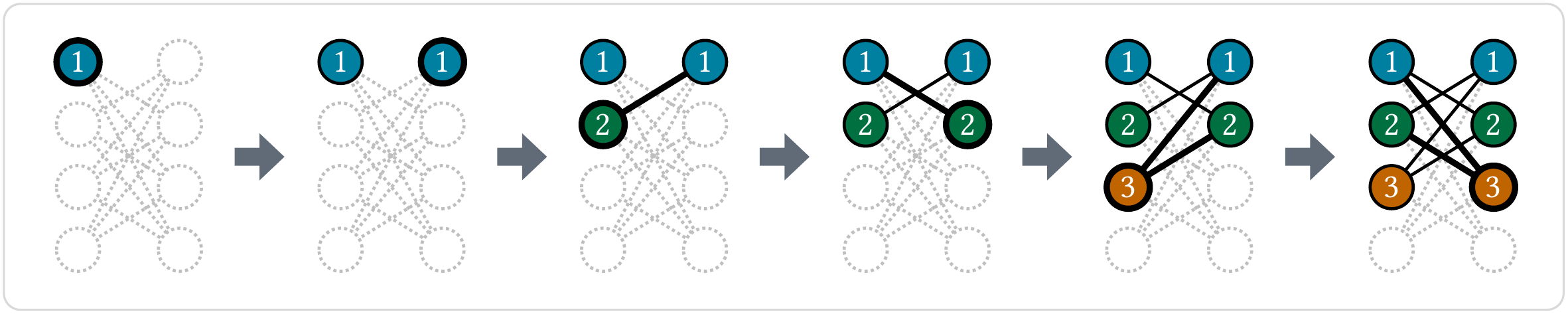}
  \caption{A worst-case bipartite graph on which First-Fit uses $n/2$ colors.}
  \label{fig:101}
\end{figure}

Historically, the central focus in the study of online algorithms has been \emph{competitive analysis}, which compares an online algorithm's performance to that of an optimal offline algorithm. This is also the case in online coloring; the competitive ratio\footnote{The competitive ratio is the maximum possible value of ``the algorithm's performance (e.g., number of colors used)'' divided by ``the optimal performance (e.g., chromatic number)''.} for online coloring was studied as early as the 1980s \cite{GL88, LST89}. However, straightforward algorithms perform poorly. For example, the First-Fit algorithm, which always chooses the smallest available color, has a competitive ratio of at least $n/4$ (for graphs with $n$ vertices), as it may use $n/2$ colors even on a bipartite graph (\autoref{fig:101}). This makes online coloring particularly challenging compared to other online problems.

\paragraph{Competitive ratio.}

As one of the most important results in online coloring, Halld\'{o}rsson and Szegedy (1994) \cite{HS94} presented a lower bound of $\Omega(n / \log^2 n)$ on the competitive ratio. This shows the inherent difficulty of online coloring. However, a central question remains: \emph{``Do randomized algorithms outperform deterministic algorithms in online coloring?''}

\paragraph{Deterministic vs. randomized.}

On the deterministic side, Lov\'{a}sz, Saks, and Trotter (1989) \cite{LST89} found an algorithm with a competitive ratio of $O(n / \log^* n)$, followed by improvements to $O(n \log \log \log n / \log \log n)$ by Kierstead (1998) \cite{Kie98b} and to $O(n / \log \log n)$ by Kawarabayashi, Yoneda, and Yoneda in 2025 \cite{KYY27} (to appear in SODA '27). On the randomized side, Vishwanathan (1992) \cite{Vis92} found an algorithm with a competitive ratio of $O(n / \log^{1/2} n)$, and improved to $O(n / \log n)$ by Halld\'{o}rsson (1997) \cite{Hal97}. We assume an oblivious adversary for randomized algorithms, where the input is fixed independently of the algorithm's output.

\begin{figure}[t]
  \centering
  \includegraphics[width=\linewidth]{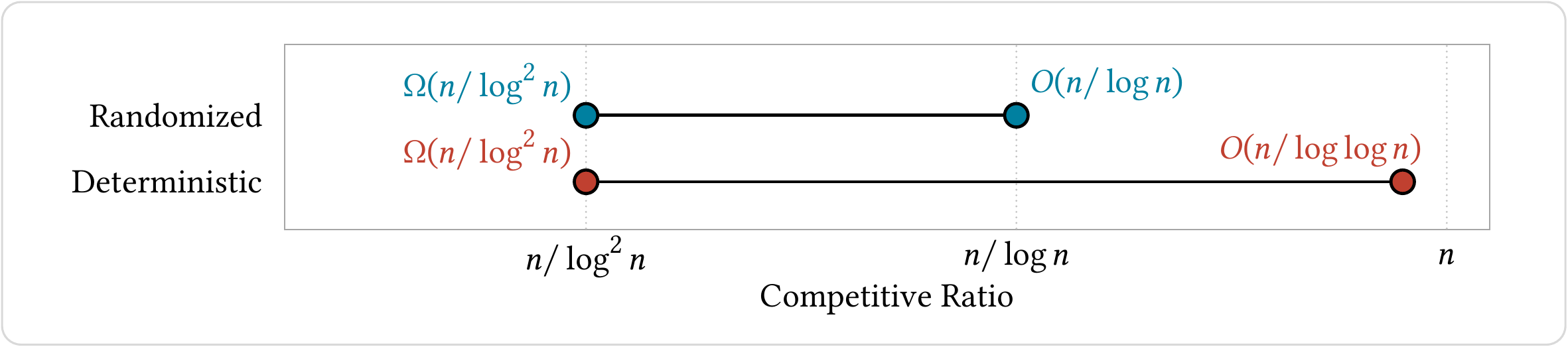}
  \caption{The previously known lower and upper bounds for randomized and deterministic algorithms. These bounds do not establish an asymptotic separation.}
  \label{fig:102}
\end{figure}

Although the best-known randomized upper bound is better than the deterministic one, the asymptotic separation between the optimal competitive ratios remains open because the best-known lower bounds for both are $\Omega(n / \log^2 n)$; see \autoref{fig:102}. Whether an asymptotic separation exists is crucial not only for online coloring but also for the broader field of online algorithms. In the worst-case analysis of deterministic algorithms, we must assume an ``adversarial attack'', which has also been studied in the context of machine learning, such as in the adversarial bandit problem \cite{ACF+02}. Therefore, settling the existence question would clarify whether the adversary is effective. Note that an asymptotic separation has been established for many online problems, such as online caching \cite{FKL+91} and the $k$-server problem in general metric spaces \cite{MMS88, BCL+18, Ope26}.

\vspace{-0.3em}

\paragraph{Positioning.}

Online coloring has a deep connection with structural graph theory in general, as many algorithmic improvements are made by exploiting novel subgraph structures \cite{Vis92, KYY27, BFS+26, YY26}. This problem has also been studied in the context of informational hardness \cite{Hal00, BBH+14} and parallel and distributed computing \cite{Hal97, AEL+23}. Moreover, online coloring can be applied to scheduling \cite{Erg25}, resource allocation \cite{Kie91}, and cloud computing \cite{RW24}.

\vspace{-0.3em}

\paragraph{Related work.}

In online coloring, researchers have studied a wide range of topics beyond the competitive ratio. The first topic is online coloring for $k$-colorable graphs (for a constant $k \geq 2$). On the deterministic side, Kierstead \cite{Kie98b} presented an algorithm that uses $\widetilde{O}(n^{1-1/k!})$ colors, and recently Kawarabayashi, Yoneda, and Yoneda \cite{KYY27} improved the bound to $\widetilde{O}(n^{1-2/(k(k-1))})$ colors. On the randomized side, Vishwanathan \cite{Vis92} found an algorithm that uses $\widetilde{O}(n^{1-1/(k-1)})$ colors. A large gap remains to the best-known lower bound of $\Omega(\log^{k-1} n)$ colors \cite{Vis92}.\footnote{A very recent preprint \cite{Li26} claims a lower bound of $L[1/3, \Omega(1)] = e^{\Omega((\log n)^{1/3} (\log \log n)^{2/3})}$ colors for the case $k = 3$.}

The second topic is online coloring for special graph classes. Kierstead and Trotter \cite{KT81} presented a 3-competitive online coloring algorithm for interval graphs. Trees \cite{Bea76, GL88}, bipartite graphs \cite{LST89, BBH+14, GKM+14}, planar graphs \cite{Ira94}, and bounded-treewidth graphs \cite{AS21} can be colored online using $\Theta(\log n)$ colors. For every fixed $\varepsilon > 0$, an $O(n^{\varepsilon})$-color algorithm exists for graphs with sufficiently large odd girth \cite{YY26}. First-Fit performs well on $d$-degenerate graphs; it uses only $O(d \log n)$ colors. For more results, see the survey by Kierstead \cite{Kie98a}.

A related, recently well-studied topic is \emph{online edge coloring}, in which we must color edges that arrive one by one. Edge coloring is easier than vertex coloring: a $(1+o(1))$-competitive algorithm is known for graphs with maximum degree $\Delta = \omega(\log n)$ \cite{BSV+24, BSV+25b}.

\subsection{Our Contributions}

In this paper, we improve the bounds on competitive ratios from three sides, as stated below.

\begin{theorem}\label{thm:main-1}
    Every deterministic online coloring algorithm is $\Omega(n / \log n)$-competitive.
\end{theorem}

\begin{theorem}\label{thm:main-2}
    There exists a polynomial-time randomized online coloring algorithm, against an oblivious adversary, that attains a competitive ratio of $O(n (\log \log n)^2 / \log^2 n)$. 
\end{theorem}

\begin{theorem}\label{thm:main-3}
    For every fixed $\varepsilon > 0$, there exists a polynomial-time deterministic online coloring algorithm that attains a competitive ratio of $O(n / (\log \log n)^{2-\varepsilon})$.
\end{theorem}

\begin{figure}[!b]
  \centering
  \includegraphics[width=\linewidth]{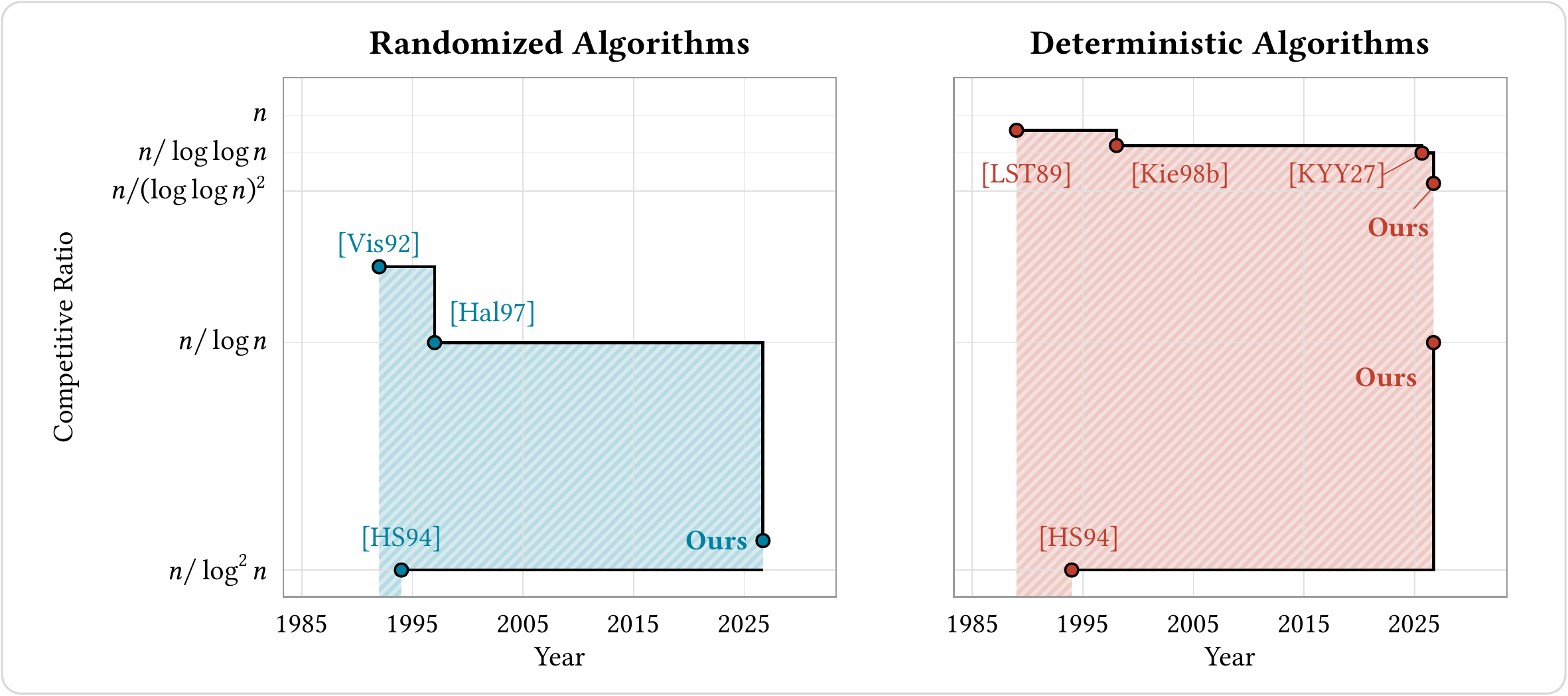}
  \caption{Historical improvements in the upper and lower bounds for randomized (left) and deterministic (right) algorithms.}
  \label{fig:103}
\end{figure}

These results represent the first substantial improvements since the 20th century. \autoref{thm:main-1} improves on Halld\'{o}rsson and Szegedy (1994) \cite{HS94}. \autoref{thm:main-2} improves on Halld\'{o}rsson (1997) \cite{Hal97}. \autoref{thm:main-3} is the first $\mathrm{poly}(\log \log n)$-factor improvement since Kierstead (1998) \cite{Kie98b}. See \autoref{fig:103} for a comparison. Consequently, the gap between the upper and lower bounds for the randomized case is almost closed; only a factor of $O((\log \log n)^2)$ remains.

Importantly, Theorems \ref{thm:main-1} and \ref{thm:main-2} show that randomization gives an asymptotic advantage in online graph coloring. Every deterministic algorithm has worst-case competitive ratio $\Omega(n/\log n)$, whereas our randomized algorithm achieves expected competitive ratio $o(n/\log n)$ against an oblivious adversary. The distinction is visible even on graphs that can be colored offline with $O(\log n)$ colors. For every deterministic algorithm, there is an input where no color is used more than twice, forcing the algorithm to use at least $n/2$ colors. In contrast, on every fixed input in this class, our randomized algorithm uses $o(n)$ colors in expectation.

In addition, to obtain \autoref{thm:main-3}, we develop a better deterministic online coloring algorithm for graphs without $r$-cliques: our algorithm uses $O(n (\log \log n / \log n)^{1/(r-2)})$ colors for every constant $r \geq 3$ (\autoref{rem:5-clique-free}). The best previously known deterministic algorithm by Brada\v{c} et al. (2026) \cite{BFS+26} uses $O(n \log^{(r-1)} n / \log^{(r-2)} n)$ colors, where $\log^{(j)} n$ denotes the $j$-fold iterated logarithm.

\subsection{Improvements on Larger Chromatic Numbers} \label{sec:1.3}

Beyond the competitive analysis, our results also improve performance and hardness for online coloring of graphs with larger chromatic numbers by a significant margin.

Historically, competitive analysis in online coloring has focused on improving algorithmic performance on graphs with small chromatic numbers, particularly $k$-colorable graphs for a constant $k \geq 3$. For the randomized side, an $\widetilde{O}(n^{1-1/(k-1)})$-color algorithm \cite{Hal97} yielded an $O(n / \log n)$-competitive algorithm. For the deterministic side, an $\widetilde{O}(2^{2^{k+3}} n^{1-2/(k(k-1))})$-color algorithm \cite{KYY27} yielded an $O(n / \log \log n)$-competitive algorithm. However, we take the opposite approach, targeting graphs with large and middle-tier chromatic numbers.

\subsubsection{Large-Chromatic Approach (\autoref{thm:main-1} and \autoref{thm:main-2})}

In \autoref{thm:main-1} and \autoref{thm:main-2}, we consider online coloring of graphs with a large chromatic number $k$, for example, when $k = n^c$ for some $c > 0$. To start, consider \autoref{alg:1-lv2}, which uses each color on at most two vertices and, whenever possible, uses a color already used exactly once. This algorithm is somewhat effective when $k$ is large; when $k = o(n)$, it uses only $n/2 + o(n)$ colors, and only an $o(1)$ fraction of the colors appear on exactly one vertex.

\begin{algorithm}[!b]
    \caption{An algorithm for online coloring.}
    \label{alg:1-lv2}
    \begin{algorithmic}[1]
        \For {each arrival of a vertex $v$}
            \If {there exists a color $c$ that is used \emph{exactly} once and is not adjacent to $v$}
                \State Color $v$ with color $c$
            \Else
                \State Color $v$ with a new color
            \EndIf
        \EndFor
    \end{algorithmic}
\end{algorithm}

\begin{proposition} \label{prop:1-lv2}
    Let $n, k \geq 1$ be integers. For a $n$-vertex graph without $(k+1)$-cliques, \autoref{alg:1-lv2} uses at most $\frac{n+k}{2}$ colors. Also, the same bound trivially holds for $k$-colorable graphs.
\end{proposition}

\begin{proof}
    Let $v_1, \dots, v_t$ be the vertices whose colors are used exactly once. These vertices must form a clique, because if $v_i$ and $v_j$ are not adjacent for some $i, j$ ($i < j$), then $v_j$ would have been colored with the same color as $v_i$, a contradiction. Hence, $t \leq k$ by assumption. Since all other colors are used exactly twice, the algorithm uses $t + \frac{n-t}{2} = \frac{n+t}{2} \leq \frac{n+k}{2}$ colors.
\end{proof}

Since \autoref{alg:1-lv2}, which uses at most two vertices per color, performs well for large chromatic numbers $k$, a natural question arises: \emph{``Can this approach be extended to three vertices per color?''} The answer is \emph{no} for deterministic algorithms but \emph{yes} for randomized algorithms.

For every deterministic algorithm, there exists an $\Theta(\log n)$-colorable graph on which the algorithm uses each color on at most two vertices and hence uses at least $n/2$ colors, as we will prove in \autoref{sec:3}. This is how we derive an $\Omega(n / \log n)$ lower bound. 

\begin{figure}[t]
    \centering
    \includegraphics[width=\linewidth]{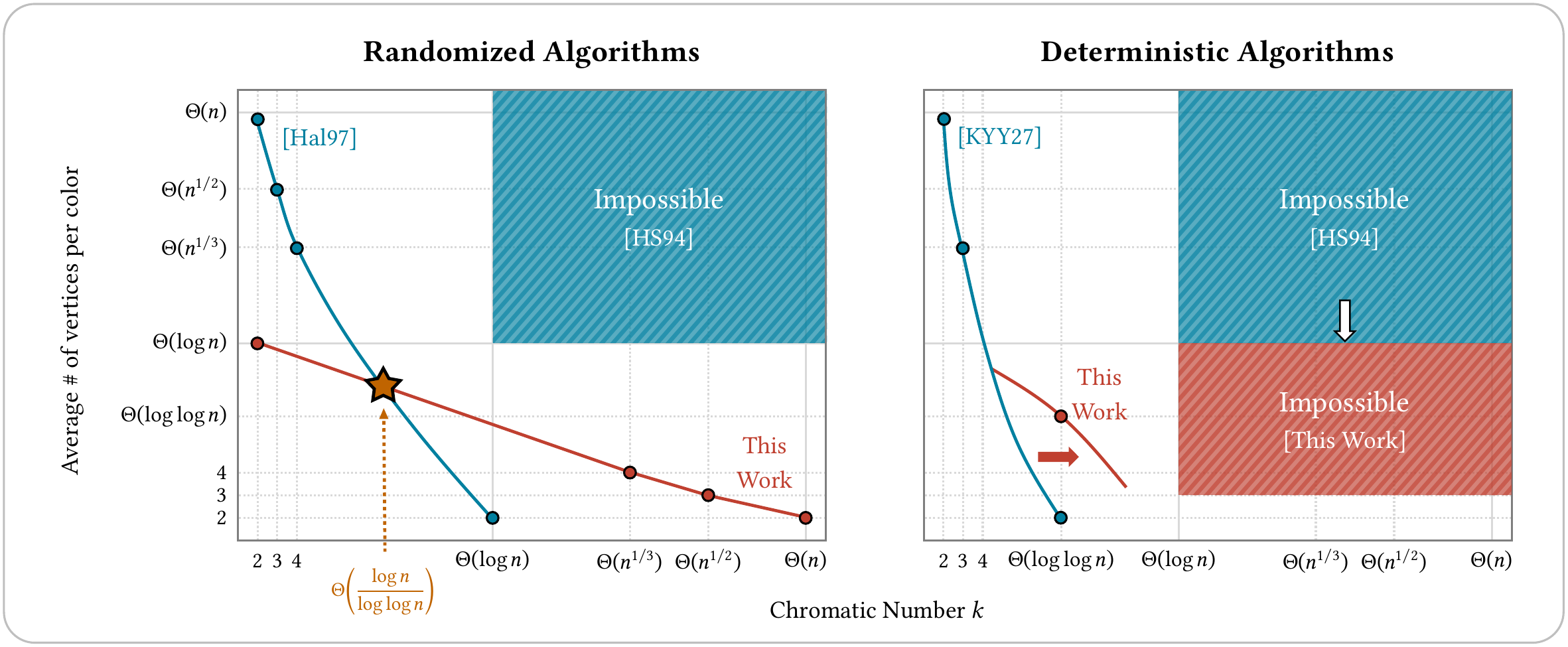}
    \caption{Performance frontiers for online coloring by chromatic number $k$, for randomized and deterministic algorithms. The previous results are shown in blue, and our results are shown in red.}
    \label{fig:104}
\end{figure}

In contrast, for randomized algorithms, our new algorithm colors any $k$-colorable graph with expected $n/3 + o(n)$ colors when $k = o(n^{1/2})$, as we prove in \autoref{sec:4.3}. In general, for every integer $t \geq 2$, there exists a randomized algorithm that colors any $k$-colorable graph with expected $n/t + o(n)$ colors when $k = o(n^{1/(t-1)})$. \autoref{fig:104} (left) shows the performance by the chromatic number $k$. Roughly speaking, the frontier of Halld\'{o}rsson's algorithm \cite{Hal97} is $t^{k-1} \approx n$, while the frontier of our algorithm is $k^{t-1} \approx n$, crossing at $k = t = \Theta(\log n / \log \log n)$. This is why combining them attains a competitive ratio of $O(n (\log \log n / \log n)^2)$.

\subsubsection{Mid-Chromatic Approach (\autoref{thm:main-3})}

To improve deterministic algorithms (\autoref{thm:main-3}), we cannot take the large-chromatic approach due to the negative lower-bound result. Instead, we take a different approach that focuses on a regime of ``moderate'' chromatic numbers $k$; see \autoref{fig:104} (right). For background, Kawarabayashi, Yoneda, and Yoneda's deterministic algorithm \cite{KYY27} uses $\Omega((\log \log n)^{1-\varepsilon/2})$ vertices per color\footnote{In fact, a stronger guarantee holds for the algorithm from \cite{KYY27}: for $k \in [1, (\log \log n)^{1-\varepsilon/2}]$, this algorithm uses $\Omega((\log n)^c)$ vertices per color for any fixed $c > 0$.} in the regime of $k \in [1, (\log \log n)^{1-\varepsilon/2}]$.

In this paper, we also achieve equivalent performance for $k \in [(\log \log n)^{1-\varepsilon/2}, (\log \log n)^{2-\varepsilon}]$. We make this possible by upgrading their algorithm with our new technical framework based on clique-free graphs; this is the main technical contribution of this paper. Using this approach, we attain a competitive ratio of $O(n / (\log \log n)^{2-\varepsilon})$.

\subsection{Organization}

\autoref{sec:1b} presents technical overviews of each algorithm. \autoref{sec:2} presents formal definitions and fundamental lemmas used in our proofs. \autoref{sec:3} presents an $\Omega(n / \log n)$ lower bound for the competitive ratio of deterministic online coloring (\autoref{thm:main-1}). \autoref{sec:4} presents an $O(n (\log \log n)^2 / \log^2 n)$-competitive algorithm for randomized online coloring (\autoref{thm:main-2}). \autoref{sec:5} presents an $O(n / (\log \log n)^{2-\varepsilon})$-competitive algorithm for deterministic online coloring (\autoref{thm:main-3}). Finally, \autoref{sec:6} concludes the paper with conjectures.

\section{Technical Overview}\label{sec:1b}

\subsection{\autoref{thm:main-1}: The Lower Bound for Deterministic Algorithms}

In the worst-case analysis of deterministic online algorithms, we can assume that an adversary determines the neighbors of each new vertex $v$ based on the algorithm's current output. To this end, we develop a strategy for the adversary that forces every (online) color to be used on at most two vertices. For a $2\ell$-colorable graph ($\ell \geq 1$), we achieve this goal through $2^{\ell}$ vertices.

\paragraph{Transparent model.}

Halld\'{o}rsson and Szegedy \cite{HS94} proved an $\Omega(n / \log^2 n)$ lower bound based on the \emph{transparent model}, in which the adversary maintains a (proper) $2\ell$-coloring $\varphi: V \to \{1, \dots, 2\ell\}$ in real time. Immediately \textbf{after} the algorithm colors a vertex $v$, the adversary must irrevocably assign $\varphi(v) \in \{1, \dots, 2\ell\}$, called a \emph{base color} (to distinguish it from the ``colors'' that the algorithm assigns). We assume that the coloring $\varphi$ is visible to the algorithm. While this model simplifies the analysis, the adversary cannot force every (online) color to appear on at most two vertices for more than $4\ell^2 + 2\ell$ vertices (\autoref{sec:3.1}).

\paragraph{New adversary.}

We modify the transparent model as follows. For the first vertex $v_1$ of a color, the adversary does not assign the base color $\varphi(v_1)$. For the second vertex $v_2$ of this color, they must assign the base colors $\varphi(v_1)$ and $\varphi(v_2)$ simultaneously. Here, we impose one rule: the adversary must assign the base color $2i-1$ to one of them and $2i$ to the other, for some $i \in \{1, \dots, \ell\}$.

\begin{figure}[!b]
  \centering
  \includegraphics[width=\linewidth]{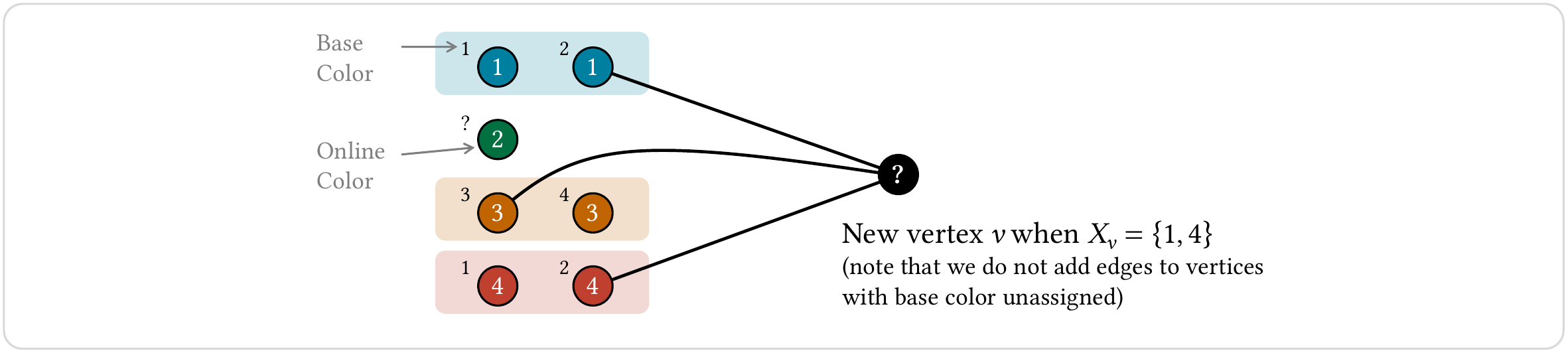}
  \caption{An example of the neighbor decision by the adversary when $X_v = \{1, 4\}$ is selected. The player cannot color $v$ with colors 1, 3, or 4, so $v$ cannot become the ``third vertex''.}
  \label{fig:201}
\end{figure}

When the adversary decides the neighbors of each vertex $v$, they choose a set $X_v \subseteq \{1, \dots, 2\ell\}$ such that exactly one of $2i-1$ and $2i$ belongs to $X_v$ for each $i \in \{1, \dots, \ell\}$, and draws edges from $v$ to all present vertices with a base color not in $X_v$. Note that there are $2^\ell$ possible sets. Then, $v$ cannot become the ``third vertex'' of any color (see \autoref{fig:201}). The problem is whether the base colors $\varphi(v_1)$ and $\varphi(v_2)$ can be validly assigned. The trick is that, as long as the sets $X_v$ ($v \in V$) are all distinct, the adversary can always assign valid base colors. Hence, the adversary can maintain the strategy for $2^\ell$ vertices. This yields an $\Omega(n / \log n)$-competitive lower bound.

\subsection{\autoref{thm:main-2}: The Upper Bound for Randomized Algorithms}

\begin{figure}[t]
  \centering
  \includegraphics[width=\linewidth]{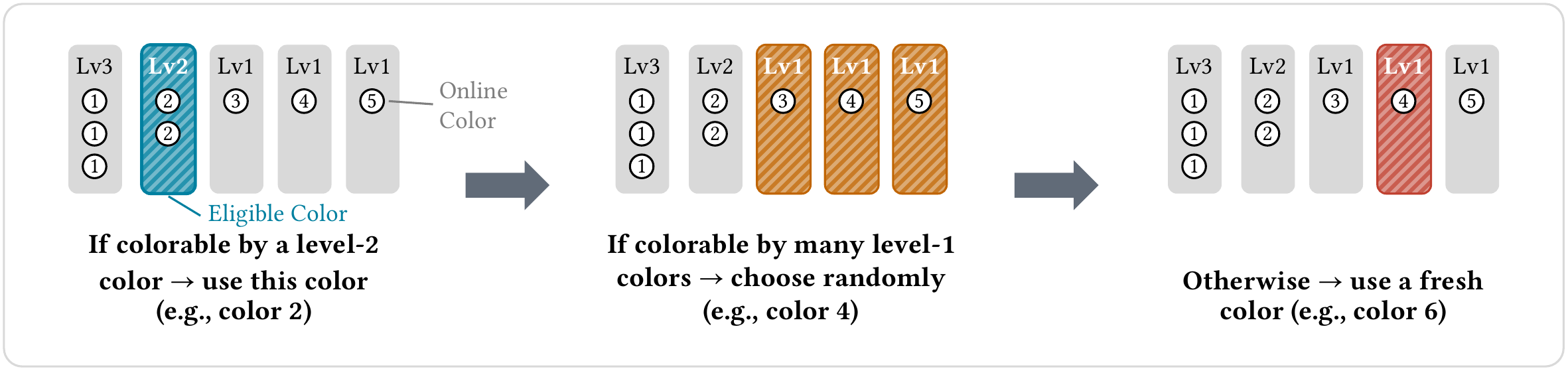}
  \caption{A sketch of our algorithm. Each bin represents the set of vertices assigned the same online color. We highlight the colors available to a new vertex $v$ (i.e., colors not used by its neighbors).}
  \label{fig:203}
\end{figure}

We extend \autoref{alg:1-lv2} to ``three vertices per color'' using randomization. For convenience, we call a color used on exactly $\ell$ vertices a ``level-$\ell$ color''. Given a parameter $\theta \geq 1$, consider the following three-layer algorithm: when each vertex $v \in V$ arrives, (1) if $v$ can be colored with a level-2 color, use that color, (2) otherwise, if $v$ can be colored with at least $\theta$ level-1 colors, randomly choose one of them, (3) otherwise, use a fresh color. See \autoref{fig:203} for an illustration of the algorithm.

\paragraph{Analysis.}

We analyze the expected number of colors used for a $k$-colorable graph. It is easy to see that there are at most $k\theta$ level-1 colors (\autoref{lem:4-lv3-onecolor}). We then analyze the expected number of level-2 colors. To this end, we bound the probability that the final number of level-2 colors is exactly $m$. Consider the scenario in which the vertices with level-2 colors are listed as
\begin{equation*}
    (a_1, b_1), \dots, (a_m, b_m) \quad (a_1 \prec b_1, \dots, a_m \prec b_m; b_1 \prec \dots \prec b_m),
\end{equation*}
where $a_i$ and $b_i$ have the same color for each $i \in \{1, \dots, m\}$. Here, $x \prec y$ means that vertex $x$ arrives earlier than $y$. This exact scenario happens with probability at most $1/\theta^m$ (\autoref{lem:4-lv3-prob}), because when each vertex $b_i$ arrives, the algorithm must choose the same color as vertex $a_i$, which happens with probability at most $1/\theta$ due to step (2) of the algorithm. On the other hand, the number of lists $(a_1, b_1), \dots, (a_m, b_m)$ that occur with non-zero probability can be shown to be at most $k^m \binom{n}{m}^2$ (\autoref{lem:4-lv3-count}). Thus, the probability that there are exactly $m$ level-2 colors is at most
\begin{equation*}
    \frac{1}{\theta^m} \cdot \left\{k^m \cdot \binom{n}{m}^2\right\} \leq \left(\frac{e^2 k n^2}{\theta m^2}\right)^m.
\end{equation*}
This bound rapidly approaches zero beyond $m = en\sqrt{k/\theta}$. Hence, the expected number of level-2 colors is $O(n\sqrt{k/\theta})$ (\autoref{lem:4-lv3-twocolor}). As a result, when $k = o(\sqrt{n})$, only an $o(1)$ fraction of colors are at levels 1 or 2 with parameter $\theta = \sqrt{n}$. In this case, we use $n/3 + o(n)$ colors in expectation.

\paragraph{Generalization.}

We further generalize the algorithm to general $d \geq 3$ layers. With an appropriate parameter setting that uses a different $\theta$ for each layer, we use $n/d + o(n)$ colors in expectation when $k = o(n^{1/(d-1)})$, for any fixed $d$. This result leads to an $O(n (\log \log n)^2 / \log^2 n)$-competitive algorithm by combining with Halld\'{o}rsson's algorithm \cite{Hal97}; see also the discussion in \autoref{sec:1.3}.

\subsection{\autoref{thm:main-3}: The Upper Bound for Deterministic Algorithms}

We present an $O(n / (\log \log n)^{2-\varepsilon})$-competitive deterministic algorithm. Let $m := (\log \log n)^{1-\varepsilon/2}$. Kawarabayashi, Yoneda, and Yoneda's algorithm \cite{KYY27} colors $m$-colorable graphs online using $O(n/m^2)$ colors. Our goal is to upgrade their algorithm to color $m^2$-colorable graphs using $O(n/m)$ colors; this suffices to show the competitive ratio. We clarify the issues that arise in the $m^2$-colorable case in their algorithm, then present our new technical framework to resolve them.

\subsubsection{The KYY27 Framework and its Limitation}

First, we describe an algorithm to color 3-colorable graphs using $O(n^{1-c})$ colors for a small $c > 0$ \cite{Kie98b}. We color each vertex $v$ using First-Fit if it can be colored with one of the first $n^{1-c}$ colors. If $v$ cannot be colored in this way, then observe that $v$ has degree at least $n^{1-c}$. We color these remaining vertices by exploiting their large degrees, as follows.

\begin{figure}[!b]
  \centering
  \includegraphics[width=\linewidth]{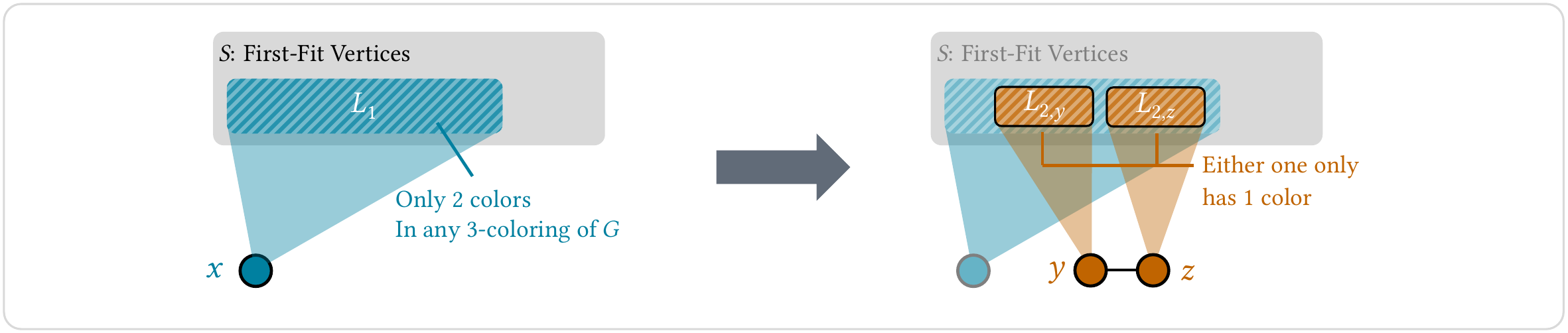}
  \caption{A sketch of an algorithm that colors 3-colorable graphs using $O(n^{1-c})$ colors.}
  \label{fig:204}
\end{figure}

\begin{enumerate}
    \item Let $S$ be the set of vertices colored by First-Fit. For a fixed $x \in V \setminus S$, let $L_1 := N_S(x)$. Then, in some 3-coloring of $G$, at most 2 colors appear in $L_1$ (\autoref{fig:204}, left).
    \item In addition, for a fixed edge $\{y, z\}$ with $y, z \in V \setminus S$, let $L_{2,y} := N_{L_1}(y)$ and $L_{2,z} := N_{L_1}(z)$. Then, in some 3-coloring of $G$, only one color appears in either $L_{2,y}$ or $L_{2,z}$ (\autoref{fig:204}, right).\footnote{This is because for any 3-coloring $\varphi : V \to \{1, 2, 3\}$, either $\varphi(x) \neq \varphi(y)$ or $\varphi(x) \neq \varphi(z)$ holds. If $\varphi(x) \neq \varphi(y)$, then $L_{2,y}$ consists of 1 color; if $\varphi(x) \neq \varphi(z)$, then $L_{2,z}$ consists of 1 color.}
    \item This means that either $G[N(L_{2,y})]$ or $G[N(L_{2,z})]$ is 2-colorable (bipartite). It is known that bipartite graphs can be colored online using only $O(\log n)$ colors \cite{LST89}. Moreover, in favorable cases, both $L_{2,y}$ and $L_{2,z}$ are large (due to the large degree of each vertex in $V \setminus S$), and hence many future vertices belong to $N(L_{2,y})$ or $N(L_{2,z})$. As a result, we use only $O(n^{1-c})$ colors. (To achieve $O(n^{1-c})$ colors in all cases, \cite{Kie98b} used some additional techniques.)
\end{enumerate}

Next, we explain how \cite{KYY27} generalizes this algorithm to $k$-colorable graphs for general $k$. The crucial observation is that in the algorithm above, we obtain a ``small'' 1-chromatic subgraph (the vertex $x$) in Step 1 and a ``small'' 2-chromatic subgraph (the edge $\{y, z\}$) in Step 2. Therefore, they introduce an online subroutine $\textsf{LOC}(k)$ that either colors a graph effectively or terminates by returning a ``small'' subset $X \subseteq V$ such that $\chi(G[X]) > k$. In \cite{KYY27}, similarly to Steps 1--3 above, $\textsf{LOC}(k)$ is a recursive algorithm that calls $\textsf{LOC}(0), \textsf{LOC}(1), \dots, \textsf{LOC}(k-1)$.

However, \cite{KYY27} cannot effectively color $m^2$-colorable graphs due to this recursive structure. Let $f_k$ be an upper bound on the size of a subgraph returned by \textsf{LOC}$(k)$. Then, $f_k \approx f_0 \times f_1 \times \dots \times f_{k-1}$, as the returned subgraph is recursively constructed using the outputs of $\textsf{LOC}(0), \dots, \textsf{LOC}(k-1)$. Therefore, $f_k$ grows double-exponentially and ultimately exceeds $n$ at $k = \Theta(\log \log n) \leq m^2$. The required number of colors for $k$-colorable graphs increases in proportion to $f_k$, so \cite{KYY27} is unsuitable for $m^2$-colorable graphs.

\subsubsection{Our Techniques}

To address this issue, we attempt to ``shorten'' the recursion: we try to solve $\textsf{LOC}(k)$ by calling only $\textsf{LOC}(0), \dots, \textsf{LOC}(k-m)$. This change would make $f_k \approx f_0 \times f_1 \times \dots \times f_{k-m}$, and thus $f_{m^2} = n^{o(1)}$, which is sufficiently small. We achieve this by introducing a new algorithm that colors $(m+1)$-clique-free graphs online with $O(n/m)$ colors (better than \cite{BFS+26}) and integrating it into the recursive framework of \cite{KYY27}.

\paragraph{Clique-free algorithm.} Our recursive algorithm for coloring $k$-clique-free graphs $(2 \leq k \leq m + 1)$ is as follows. First, prepare a limited number of bins $B_1, B_2, \dots$, ensuring each bin is $(k-1)$-clique-free. Then, we can color the vertices in each bin using our (recursively constructed) algorithm for $(k-1)$-clique-free graphs, using disjoint colors for different bins.

\begin{figure}[!b]
  \centering
  \includegraphics[width=\linewidth]{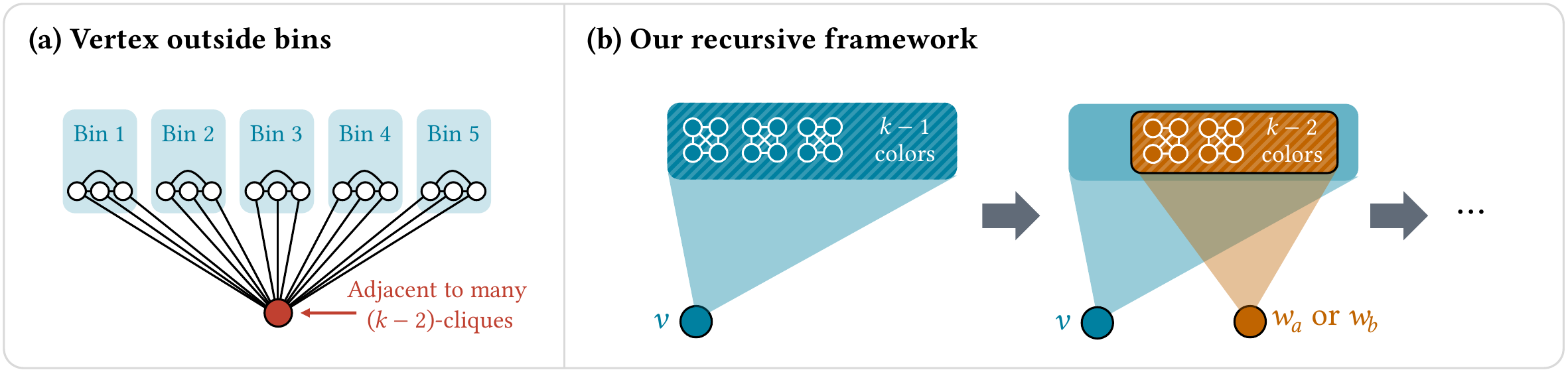}
  \caption{(a) The state when a new vertex $v$ cannot be placed in any bin. (b) The framework of our new algorithm, which iteratively restricts the number of colors allowed across a collection of cliques.}
  \label{fig:205}
\end{figure}

When a new vertex $v$ arrives, we place $v$ in a bin if a suitable bin exists. Otherwise, we color $v$ using the fact that $v$ is adjacent to many $(k-2)$-cliques; see \autoref{fig:205} (a). Consider two different vertices $x$ and $y$ outside the bins. If there is at least one $(k-2)$-clique such that both $x$ and $y$ are adjacent to all its vertices, then $x$ and $y$ are not adjacent, because otherwise the graph would not be $k$-clique-free. We exploit this no-edge property to color vertices outside the bins effectively (see \autoref{sec:5.2}--\autoref{sec:5.3} for details).

\paragraph{Shortening the recursion.} To use the clique-free algorithm in the main problem, instead of First-Fit, we begin with a greedy-bin algorithm similar to the clique-free algorithm, in which each bin is $(m+1)$-clique-free. Then, we can force any arriving vertices outside bins to be adjacent to many $m$-cliques. Using a recursive framework similar to that of Steps 1--3, we iteratively restrict the number of colors allowed across a collection of cliques (\autoref{fig:205} (b)).

Ultimately, we aim to create a structure in which a collection of $m$-cliques, $S_1, S_2, \dots, S_p$, shares the same set of $m$ colors in some $k$-coloring of $G$. Then, every future vertex that is adjacent to (all $m$ vertices of) at least one of these cliques must use the remaining $k-m$ colors in a $k$-coloring of $G$, which is why the recursion stops at \textsf{LOC}$(k-m)$. We give the details in \autoref{sec:5.6}--\autoref{sec:5.11}.

\section{Preliminaries}\label{sec:2}

We present notation, formal definitions, and a fundamental tool required in the main proof.

\paragraph{Notation.}

Let $G = (V, E)$ be a graph. The set of vertices adjacent to a vertex $v \in V$ is denoted by $N(v)$. The set of vertices adjacent to at least one vertex in a vertex set $S$ is denoted by $N(S)$, i.e., $N(S) = \bigcup_{v \in S} N(v)$. For a set $X \subseteq V$, we also denote $N_X(v) = N(v) \cap X$ and $N_X(S) = N(S) \cap X$. We denote $G[S]$ as the subgraph induced by $S \subseteq V$. We denote the chromatic number by $\chi(G)$. For a positive integer $n$, we denote $[n] = \{1, \dots, n\}$. We generally assume base 2 for the logarithm.

\begin{definition}[Online Coloring]\label{def:2-online-coloring}
    In the online coloring problem, an undirected graph $G = (V, E)$ starts empty; i.e., it has no vertices. The vertices of $G$ are given one by one. Immediately after a vertex $v$ arrives (along with all edges incident to previously arrived vertices), we must assign a color $c(v) \in \mathbb{N}$ that differs from the colors of all adjacent vertices. The goal is to minimize the number of colors used, $|\{c(v) : v \in V\}|$.
\end{definition}

\begin{definition}[Competitive Ratio]
    Consider a minimization problem. For an instance $I$ of this problem, let $\mathrm{ALG}(I)$ be the performance of an online algorithm and let $\mathrm{OPT}(I)$ be the optimal (offline) performance. Then, the competitive ratio of this online algorithm is $\max_I \left\{\mathrm{ALG}(I) / \mathrm{OPT}(I)\right\}$ for deterministic algorithms, and $\max_I \left\{\mathbb{E}[\mathrm{ALG}(I)] / \mathrm{OPT}(I)\right\}$ for randomized algorithms against an oblivious adversary.
\end{definition}

\paragraph{Assuming $n$ and $k$ are known in advance.}

To compute the competitive ratio of the online coloring algorithms, we must consider the case where the number of vertices $n$, and the chromatic number $k$, are not known in advance. However, by the following lemma, assuming that $n$ and $k$ are known in advance does not change the asymptotic order of the competitive ratio. Note that this lemma is used in \autoref{sec:4} and \autoref{sec:5}, but not in \autoref{sec:3}.

\begin{lemma}
    \label{lem:2-competitive-ratio} 
    Let $f: \mathbb{N} \to \mathbb{R}_{\geq 0}$ be a non-decreasing function that satisfies $f(n) = \Omega(n^{1/100})$ and $f(n) \leq n$.
    \begin{itemize}
        \item[(i)] For every $n, k \geq 1$, suppose that there is a deterministic online coloring algorithm $\mathcal{A}_{n, k}$ that colors any $n$-vertex $k$-colorable graphs within $k \cdot f(n)$ colors when $n$ and $k$ are known in advance. Then, there is a $96 f(n)$-competitive deterministic algorithm for online coloring. Moreover, the algorithm runs in polynomial time if each algorithm $\mathcal{A}_{n, k}$ is run in polynomial time (with respect to $n$).
        \item[(ii)] The same claims also apply for randomized algorithms against an oblivious adversary, with the color bounds in both the assumption and the conclusion interpreted in expectation.
    \end{itemize}
\end{lemma}

\begin{proof}
    Consider (ii) the randomized case. For simplicity, assume that the final number of vertices $n$ is known in advance at this point. Consider the following algorithm.
    
    \begin{quote}
        For $d = 0, 1, 2, \dots$, we process the algorithm $\mathcal{A}_{n, 2^d}$ while it uses at most $2^{d+2}  \cdot f(n)$ colors. If coloring $v$ would cause the number of colors exceeding $2^{d+2} \cdot f(n)$, we finish $\mathcal{A}_{n, 2^d}$ and process later vertices with $\mathcal{A}_{n, 2^{d+1}}$, starting from $v$. We use disjoint colors for different $d$.
    \end{quote}
    
    \noindent
    We analyze the number of colors used. Let $q := \lceil \log_2 \chi(G) \rceil$ for the final graph $G$. Then, for $i \geq 0$, the algorithm $\mathcal{A}_{n, 2^{q+i}}$ uses expected $2^{q+i} f(n)$ colors, because $\chi(G) \leq 2^{q+i}$. By Markov's inequality, it uses more than $2^{q+i+2} f(n)$ colors with probability at most $1/4$. Therefore, the algorithm $\mathcal{A}_{n, 2^{q+i}}$ starts with probability at most $1/4^i$. The total expected number of colors is
    \begin{align*}
        \sum_{i=0}^{q-1} 2^{i+2} f(n) + \sum_{i=0}^\infty \left(\frac{1}{4^i} \cdot 2^{q+i+2} f(n)\right)
        & \leq 2^{q+2} f(n) + 2^{q+3} f(n) \\
        & = 12 \cdot 2^q f(n) \leq 24 \chi(G) f(n).
    \end{align*}

    \noindent
    Now, consider the case where $n$ is not known in advance. Let $\mathcal{B}_n$ be the algorithm above when $n$ is known in advance. Consider the following algorithm:
    
    \begin{quote}
        Let $t_i$ be the largest integer such that $f(t_i) \leq 2^i$ (or $t_i = 0$ if no such integer exists). For the next $t_i$ vertices or until the last vertex arrives, we use the algorithm $\mathcal{B}_{t_i}$. We use completely disjoint colors for different layer $i$.
    \end{quote}
    
    \noindent
    Let $i' = \lceil \log_2 f(n) \rceil$. Since $f(n) \leq 2^{i'}$, we have $t_{i'} \geq n$, so the process $i > i'$ never occurs. Let $c_i$ be the number of colors used in the layer $i$. Since $\mathbb{E}[c_i] \leq 24\chi(G)f(t_i)$, the expected total number of colors used is
    \begin{equation*}
        \mathbb{E}\left[\sum_{i=0}^{i'} c_i\right] \leq \sum_{i=0}^{i'} 24\chi(G)f(t_i) \leq 24\chi(G) \sum_{i=0}^{i'} 2^i \leq 24\chi(G) \cdot 2^{i'+1} \leq 96\chi(G)f(n).
    \end{equation*}
    \noindent
    Therefore, we attain an (expected) competitive ratio of $96f(n)$. The deterministic case can also be proved analogously. Finally, this algorithm run in polynomial time because $\mathcal{B}_k$ is obviously polynomial time and $t_i = O(n^{100})$ when $n$ is large enough.
\end{proof}

\section{The Lower Bound for Deterministic Algorithms}\label{sec:3}

In this section, we prove that any deterministic online coloring algorithm has competitive ratio $\Omega(n / \log n)$. To this end, we consider to what extent we can ``force'' the algorithm to use each color in at most two vertices.

\subsection{Halld\'{o}rsson and Szegedy's Model and its Limitations}\label{sec:3.1}

The worst-case performance of deterministic online algorithms can be discussed in the context of an \emph{adversarial attack} against a player. In online coloring, the \emph{adversary} decides the set of incident edges of each new vertex $v$, adaptively, by looking at the player's coloring so far.

\begin{figure}[!b]
  \centering
  \includegraphics[width=\linewidth]{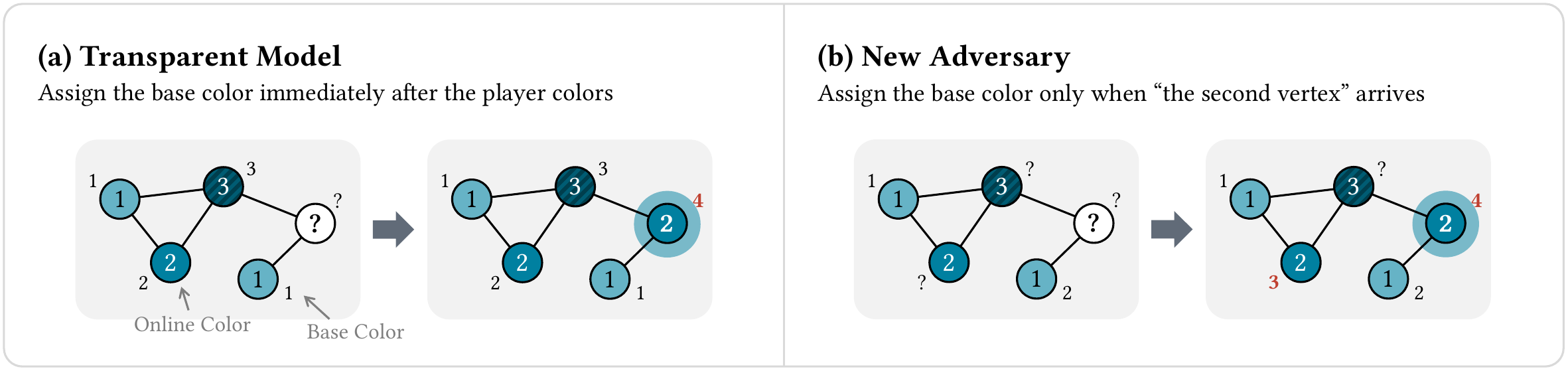}
  \caption{Examples of (a) transparent model and (b) our new model. We write \texttt{?} for the vertices whose base color is currently unassigned. For the new model, we leave the base color unassigned for vertices whose online colors are currently used in just one vertex.}
  \label{fig:301}
\end{figure}

\paragraph{Transparent model.}
We assume that the graph is $k$-colorable. To obtain a lower bound on the competitive ratio, Halld\'{o}rsson and Szegedy (1994) \cite{HS94} considered a model in which the adversary maintains a $k$-coloring of the graph, $\varphi: V \to \{1, \dots, k\}$, online. To distinguish it from the colors that the player assigns, we call the colors used in $\varphi$ the \emph{base colors}.
\begin{description}
    \item[Step 1.] The adversary decides the set of edges incident to a new vertex $v$.
    \item[Step 2.] The player assigns its color $c(v) \in \mathbb{N}$.
    \item[Step 3.] The adversary decides its base color $\varphi(v) \in \{1, \dots, k\}$ in response to the player. This must differ from the base colors of all its neighbors. $\varphi(v)$ \textbf{cannot be changed} in the future.
\end{description}
This is called the \emph{transparent model} because we assume that \textbf{the player can also see the $k$-coloring $\varphi$}. Apparently, the best strategy in Step 1 is to choose an admissible set of base colors
\begin{equation*}
    X_v \subseteq \{1, \dots, k\}: \text{the set of base colors that can be assigned to $v$ (in Step 3)}
\end{equation*}
and make all the previous vertices with base colors in $\{1, \dots, k\} \setminus X_v$ adjacent to $v$, because player can also see $\varphi$. See \autoref{fig:301} (a) for the illustration of this model.

Compared to ordinary online coloring, the transparent model is \emph{handicapped} for the adversary; indeed, the adversary must irrevocably assign each base color in Step 3, and the player is given additional information about base coloring. However, it can still force the player to use many colors, as shown below, implying an $\Omega(n / \log^2 n)$ lower bound on the competitive ratio.

\begin{theorem}[\cite{HS94}]
    \label{thm:3-hs94}
    Let $k \geq 1$ be an integer. In the transparent model, the adversary has a strategy that forces the player to use $2^k - 1$ colors on a $k$-colorable graph with $k \cdot 2^{k-1}$ vertices.
\end{theorem}

\paragraph{Limitations.}
On the other hand, the player has a strategy that uses at most $2^k - 1$ colors for any $k$-colorable input in the transparent model, which shows the tightness of the result above. In particular, \cite[Section 4.5]{HS94} discusses a player strategy that yields the following outcome.

\begin{theorem}[\cite{HS94}, restated] \label{thm:3-hs94-player}
    For each color $c \in \mathbb{N}$, let $T_c \subseteq \{1, \dots, k\}$ be the current set of base colors used on vertices of an (online) color $c$. There exists a strategy for the player that maintains the condition that ``the sets $T_c$ are distinct over all present (online) colors $c$''.
\end{theorem}

This also implies that $\Theta(n / \log^2 n)$ is indeed the optimal competitive ratio for the transparent model \cite{HS94}. More importantly, under the transparent model, the adversary cannot force the player to use each color in at most two vertices for more than $k(k+1)$ vertices. This is due to \autoref{thm:3-hs94-player} and that there are $\frac{1}{2} k(k+1)$ distinct non-empty subsets of $\{1, \dots, k\}$ with size at most two. This $O(k^2)$-vertex limit is far from our exponential goal. To obtain a better bound, we need a new model that does not decide the $k$-coloring $\varphi$ immediately.

\subsection{New Adversary Model}\label{sec:3.2}

We consider an adversary strategy to avoid creating the ``third vertex'' of the same color. The goal of this subsection is to prove the following lemma.

\begin{lemma} \label{lem:3-main}
    Let $\ell \geq 1$ be an integer. The adversary has a strategy that forces the player to use each color in at most two vertices on a $2\ell$-colorable graph with $2^\ell$ vertices.
\end{lemma}

\begin{proof}
    Below, we give a simple proof. Assuming each color is used in at most two vertices, we say that a vertex is \emph{paired} if another vertex has the same color, and otherwise \emph{unpaired}.
    
    \paragraph{The new model.} We modify the transparent model by postponing the assignment of base colors while constructing a $2\ell$-coloring $\varphi$. When the first vertex $v_1$ of a certain color arrives, we do not fix its base color $\varphi(v_1)$. When the second vertex $v_2$ of this color arrives, we fix the base colors of both paired vertices, $\varphi(v_1)$ and $\varphi(v_2)$, simultaneously. When deciding the base colors of two newly paired vertices, we impose a new rule: the adversary must choose $s \in \{1, \dots, \ell\}$ and assign base color $2s-1$ to one of them and $2s$ to the other (e.g., $s = 2$ for \autoref{fig:301} (b)).
    
    \begin{figure}[!b]
      \centering
      \includegraphics[width=\linewidth]{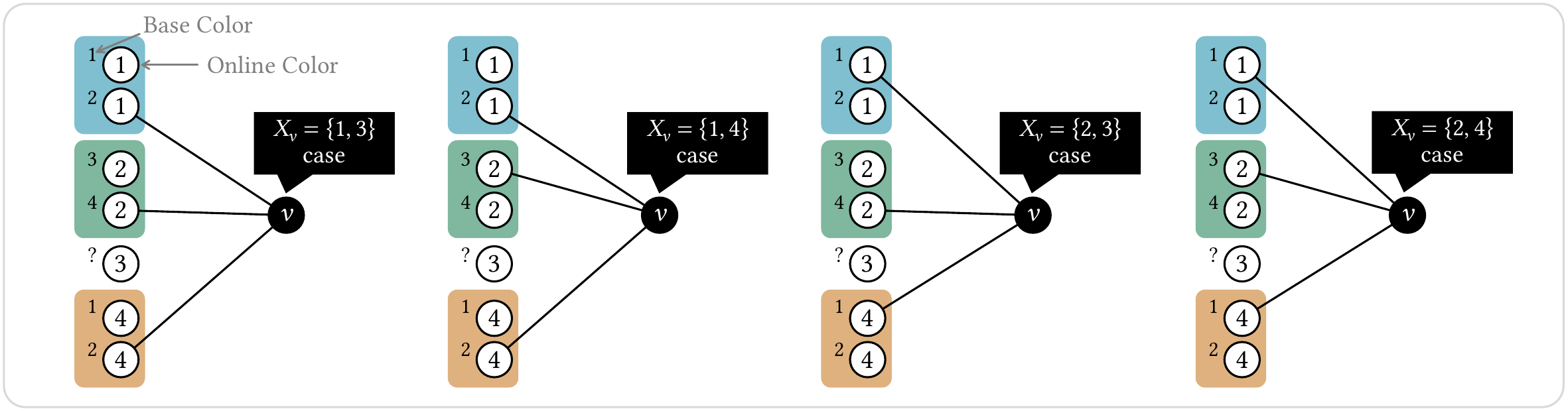}
      \caption{The edges incident to $v$ in four cases of $X_v$. In all four cases, the player cannot select online colors already used in two vertices (colors 1, 2, 4), which makes impossible to create the ``third vertex''. The graphs are created for explanation and contains more than $2^\ell$ vertices.}
      \label{fig:302}
    \end{figure}

    \paragraph{Adversary algorithm.} We consider the adversary proceeding as follows:

    \begin{itemize}
        \item When deciding the set of edges incident to each vertex $v$, similar to the transparent model, they first choose the set of its \emph{admissible base colors} $X_v \subseteq \{1, \dots, 2\ell\}$, where for each $s \in \{1, \dots, \ell\}$, exactly one of $\{2s-1, 2s\}$ must be in $X_v$. Then, they make all the previous \textbf{paired} vertices with base colors in $\{1, \dots, 2\ell\} \setminus X_v$ adjacent to $v$. No edges are added to unpaired vertices. With this method, it is impossible for the player to select an online color that has already been used on two vertices (see \autoref{fig:302} for examples). Importantly, the adversary must choose different $X_v$ for each vertex $v$; however, this is possible because $X_v$ has $2^\ell$ possible subsets.
        
        \item When deciding the base color of a new pair of vertices $v_1$ and $v_2$, we choose $s$ such that $2s-1$ is in exactly one of $\{X_{v_1}, X_{v_2}\}$, and $2s$ is in the other. Then, if $2s-1 \in X_{v_1}$, we assign $\varphi(v_1) \leftarrow 2s-1$ and $\varphi(v_2) \leftarrow 2s$ (and the reverse if $2s \in X_{v_1}$). Note that such an $s$ always exists due to $X_{v_1} \neq X_{v_2}$ (by the distinctness of $X_v$). This method always satisfies the condition written at the end of ``The new model’’ section.
    \end{itemize}

    \paragraph{Correctness.} Finally, we show the constructed graph is always $2\ell$-colorable. For analysis, for each vertex $v$ whose base color remains unassigned, consider assigning $\varphi(v)$ to an arbitrary element of $X_v$. Then, $\varphi(u_1) \neq \varphi(u_2)$ always holds for every two adjacent vertices $u_1, u_2$ (where $u_1$ arrives earlier than $u_2$) whose base colors are assigned. The reason is as follows:

    \begin{quote}
        First, $\varphi(u_2) \in X_{u_2}$ holds by our assignment rule. Second, $\varphi(u_1) \notin X_{u_2}$ holds. This is because $u_1$ must already be paired when $u_2$ arrives (due to ``The new model’’) and our edge-construction rule. Thus, $\varphi(u_1) \neq \varphi(u_2)$.
    \end{quote}

    \noindent
    Since $\varphi(v) \in \{1, \dots, 2\ell\}$ for all $v$, the constructed graph is $2\ell$-colorable. The player cannot use the same online color on three vertices (see ``Adversary algorithm’’ paragraph), so \autoref{lem:3-main} holds.
\end{proof}

\subsection{Competitive Ratio}\label{sec:3.3}

With \autoref{lem:3-main}, we can immediately derive an $\Omega(n / \log n)$ lower bound on the competitive ratio.

\begingroup
\def\thetheorem{\ref{thm:main-1}}
\begin{theorem}
    Every deterministic online coloring algorithm is $\Omega(n / \log n)$-competitive.
\end{theorem}
\addtocounter{theorem}{-1}
\endgroup

\begin{proof}
    We prove that the competitive ratio is $\Omega(n / \log n)$ even if the final number of vertices $n \geq 2$ is known. We apply \autoref{lem:3-main} with $\ell = \lceil \log_2 n \rceil$; note that $n \leq 2^{\ell}$. Then, for every deterministic algorithm, there exists an input of an $n$-vertex $2\ell$-colorable graph such that this algorithm uses at least $n/2$ colors. This gives a lower bound of $\frac{n/2}{2 \lceil \log_2 n \rceil} = \Omega(n / \log n)$ on the competitive ratio.
\end{proof}

\section{The Upper Bound for Randomized Algorithms}\label{sec:4}

In this section, we present a randomized online coloring algorithm that attains a competitive ratio of $O(n (\log \log n)^2 / (\log n)^2)$, assuming an oblivious adversary. This significantly closes the gap to an $\Omega(n / \log^2 n)$ lower bound given by Halld\'{o}rsson and Szegedy \cite{HS94}.

\subsection{Can We Use the Third Vertex?}

In the introduction, we showed that \autoref{alg:1-lv2} uses only $n/2 + o(n)$ colors, with each color on at most two vertices, if the chromatic number is not too large (\autoref{prop:1-lv2}). This raises a natural question: \emph{``if the chromatic number is not too large, can we use only $n/3 + o(n)$ colors, each on at most three vertices?''} This means that most colors must be used exactly three times.

In \autoref{sec:3}, we have proved that this is impossible for deterministic algorithms: even for $O(\log n)$-colorable graphs, the adversary can force all colors to be used on at most two vertices. However, for randomized algorithms, the answer is yes for graphs with chromatic number $o(n^{1/2})$, and we show our intuition below.

\paragraph{The power of randomization.} First, consider the most natural extension of \autoref{alg:1-lv2}. For each vertex $v \in V$, (1) use a color used in two vertices if possible; (2) otherwise, use a color used in one vertex if possible; (3) otherwise, use a fresh color. Unfortunately, this algorithm fails on a bipartite graph example given in the Introduction section (\autoref{fig:101}), where no color is used on three vertices.

However, with the power of randomization, this barrier can be addressed. Consider a new algorithm that occasionally uses a fresh color even if (2) is possible. Then, in the previous example (\autoref{fig:101}), many colors can be used on three vertices. This is because the randomization changes ``which pairs of two vertices end up sharing the same color'' (note that, in the previous deterministic case, the pairing is always between the $(2i-1)$-th arrived vertex and the $2i$-th arrived vertex, and this is exactly why we cannot use the same color in three vertices). This is the intuition for why randomization is strong.

In the following two subsections, we introduce an algorithm that achieves $n/3+o(n)$ colors in expectation when the chromatic number is $o(n^{1/2})$.

\subsection{Notation and Assumptions} \label{sec:4.2}

We use $G^* = (V^*, E^*)$ to denote the final graph, distinguishing it from the current graph used for online coloring. For $v_1, v_2 \in V^*$, we write $v_1 \prec v_2$ if $v_1$ arrives before $v_2$. We use the constant $e \approx 2.718$. The following definitions are extensively used in the following subsections.
\begin{quote}
    Let $\ell \geq 1$ be an integer. We call a color that is (currently) used in exactly $\ell$ vertices ``a level-$\ell$ color''. We call a vertex with a level-$\ell$ color ``a level-$\ell$ vertex''.
\end{quote}
We assume that $G^*$ is $k$-colorable. For the analysis, we fix a $k$-coloring $\varphi: V^* \to \{1, \dots, k\}$ of $G^*$. To distinguish it from the (online) colors used in the algorithm, we call the colors used in $\varphi$ the ``base colors''. In addition, we say that an (online) color $c$ is ``eligible'' for a new vertex $v$ if and only if $v$ is not adjacent to any previously arrived vertex whose online color is $c$.

\subsection{Three-Level Algorithm}\label{sec:4.3}

We consider the following ``three-level'' algorithm, which upgrades the ``two-level'' algorithm of \autoref{alg:1-lv2}. We use an integer parameter $\theta \geq 1$ as a threshold. For each vertex $v \in V^*$:
\begin{enumerate}
    \item If there is an eligible level-2 color, we assign this color to $v$.
    \item Otherwise, if there are at least $\theta$ eligible level-1 colors, we choose one of them uniformly at random and assign this color to $v$.
    \item Otherwise, we assign a fresh color to $v$.
\end{enumerate}
See \autoref{fig:203} for the illustration. With an appropriate setting of parameter $\theta$, this algorithm uses only $n/3 + o(n)$ colors in expectation for $k$-colorable graphs with $k = o(\sqrt{n})$. Below, we analyze this algorithm by estimating the expected number of level-1 and level-2 colors.

\begin{lemma} \label{lem:4-lv3-onecolor}
    In the final state, there are at most $k\theta$ level-1 colors.
\end{lemma}

\begin{proof}
    If there are $k\theta + 1$ or more level-1 vertices, the pigeonhole principle implies there are $\theta+1$ level-1 vertices $v_1, \dots, v_{\theta+1}$ (in arriving order) with the same base color $\varphi(v_i)$. Since $v_{\theta+1}$ is not adjacent to $v_1, \dots, v_{\theta}$, the vertex $v_{\theta+1}$ should have been colored by Step 2 (or Step 1), which is a contradiction. Therefore, \autoref{lem:4-lv3-onecolor} holds.
\end{proof}

\paragraph{Bounding level-2 colors.} Next, we bound the number of level-2 colors. First, we list the vertices of the (final) level-2 colors as
\begin{equation*}
    (a_1, b_1), \dots, (a_m, b_m) \quad (a_1 \prec b_1, \dots, a_m \prec b_m; b_1 \prec \dots \prec b_m),
\end{equation*}
where $a_i$ and $b_i$ have the same color for each $i \in \{1, \dots, m\}$. Below, we refer to this list as ``the pairing''. We upper bound the probability that there are exactly $m$ level-2 colors in the following way: (i) bounding the probability that a fixed pairing of size $m$ occurs (\autoref{lem:4-lv3-prob}), and (ii) bounding the number of ``valid'' pairings of size $m$ (\autoref{lem:4-lv3-count}).

\begin{lemma} \label{lem:4-lv3-prob}
    The probability that the pairing becomes $(a_1, b_1), \dots, (a_m, b_m)$ is at most $1/\theta^m$.
\end{lemma}

\begin{figure}[!b]
    \centering
    \includegraphics[width=\linewidth]{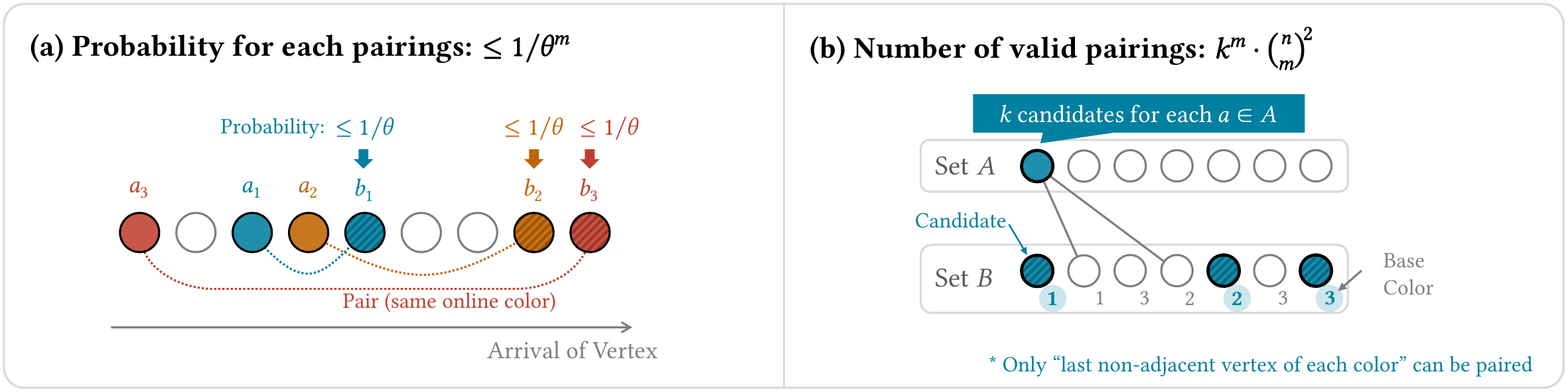}
    \caption{A sketch of our proof deriving the probability that there are exactly $m$ level-2 colors. In (b), each line between two vertices represents an edge of the graph.}
    \label{fig:401}
\end{figure}

\begin{proof}
    When vertex $b_i$ arrives, the probability that the same color as $a_i$ is chosen to become a level-2 vertex is at most $1/\theta$, since there are at least $\theta$ color choices (see \autoref{fig:401} (a)). Therefore, this happens to all $i = 1, \dots, m$ with probability at most $1/\theta^m$.
\end{proof}

\begin{lemma} \label{lem:4-lv3-count}
    The number of pairings of size $m$ that occur with non-zero probability is at most $k^m \cdot \binom{n}{m}^2$.
\end{lemma}

\begin{proof}
    We prove that, for fixed sets $A, B \subseteq V^*$ with $|A| = |B| = m$, there are at most $k^m$ pairings with $\{a_1, \dots, a_m\} = A$ and $\{b_1, \dots, b_m\} = B$ (out of $m!$ possibilities) that occur with non-zero probability. This is enough to prove \autoref{lem:4-lv3-count} because there are at most $\binom{n}{m}^2$ choices for $(A, B)$.

    Each vertex $a \in A$ can be paired with $b \in B$ only if there exists some $i \in \{1, \dots, k\}$ such that $b$ is the last vertex with base color $i$ among the vertices that are not adjacent to $a$ in $B$. The reasons are as follows. If no such $i$ exists, this means that there exists a vertex $b'$ ($b \prec b', \varphi(b) = \varphi(b')$) such that $b'$ is not adjacent to $a$. Here, we observe that $b'$ is also not adjacent to $b$, because of $\varphi(b) = \varphi(b')$. Therefore, since $a$ and $b$ have the same level-2 color $c$, the color $c$ is eligible for $b'$, and our algorithm makes $b'$ a level-3 vertex, a contradiction. Hence, there are at most $k$ choices for each $a \in A$, and at most $k^m$ valid pairings (see \autoref{fig:401} (b)).
\end{proof}

\begin{lemma} \label{lem:4-lv3-twocolor}
    In the final state, there are at most $en \sqrt{2k/\theta} + 2$ level-2 colors in expectation.
\end{lemma}

\begin{proof}
    Let $X$ be the random variable describing the size of pairings (i.e., the number of level-2 colors). By \autoref{lem:4-lv3-prob} and \autoref{lem:4-lv3-count},
    \begin{equation}\label{eq:4-lv3-count}
        \Pr[X = m] \leq \frac{1}{\theta^m} \cdot \left\{k^m \cdot \binom{n}{m}^2\right\} \leq \left(\frac{e^2 kn^2}{\theta m^2}\right)^m
    \end{equation}
    due to $\binom{n}{m} \leq (en/m)^m$. By \autoref{eq:4-lv3-count}, $\Pr[X = m] \leq 1/2^m$ holds for all integers $m \geq en \sqrt{2k/\theta}$. Therefore, the expected value of $X$ is
    \begin{align*}
        \mathbb{E}[X]
        & = \sum_{i=1}^{\lceil en\sqrt{2k/\theta} \rceil - 1} i \cdot \Pr[X = i] + \sum_{i = \lceil en\sqrt{2k/\theta} \rceil}^{\infty} i \cdot \Pr[X = i] \\
        & \leq en \sqrt{\frac{2k}{\theta}} + \sum_{i=1}^\infty \left(i \cdot \frac{1}{2^i}\right) = en \sqrt{\frac{2k}{\theta}} + 2.
    \end{align*}
    \noindent
    This proves that the expected number of level-2 colors is at most $en \sqrt{2k/\theta} + 2$.
\end{proof}

With these bounds on the number of level-1 and level-2 colors, we finally complete the analysis for the number of colors used, with an appropriate parameter setting of $\theta$. The theorem below also implies that we use at most $(1+o(1)) n/3$ colors when $k = o(\sqrt{n})$.

\begin{theorem} \label{thm:4-lv3}
    The algorithm uses at most $n/3 + O(n^{2/3} k^{2/3})$ colors in expectation, when $n$ and $k$ are known in advance, with an appropriate setting of the parameter $\theta$.
\end{theorem}

\begin{proof}
    If there are $c_1$ level-1 colors and $c_2$ level-2 colors, the total number of colors is $c_1 + c_2 + (n - c_1 - 2c_2)/3 = \frac{1}{3} n + \frac{2}{3} c_1 + \frac{1}{3} c_2$. By \autoref{lem:4-lv3-onecolor} and \autoref{lem:4-lv3-twocolor}, in expectation, this is at most
    \begin{equation*}
        \frac{n}{3} + \frac{2}{3} k\theta + \frac{1}{3} \left(en \sqrt{\frac{2k}{\theta}} + 2\right).
    \end{equation*}
    This value is $n/3 + (enk)^{2/3} + 2/3$ with the optimal parameter setting $\theta = \{e^2 n^2/(8k)\}^{1/3}$. Although $\theta$ must be an integer, the asymptotic order is unchanged.
\end{proof}

\subsection{General-Level Algorithm}

The next task is to increase the number of vertices per color from $3$ to a larger fixed $d$ ($d = 4, 5, \dots$), generalizing the ``three-level'' algorithm in \autoref{sec:4.3}.

\paragraph{Algorithm.} We set the threshold parameters $\theta_1, \dots, \theta_{d-1}$, which are all positive integers. For each vertex $v \in V^*$, we select the largest integer $i \in \{d-1, \dots, 2, 1\}$ such that there are at least $\theta_i$ eligible level-$i$ colors (see \autoref{sec:4.2} for ``eligible''). Then we choose one eligible level-$i$ color uniformly at random, and assign it to $v$. If no such $i$ exists, we assign a fresh color for $v$ (or more formally, the smallest unused color). The pseudocode is provided in \autoref{alg:4-main}.

\begin{algorithm}[t]
    \caption{A $d$-level algorithm for online coloring, with parameters $\theta_{d-1}, \dots, \theta_2, \theta_1 \geq 1$. In the description below, the level-$i$ colors are the colors that are currently used on exactly $i$ vertices, and a color is ``eligible'' if $v$ is not adjacent to any vertex of this color.}
    \label{alg:4-main}
    \begin{algorithmic}[1]
        \For {each arrival of a vertex $v$}
            \For {$i = d-1, \dots, 2, 1$}
                \If {there are at least $\theta_i$ eligible level-$i$ colors}
                    \State $c(v) \gets$ a color among eligible level-$i$ colors, selected uniformly at random
                    \State Color $v$ with color $c(v)$
                    \State \textbf{break}
                \EndIf
            \EndFor
            \If {$v$ is not yet colored}
                \State Color $v$ with the smallest unused color
            \EndIf
        \EndFor
    \end{algorithmic}
\end{algorithm}

\subsection{Analysis Part I: Grouping the Colorings}

In the following two subsections, we aim to prove that when the chromatic number is $k = o(n^{\frac{1}{d-1}})$, \autoref{alg:4-main} uses $(1 + o(1))n/d$ colors with a parameter setting $\theta_t = \lfloor n^{1 - \frac{t}{d-1}} \rfloor$ for each $t$. First, we analyze the number of level-1 colors.

\begin{lemma} \label{lem:4-main-onecolor}
    In the final state of \autoref{alg:4-main}, there are at most $k\theta_1$ level-1 colors.
\end{lemma}

\begin{proof}
    This follows from the same argument as \autoref{lem:4-lv3-onecolor}, utilizing the pigeonhole principle.
\end{proof}

Next, we analyze the number of level-$t$ colors for a \textbf{fixed} $t \in \{2, \dots, d-1\}$. To this end, for each \textbf{fixed} $m$, we bound the probability that there are exactly $m$ level-$t$ colors, and show that it is extremely rare when $m$ is sufficiently large. To show the rarity of $m$ being large, we bound the probability that the ``last vertex'' of each level-$t$ colors is listed as $b_1, \dots, b_m$, for each \textbf{fixed} $\{b_1, \dots, b_m\} \subseteq V \ (b_1 \prec \dots \prec b_m)$.

\paragraph{Preliminaries.}

As mentioned in \autoref{sec:4.2}, we fix the ``base coloring'' $\varphi: V^* \to \{1, \dots, k\}$, a $k$-coloring of $G^*$, for the purpose of the analysis. Meanwhile, the final (online) coloring is denoted by $c: V^* \to \mathbb{N}$, which is a random variable based on the choices in \autoref{alg:4-main}, and we denote
\begin{equation*}
    \mathcal{C}_{\mathrm{all}} := \left\{c' : V^* \to \mathbb{N} \mid \mathrm{Pr}[c = c'] > 0, \text{the last vertex of each level-$t$ color is $b_1, \dots, b_m$ in $c'$}\right\}
\end{equation*}
which represents the set of ``possible'' colorings where the last vertex of each level-$t$ colors is listed as $b_1, \dots, b_m$. We also denote the set of vertices earlier than $b_i$ by $V_{\prec b_i} := \{v \in V^* : v \prec b_i\}$.

\begin{definition} \label{def:4-good-set}
    A subset $\mathcal{C} \subseteq \mathcal{C}_{\mathrm{all}}$ is called a ``good set'' if it satisfies the following for all $i = 1, \dots, m$:
    \begin{quote}
        \textbf{Condition:} For every (partial) coloring $c^-: V_{\prec b_i} \to \mathbb{N}$ that can be extended to some colorings in $\mathcal{C}$, all such extensions $c \in \mathcal{C}$ assign the same color to the vertex $b_i$.
    \end{quote}
    Note that for any partial coloring $c^-$ that can be extended to some coloring in $\mathcal{C}$, \autoref{alg:4-main} must choose a level-$(t-1)$ color when coloring $b_i$ due to $\mathcal{C} \subseteq \mathcal{C}_{\mathrm{all}}$.
\end{definition}

\begin{lemma} \label{lem:4-main-prob1}
    For every good set $\mathcal{C} \subseteq \mathcal{C}_{\mathrm{all}}$, we have $\mathrm{Pr}[c \in \mathcal{C}] \leq 1/\theta_{t-1}^m$.
\end{lemma}

\begin{proof}
    Let $i \in \{1, \dots, m\}$. Suppose that, just before the vertex $b_i$ arrives, there is a (non-zero) possibility that the final coloring $c : V^* \to \mathbb{N}$ satisfies $c \in \mathcal{C}$.
    
    Then, the probability that $c \in \mathcal{C}$ remains possible after coloring $b_i$ is at most $1/\theta_{t-1}$, conditional on the state before $b_i$ arrives. This is because the algorithm always chooses a level-$(t-1)$ color for the vertex $b_i$ (as noted in \autoref{def:4-good-set}), and to preserve the possibility of $c \in \mathcal{C}$, the algorithm must choose the \emph{correct one} among at least $\theta_{t-1}$ choices. Finally, \autoref{lem:4-main-prob1} holds by applying these bounds successively for $i = 1, \dots, m$.
\end{proof}

\begin{figure}[t]
    \centering
    \includegraphics[width=\linewidth]{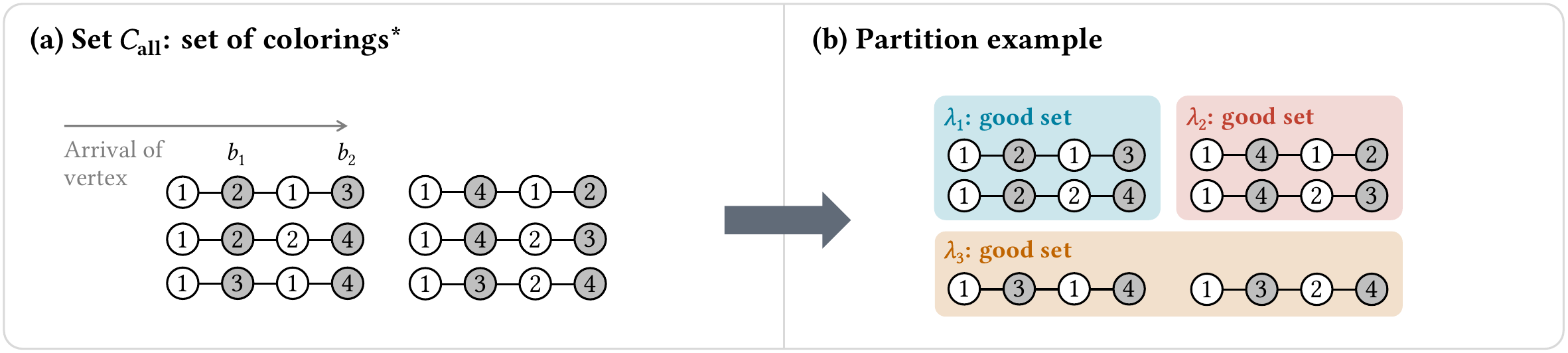}
    \caption{An illustration of our strategy. (*) The colorings in (a) are toy examples created just for illustration, and do not necessarily represent possible outputs of \autoref{alg:4-main}. In particular, some colorings (e.g., 1--3--1--4) have zero probability.}
    \label{fig:403}
\end{figure}

\paragraph{Our strategy.} Now, we explain the proof strategy. Our core strategy is to partition $\mathcal{C}_{\mathrm{all}}$ into a finite number of good sets $\mathcal{C}_{\lambda} \ (\lambda \in \Lambda)$ (where $\Lambda$ is a finite set); see \autoref{fig:403} for an illustration. Due to \autoref{lem:4-main-prob1}, if we successfully partitioned $\mathcal{C}_{\mathrm{all}}$ into $|\Lambda|$ good sets, then the probability of being exactly $m$ level-$t$ colors is at most $\binom{n}{m} \cdot |\Lambda| \cdot (1/\theta_{t-1}^m)$, where the first $\binom{n}{m}$ comes from the number of possible sets $\{b_1, \dots, b_m\}$ (note that we fix the set $\{b_1, \dots, b_m\}$ to define $\mathcal{C}_{\mathrm{all}}$). Indeed, a similar strategy was already used in the previous three-level case, as follows.

\begin{itemize}
    \item \autoref{sec:4.3} considered situations that the ``pairing'' is $(a_1, b_1), \dots, (a_m, b_m)$. Let $\Lambda$ be the set of all possible tuples $(a_1, \dots, a_m)$, and for each $\lambda = (a_1, \dots, a_m) \in \Lambda$, let $\mathcal{C}_\lambda$ be the set of final possible colorings whose pairing is $(a_1, b_1), \dots, (a_m, b_m)$. Then, it is (virtually) proved that $\mathcal{C}_\lambda$ is a good set in \autoref{lem:4-lv3-prob}, and that $|\Lambda| \leq k^m \cdot \binom{n}{m}$ in \autoref{lem:4-lv3-count}. These properties yield $\Pr[X = m] \leq \binom{n}{m} \cdot \{k^m \cdot \binom{n}{m}\} \cdot (1/\theta_{t-1}^m)$ as shown in the beginning of \autoref{lem:4-lv3-twocolor}.
\end{itemize}

\noindent
To construct the partition for the general-level case, we define the ``profile tuple'' as follows.

\begin{figure}[t]
    \centering
    \includegraphics[width=\linewidth]{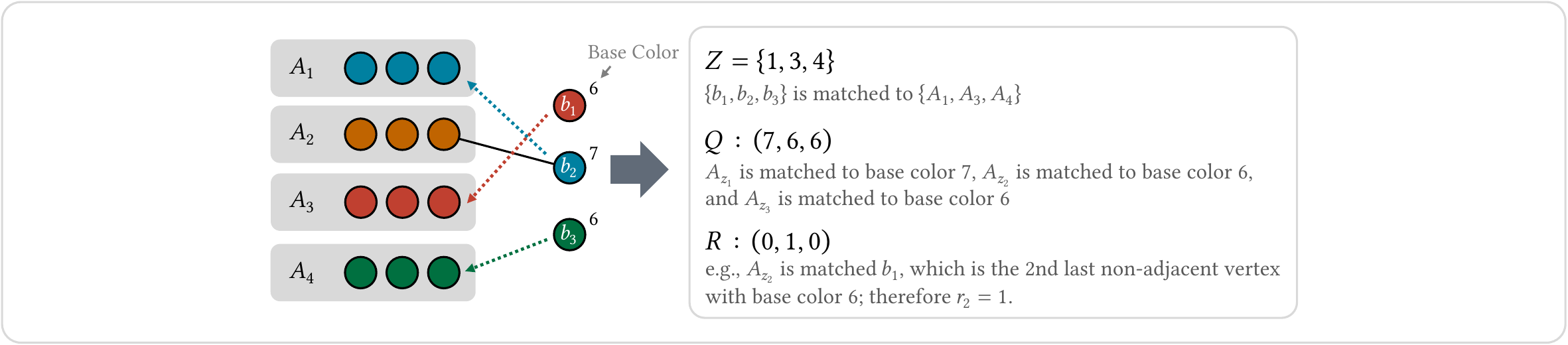}
    \caption{An example of the profile tuple $(Z, Q, R)$.}
    \label{fig:404}
\end{figure}

\begin{definition}[Profile tuple]
    Consider a coloring $c' \in \mathcal{C}_{\mathrm{all}}$. Let $A_1, A_2, \dots$ be the set of the ``first $t-1$ vertices'' for each color that is used at least $t-1$ times in $c'$, listed in ascending order of the arrival time of the $(t-1)$-th vertex. If $A_i$ and $b_j$ share the same color in $c'$, we say that ``$A_i$ and $b_j$ are matched''. Then, we define the following tuple $(Z, Q, R)$:

    \begin{itemize}
        \item $Z = \{z_1, \dots, z_m\} \ (z_1 < \dots < z_m)$: The set of indices such that each $A_{z_i}$ matches to some $b_j$. Let $\pi: [m] \to [m]$ be the (unique) permutation such that each $A_{z_i}$ is matched to $b_{\pi(i)}$.
        \item $Q = (q_1, \dots, q_m)$: Each $A_{z_i}$ is matched to a vertex of base color $q_i$, i.e., $\varphi(b_{\pi(i)}) = q_i$.
        \item $R = (r_1, \dots, r_m)$: Each $A_{z_i}$ is matched to the $(r_i+1)$-th last vertex with base color $q_i$ among $b_1, \dots, b_m$ that are not adjacent to any vertex in $A_{z_i}$. In other words,
        \begin{equation*}
            r_i := \left|\left\{j \in \{\pi(i)+1, \dots, m\} : \varphi(b_j) = q_i, \text{and $b_j$ is not adjacent to any vertex in $A_{z_i}$}\right\}\right|.
        \end{equation*}
    \end{itemize}

    \noindent
    We call $(Z, Q, R)$ the ``profile tuple'' for the coloring $c' \in \mathcal{C}_{\mathrm{all}}$; see \autoref{fig:404} for an illustration.
\end{definition}

We group the colorings in $\mathcal{C}_{\mathrm{all}}$ solely by the profile tuple. Specifically, we let $\Lambda$ be the set of all possible profile tuples $(Z, Q, R)$, and for each $\lambda = (Z, Q, R) \in \Lambda$, we let $\mathcal{C}_\lambda$ be the set of colorings in $\mathcal{C}_{\mathrm{all}}$ such that the profile tuple is $(Z, Q, R)$. Then, each $\mathcal{C}_\lambda$ is a good set, as proven below.

\begin{lemma} \label{lem:4-main-prob2}
    For each $\lambda = (Z, Q, R) \in \Lambda$, $\mathcal{C}_\lambda$ is a good set.
\end{lemma}

\begin{proof}
    Suppose that $\mathcal{C}_\lambda$ is not a good set, that is, there exists an $i \in \{1, \dots, m\}$ and a partial coloring $c^-: V_{\prec b_i} \to \mathbb{N}$ that violates the condition in \autoref{def:4-good-set}. We give two important observations:
    \begin{enumerate}
        \item Let $\ell$ be the number of colors that are used at least $t-1$ times in $c^-$. Then, $A_1, \dots, A_\ell$ is unchanged regardless of how we extend $c^-$ to a coloring in $\mathcal{C}_\lambda$.
        \item Also, each $\pi(j)$ is uniquely determined by $A_{z_j}$, $q_j$, and $r_j$: indeed, $b_{\pi(j)}$ is the $(r_j+1)$-th last vertex with base color $q_j$ among $b_1, \dots, b_m$ that are not adjacent to any vertex in $A_{z_j}$. Therefore, for $j$ with $z_j \leq \ell$, $\pi(j)$ is unchanged regardless of how we extend $c^-$ to a coloring in $\mathcal{C}_\lambda$.
    \end{enumerate}
    Since it violates the condition in \autoref{def:4-good-set}, there are two (or more) possible choices for the color of $b_i$ in the extensions of $c^-$. This means that for some $j_1, j_2$ ($j_1 \neq j_2, z_{j_1}, z_{j_2} \leq \ell$), both $A_{z_{j_1}}$ and $A_{z_{j_2}}$ can be matched to $b_i$. However, due to the observations above, $\pi(j_1)$ and $\pi(j_2)$ are unchanged across these extensions (i.e., uniquely determined by $A_{z_j}$, $q_j$, and $r_j$), and therefore $\pi(j_1) = i$ and $\pi(j_2) = i$ hold at the same time. This contradicts that $\pi$ is a permutation. Thus, $\mathcal{C}_\lambda$ is a good set.
\end{proof}

\begin{lemma}\label{lem:4-main-backward}
    For each coloring $c' \in \mathcal{C}_{\mathrm{all}}$, its profile tuple $(Z, Q, R)$ satisfies $r_1 + \dots + r_m \leq (\theta_t - 1)m$.
\end{lemma}

\begin{proof}
    We count the number of pairs $(i, j)$ such that (1) $i \in \{1, \dots, m\}, j \in \{\pi(i)+1, \dots, m\}$, (2) $\varphi(b_j) = q_i$, and (3) $b_j$ is not adjacent to any vertex in $A_{z_i}$.
    \begin{itemize}
        \item For each $i$, there are exactly $r_i$ such pairs (by definition of $r_i$). Therefore, the total number of pairs is $r_1 + \dots + r_m$.
        \item For each $j$, there are at most $\theta_t - 1$ such pairs. The reason is as follows: for a counted pair $(i, j)$, vertex $b_j$ is not adjacent to any vertex in $A_{z_i} \cup \{b_{\pi(i)}\}$, due to $\varphi(b_j) = \varphi(b_{\pi(i)}) \ (= q_i)$. Hence, if there are $\theta_t$ such $i$'s, then $b_j$ should have become a level-$(t+1)$ or higher vertex, which contradicts $c' \in \mathcal{C}_{\mathrm{all}}$. Therefore, the total number of pairs is at most $(\theta_t - 1) m$.
    \end{itemize}
    Therefore, $r_1 + \dots + r_m \leq (\theta_t - 1) m$.
\end{proof}

\begin{lemma} \label{lem:4-main-count}
    The number of possible profile tuples $(Z, Q, R)$, i.e., $|\Lambda|$, is at most $\binom{n}{m} \cdot k^m \cdot \binom{\theta_t m}{m}$.
\end{lemma}

\begin{proof}
    There are at most $\binom{n}{m}$ possibilities for $Z$ because it is a size-$m$ subset of $\{1, \dots, n\}$. There are at most $k^m$ possibilities for $Q = (q_1, \dots, q_m)$ because each $q_i$ is chosen from $\{1, \dots, k\}$. For $R = (r_1, \dots, r_m)$, we have $r_1 + \dots + r_m \leq (\theta_t-1) m$ by \autoref{lem:4-main-backward}. Hence, by the so-called ``stars and bars'' argument, there are at most $\binom{\theta_t m}{m}$ possibilities for $R$. Therefore, \autoref{lem:4-main-count} follows.
\end{proof}

\subsection{Analysis Part II: Number of Colors Used}

We are now ready to show an upper bound on the expected number of level-$t$ colors. Then we estimate the number of colors used in \autoref{alg:4-main} with an optimal parameter setting.

\begin{lemma} \label{lem:4-main-twocolor}
    Let $t \in \{2, \dots, d-1\}$. In the final state, the expected number of level-$t$ colors is at most $e^{3/2} n \sqrt{2k\theta_t/\theta_{t-1}} + 2$.
\end{lemma}

\begin{proof}
    Let $X$ be the random variable that represents the final number of level-$t$ colors. By \autoref{lem:4-main-prob1}, \autoref{lem:4-main-prob2}, \autoref{lem:4-main-count}, and that there are $\binom{n}{m}$ combinations for $b_1, \dots, b_m$,
    \begin{equation*}
        \mathrm{Pr}[X = m] = \frac{1}{\theta_{t-1}^m} \cdot \left\{\binom{n}{m} \cdot k^m \cdot \binom{\theta_t m}{m}\right\} \cdot \binom{n}{m} \leq \left(\frac{e^3 \theta_t kn^2}{\theta_{t-1} m^2}\right)^m
    \end{equation*}
    \noindent
    due to $\binom{n}{m} \leq (en/m)^m$ and $\binom{\theta_t m}{m} \leq (e\theta_t)^m$. Since
    \begin{equation*}
        \frac{e^3 \theta_t kn^2}{\theta_{t-1} m^2} \leq \frac{1}{2}
    \end{equation*}
    \noindent
    holds for all $m \geq e^{3/2} n \sqrt{2k\theta_t/\theta_{t-1}}$, the probability that the number of level-$t$ colors is exactly $m$ is at most $1/2^m$ for all $m \geq e^{3/2} n \sqrt{2k\theta_t/\theta_{t-1}}$. Therefore,
    \begin{equation*}
        \mathbb{E}[X] \leq e^{3/2} n \sqrt{\frac{2k\theta_t}{\theta_{t-1}}} + \sum_{i=1}^\infty \left(i \cdot \frac{1}{2^i}\right) = e^{3/2} n \sqrt{\frac{2k\theta_t}{\theta_{t-1}}} + 2. \qedhere
    \end{equation*}
\end{proof}

\begin{lemma} \label{lem:4-main-allcolor}
    The expected number of colors used in \autoref{alg:4-main} is at most
    \begin{equation*}
        \frac{n}{d} + 2(d-2) + k\theta_1 + \sum_{t=2}^{d-1} e^{3/2} n \sqrt{\frac{2k\theta_t}{\theta_{t-1}}}.
    \end{equation*}
\end{lemma}

\begin{proof}
    If there are $c_t$ level-$t$ colors for each $t = 1, \dots, d-1$, the expected total number of colors is
    \begin{align*}
        \sum_{t=1}^{d-1} \mathbb{E}[c_t] + \frac{1}{d} \left\{n - \sum_{t=1}^{d-1} (t \cdot \mathbb{E}[c_t])\right\}
        & \leq \frac{n}{d} + \sum_{t=1}^{d-1} \mathbb{E}[c_t] \\
        & \leq \frac{n}{d} + k\theta_1 + \sum_{t=2}^{d-1} \left(e^{3/2} n \sqrt{\frac{2k\theta_t}{\theta_{t-1}}} + 2\right).
    \end{align*}
    where the last inequality is due to \autoref{lem:4-main-onecolor} and \autoref{lem:4-main-twocolor}.
\end{proof}

\begin{theorem} \label{thm:4-main}
    Let $d \geq 2$ be an integer. \autoref{alg:4-main} uses at most $n/d + 11dn^{(2d-4)/(2d-3)} k^{(d-1)/(2d-3)}$ colors in expectation for an $n$-vertex $k$-colorable graph, when $n$ and $k$ are known in advance, with an appropriate setting of parameters $\theta_{d-1}, \dots, \theta_2, \theta_1$.
\end{theorem}

\begin{proof}
    Consider setting the parameters to $\theta_t = \lfloor (n^2/k)^{(d-t-1)/(2d-3)} \rfloor$ for each $t \in \{1, \dots, d-1\}$. Let $c$ be the random variable that represents the total number of colors \autoref{alg:4-main} used. Due to \autoref{lem:4-main-allcolor}, the following holds:
    \begin{align*}
        \mathbb{E}[c]
        & \leq \frac{n}{d} + 2(d-2) + \textcolor{darkgreen}{k\theta_1} + \textcolor{purple}{\sum_{t=2}^{d-1} e^{3/2} n \sqrt{\frac{2k\theta_t}{\theta_{t-1}}}}. \\
        & \leq \frac{n}{d} + 2(d-2) + \textcolor{darkgreen}{k\theta_1} + \textcolor{purple}{\sum_{t=2}^{d-1} \left(e^{3/2} n\cdot 2n^{-\frac{1}{2d-3}} k^{\frac{d-1}{2d-3}}\right)} \\
        & \leq \frac{n}{d} + 2(d-2) + \textcolor{darkgreen}{n^{\frac{2d-4}{2d-3}} k^{\frac{d-1}{2d-3}}} + \textcolor{purple}{2e^{3/2} \sum_{t=2}^{d-1} n^{\frac{2d-4}{2d-3}} k^{\frac{d-1}{2d-3}}} \\
        & \leq \frac{n}{d} + (2e^{3/2} + 2) \cdot dn^{\frac{2d-4}{2d-3}} k^{\frac{d-1}{2d-3}} \\
        & \leq \frac{n}{d} + 11dn^{\frac{2d-4}{2d-3}} k^{\frac{d-1}{2d-3}}.
    \end{align*}
    \noindent
    We highlight each term with different colors for clarity. Note that the second inequality holds due to $\theta_t'/\theta_t \leq 2$ when $\theta_t'$ is a real number that removes the floor function from $\theta_t$.
\end{proof}

\subsection{The Competitive Ratio}

\autoref{thm:4-main} obtains an algorithm with a non-trivial performance guarantee for graphs with large chromatic number $k$. On the other hand, in 1997, Halld\'{o}rsson \cite{Hal97} presented a randomized online coloring algorithm for graphs with small $k$, summarized as follows.

\begin{theorem}[\cite{Hal97}] \label{thm:4-hal97}
    There is a polynomial-time randomized online coloring algorithm for $k$-colorable graphs such that the expected number of colors used is at most $39 n^{(k-2)/(k-1)} (\log n)^{1/(k-1)}$ for any $k \geq 2$.
\end{theorem}

Combining our main result with \autoref{thm:4-hal97}, we obtain the desired $O(n (\log \log n)^2 / \log^2 n)$-competitive randomized algorithm as follows.

\begingroup
\def\thetheorem{\ref{thm:main-2}}
\begin{theorem}
    There exists a polynomial-time randomized online coloring algorithm, against an oblivious adversary, that attains a competitive ratio of $O(n (\log \log n)^2 / \log^2 n)$. 
\end{theorem}
\addtocounter{theorem}{-1}
\endgroup

\begin{proof}
    Let $\alpha = \frac{\log n}{2 \log \log n}$. First, by applying \autoref{thm:4-main} with $d = \max(\lceil \log n / (6 \log \log n) \rceil, 2)$, there exists an algorithm that uses at most
    \begin{equation}
       \frac{n}{d} + 11 d n^{1-\frac{1}{2d-3}} \cdot k \leq \frac{6 n \log \log n}{\log n} + O\left(\frac{\log n}{\log \log n} \cdot \frac{n}{\log^3 n}\right) \cdot k \label{eq:4-all}
    \end{equation}
    \noindent
    when both $n$ and $k$ are known in advance. This implies that (a) for $\alpha < k$, there is an algorithm that uses $O(n (\log \log n)^2 / \log^2 n) \cdot k$ colors. (b) Second, by \autoref{thm:4-hal97}, for $2 \leq k \leq \alpha$, there exists an algorithm that uses $O(n / \log^2 n)$ colors. (c) Third, for $k = 1$, there exists an algorithm that uses one color. Applying \autoref{lem:2-competitive-ratio} for (a)--(c), we obtain an $O(n (\log \log n)^2 / \log^2 n)$-competitive algorithm. Finally, this runs in polynomial time, since both \autoref{thm:4-main} and \autoref{thm:4-hal97} run in polynomial time and $n (\log \log n)^2 / \log^2 n \geq n^{1/100}$ (which is required to apply \autoref{lem:2-competitive-ratio}).
\end{proof}

\section{The Upper Bound for Deterministic Algorithms}\label{sec:5}

In this section, we present a deterministic online coloring algorithm that attains a competitive ratio of $O(n / (\log \log n)^{2-\varepsilon})$, for an arbitrary fixed $\varepsilon$ ($0 < \varepsilon < 2$).

\paragraph{Proof framework.}

In the main problem, we consider online coloring for $N$-vertex graphs. The following parameters are used throughout this section:
\begin{equation*}
    M := \left\lfloor (\log \log N)^{1-\varepsilon/2} \right\rfloor, \quad L := \left\lfloor N^{1/M^{11}} \right\rfloor.
\end{equation*}
The goal of the entire section is to develop a deterministic algorithm that colors $N$-vertex $M^2$-colorable graphs using $O(N/M)$ colors, where $N$ is known in advance. In the analysis, we always assume that $N$ is \textbf{sufficiently large}. Note that this means $M$ and $L$ are also sufficiently large.

This suffices to prove the desired competitive ratio of $O(N/M^2)$ because of \autoref{lem:2-competitive-ratio} and the fact that Kawarabayashi, Yoneda, and Yoneda's algorithm \cite{KYY27} already colors $M$-colorable graphs using $O(N/M^2)$ colors, as we discuss in \autoref{sec:5.12} (note that we use uppercase $N$ to distinguish it from $n$, the number of vertices in ``subproblems'' we discuss in later subsections, which is often unknown). This section is divided into three parts.

\begin{description}
    \item[\textbf{Part A.}] First, we construct a deterministic online algorithm that colors $L$-vertex $(M+1)$-clique-free graphs using at most $L/M$ colors. Note that this improves upon the best-known algorithm by Brada\v{c} et al. \cite{BFS+26}. We discuss this in \autoref{sec:5.2} and \autoref{sec:5.3}.
    \item[\textbf{Part B.}] Then, we present a new technical framework that adapts the algorithm for clique-free graphs to the framework of \cite{KYY27}, and show that $M^2$-colorable graphs can be colored online with $O(N/M)$ colors. \autoref{sec:5.4} gives an overview of the idea, and \autoref{sec:5.5} through \autoref{sec:5.11} present the algorithm and its analysis.
    \item[\textbf{Part C.}] Finally, we prove a competitive ratio of $O(n / (\log \log n)^{2-\varepsilon})$ in \autoref{sec:5.12}.
\end{description}

\subsection{Clique-Free Algorithm Part I: Online Object Coloring} \label{sec:5.2}

In this subsection, we define a novel problem, \emph{online object coloring}, which serves as a key tool in our clique-free algorithm. (A similar idea also appears in the work of Brada\v{c} et al. \cite{BFS+26}.)

\begin{definition}[Online Object Coloring] \label{def:5-object-coloring}
    There is a finite non-empty ground set $U$. In the \emph{online object coloring} problem, the sets $A_1, \dots, A_n \subseteq U$ arrive sequentially. When $A_i$ arrives, we must immediately assign a color to it, so that the following requirement is satisfied:
    \begin{quote}
        \textbf{Requirement:} For each color $c$ assigned to at least one set, the sets with color $c$ must have a common element. In other words, $\bigcap_{i \in T(c)} A_i \neq \emptyset$ must hold, where $T(c)$ is the set of indices $i$ for which color $c$ is assigned to the set $A_i$.
    \end{quote}
    This problem specifies a real parameter $\mu \geq 1$ that is known from the beginning. It is guaranteed that $|A_i| \geq |U|/\mu$ for each $i$. Note that $n$ is not known in advance. The goal is to minimize the number of colors used. We assume that the ground set $U$ is not known in advance; see \autoref{rem:5-object-coloring} for clarification.
\end{definition}

\begin{remark} \label{rem:5-object-coloring}
    We assume that the ground set $U$, or even its size $|U|$, is not known in advance. This means that an online object coloring algorithm can only ``see'' the elements in $U$ that belong to the sets $A_1, \dots, A_i$ that have already arrived. Since the parameter $\mu \geq 1$ is known to the algorithm, the algorithm also knows that $|A_i| \geq |U|/\mu$ holds (which also implies that $|A_i| \geq |A_1 \cup \dots \cup A_n| / \mu$ will definitely hold).
    
    Consequently, for a ``parent'' algorithm that uses online object coloring as a subroutine (such as \autoref{alg:5-clique}), \textbf{it is not required to specify $U$ from the beginning}, because an online object coloring algorithm cannot use information about $U$. Hence, it is possible for the parent algorithm to specify the sets $A_1, A_2, \dots$ online, and finally fix $U = A_1 \cup \dots \cup A_n$ only for the purpose of analysis. This is valid as long as $|A_i| \geq |U|/\mu$ holds for every $i$.
\end{remark}

\begin{algorithm}[!b]
    \caption{An algorithm for online object coloring.}
    \label{alg:5-object-coloring}
    \begin{algorithmic}[1]
        \State $t \gets 0$
        \For {each arrival of a set $A_i$}
            \If {there exists $c \in \{1, \dots, t\}$ such that $|X_c \cap A_i| \geq |X_c| / (2\mu)$}
                \State Assign color $c$ to the set $A_i$ 
                \State $X_c \gets X_c \cap A_i$
            \Else
                \State Assign color $t+1$ to the set $A_i$
                \State $X_{t+1} \gets A_i \setminus (X_1 \cup \dots \cup X_t)$
                \State $t \gets t+1$
            \EndIf
        \EndFor
    \end{algorithmic}
\end{algorithm}

\begin{lemma} \label{lem:5-object-coloring}
    Let $\mu \geq 1$ be a real number. There exists a deterministic algorithm for online object coloring with parameter $\mu$ such that the number of colors used is at most
    \begin{equation*}
        \min_{h \in \{1, 2, 3, \dots\}} \left\{\frac{n}{h} + (2\mu)^{h-1}\right\}.
    \end{equation*}
\end{lemma}

\begin{proof}
    Consider the following algorithm. We maintain a list of sets $X_1, \dots, X_t$; initially $t = 0$. When a new set $A_i$ arrives, we perform the following steps:
    
    \begin{enumerate}
        \item[(1)] If there exists $c \in \{1, \dots, t\}$ such that $|X_c \cap A_i| \geq |X_c| / (2\mu)$, then we choose any such $c$ and assign color $c$ to the set $A_i$. We then update $X_c \gets X_c \cap A_i$.
        \item[(2)] Otherwise, we assign color $t+1$ to the set $A_i$. Then, we set $X_{t+1} \gets A_i \setminus (X_1 \cup \dots \cup X_t)$ and update $t \gets t+1$.
    \end{enumerate}

    The pseudocode of this algorithm is given in \autoref{alg:5-object-coloring}. It is easy to see that the sets $X_1, \dots, X_t$ are disjoint. The coloring satisfies the requirement, because for each color $c \in \{1, \dots, t\}$, the set $X_c$ is a subset of ``the intersection of all sets $A_i$ with color $c$'', and $X_c$ will not be empty. When Case (1) occurs, the size of the set $X_c$ decreases by a factor of at most $2\mu$. When Case (2) occurs, we can lower bound the size of the new set $X_{t+1}$ as follows:
    \begin{equation*}
        |X_{t+1}| = |A_i| - \sum_{c=1}^t |X_c \cap A_i| \geq |A_i| - \frac{1}{2\mu} \sum_{c=1}^t |X_c| \geq \frac{|U|}{\mu} - \frac{|U|}{2\mu} = \frac{|U|}{2\mu}.
    \end{equation*}
    Hence, for each color $c$, we have $|X_c| \geq |U| / (2\mu)^{|T(c)|}$, where $T(c)$ is the set of indices $i$ for which color $c$ is assigned to the set $A_i$. Therefore, for any integer $h \geq 1$, the number of colors $c$ with $|T(c)| \leq h-1$ is at most $(2\mu)^{h-1}$, because such colors $c$ satisfy $|X_c| \geq |U| / (2\mu)^{h-1}$. Also, the number of colors $c$ with $|T(c)| \geq h$ is at most $n/h$. Therefore, \autoref{lem:5-object-coloring} follows.
\end{proof}

\subsection{Clique-Free Algorithm Part II: Main}\label{sec:5.3}

We now solve the problem of coloring $r$-clique-free graphs online, denoted by $\textsf{CliqueFree}(r)$. In particular, we show the following lemma.

\begin{lemma} \label{lem:5-clique-free}
    Let $A \geq 1$ be a (fixed) integer parameter. Define the sequence $n_2, n_3, \dots$ as follows:
    \begin{equation*}
        n_2 = A, \quad n_r = 3A \cdot \left\{6r (n_{r-1})^{r-1}\right\}^{3A} \ (r = 3, 4, \dots)
    \end{equation*}
    Then, for every $r \in \{2, 3, \dots\}$, there exists a deterministic algorithm for $\textsf{CliqueFree}(r)$ on $n_r$ vertices that uses at most $\frac{r-1}{A} \cdot n_r$ colors.
\end{lemma}

We construct an algorithm by induction on $r$. For $r=2$, the graph has no edges, so we color all vertices with the same color. This meets the requirement for \autoref{lem:5-clique-free} because $\frac{1}{A} \cdot n_2 = 1$. Below, we describe the algorithm for $\textsf{CliqueFree}(r)$ when $r \geq 3$; its pseudocode is given in \autoref{alg:5-clique}.

\begin{algorithm}[!b]
    \caption{$\textsf{CliqueFree}(r)$: A recursive algorithm to color an $n_r$-vertex $r$-clique-free graph online, using the object coloring procedure.}
    \label{alg:5-clique}
    \begin{algorithmic}[1]
        \State Initiate an instance of online object coloring with parameter $\mu = 3r \cdot (n_{r-1})^{r-1}$
        \State $B_1, B_2, \dots \leftarrow \emptyset$
        \State $J \gets n_r / (3r \cdot n_{r-1})$
        \For {each arrival of a vertex $v$}
            \State $j \leftarrow$ the smallest $j$ such that $|B_j| < n_{r-1}$ and $G[B_j \cup \{v\}]$ is $(r-1)$-clique-free
            \If {$B_j = \emptyset$ and $|\{\ell : 1 \leq |B_\ell| < n_{r-1}\}| = J$}
                \State $B \gets B_1 \cup \dots \cup B_{j-1}$ (the set of vertices contained in any bin)
                \State Present the set $\textsf{CLIQUE}(N_B(v))$ to online object coloring
                \State Color $v$ with the color assigned to this set in online object coloring
            \Else
                \State $B_j \leftarrow B_j \cup \{v\}$
                \State Color $v$ with the instance of $\textsf{CliqueFree}(r-1)$ created for $B_j$
            \EndIf
        \EndFor
    \end{algorithmic}
\end{algorithm}

\paragraph{Algorithm.}

First, we prepare bins $B_1, B_2, \dots \ (\subseteq V)$, initially empty. It is assumed that each $B_i$ induces a $(r-1)$-clique-free subgraph and contains at most $n_{r-1}$ vertices. Roughly speaking, the strategy is to assign each vertex to a bin, and then color the vertices of each $B_i$ using our algorithm for $\textsf{CliqueFree}(r-1)$ (defined by the induction hypothesis) because of this assumption. We use disjoint colors for different bins.

\begin{figure}[t]
    \centering
    \includegraphics[width=\linewidth]{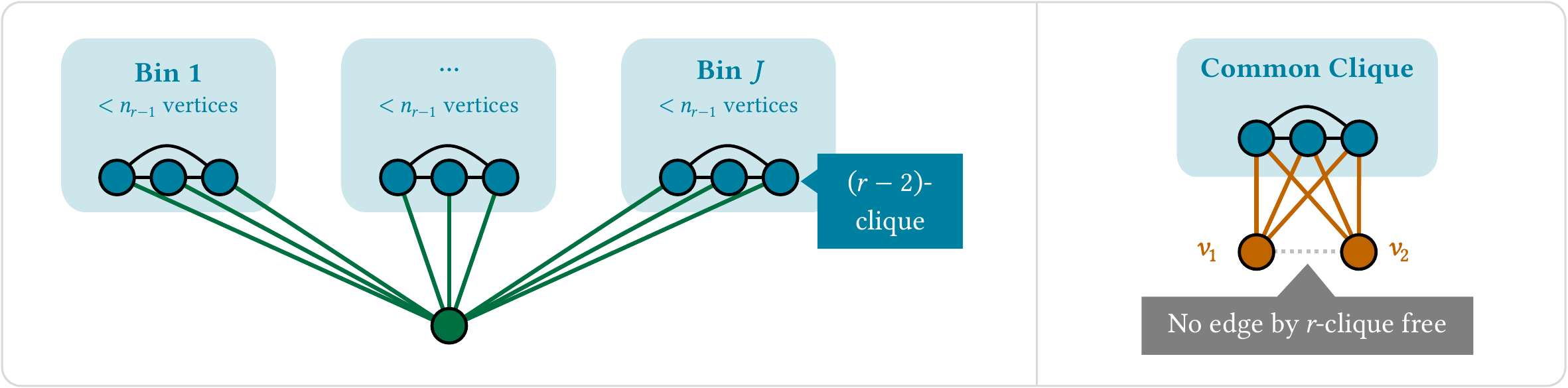}
    \caption{(Left) An example in which a vertex $v$ cannot be put into any bin when $r = 5$. (Right) An illustration of two vertices $v_1, v_2$ outside the bins that are assigned the same color. If there were an edge between $v_1$ and $v_2$, this would contradict the fact that the entire graph is $r$-clique-free, since there must be at least one common clique.}
  \label{fig:502}
\end{figure}

When a new vertex $v$ arrives, we try to put $v$ into a bin. Precisely, we compute the smallest $j$ such that $|B_j| < n_{r-1}$ and $G[B_j \cup \{v\}]$ is $(r-1)$-clique-free, and try to put $v$ into $B_j$. However, this approach might cause many ``small'' bins and exceed the limit of $\frac{r-1}{A} \cdot n_r$ colors. Therefore, we let
\begin{equation*}
    J := \frac{n_r}{3r \cdot n_{r-1}},
\end{equation*}
require that the number of ``incomplete'' bins (i.e., bins of size between $1$ and $n_{r-1}-1$) should be at most $J$ (note that $J$ is an integer). Hence, if $B_j$ is empty and there are $J$ incomplete bins, then we do not put $v$ into any bin. The good news is that, in this case, there are many $(r-2)$-cliques whose vertices are all adjacent to $v$ (such a situation is illustrated in \autoref{fig:502}, left). Indeed, each incomplete bin $B_j$ contains at least one such $(r-2)$-clique because $v$ was not added to the bin $B_j$. Thus, across all bins, there are at least $J$ such $(r-2)$-cliques. We color such vertices $v$ by exploiting this property, as described below.

\paragraph{Coloring outside bins.} We denote by $\textsf{CLIQUE}(S)$ the set of $(r-2)$-cliques contained in $S \subseteq V$ whose vertices all belong to the same bin. Let $B := B_1 \cup \dots \cup B_{j-1}$ be the set of vertices contained in any bin. We then present the set $\textsf{CLIQUE}(N_B(v))$ to online object coloring (as an input), and color $v$ with the color assigned to this set in online object coloring.
\begin{itemize}
    \item Technically, we consider an instance of online object coloring where the ground set $U$ is the ``final'' set of $(r-2)$-cliques whose vertices all belong to the same bin. (Some readers may wonder if this is valid in the online setting, but this is justified in \autoref{rem:5-object-coloring}.)
    \item For the parameter $\mu$ in online object coloring, since each set has size at least $J$ and we have $|U| \leq n_r \cdot (n_{r-1})^{r-2}$ (because there are at most $n_r$ bins), we can set
    \begin{equation*}
        \mu = 3r \cdot (n_{r-1})^{r-1} \quad \left(= \frac{n_r \cdot (n_{r-1})^{r-2}}{J}\right).
    \end{equation*}
    \item We assume that the colors used in online object coloring are disjoint from those used in bins.
\end{itemize}
This algorithm still gives a proper coloring. Indeed, for any two vertices $v_1, v_2$ ($v_1 \neq v_2$) assigned the same color in online object coloring, there is an $(r-2)$-clique $S \subseteq V$ that belongs to both $\textsf{CLIQUE}(N_B(v_1))$ and $\textsf{CLIQUE}(N_B(v_2))$, because of the requirement in \autoref{def:5-object-coloring}. If there were an edge between $v_1$ and $v_2$, then $S \cup \{v_1, v_2\}$ would form an $r$-clique, contradicting the fact that the graph is $r$-clique-free (\autoref{fig:502}, right).

\paragraph{Analysis.}

Next, we analyze the number of colors used in \autoref{alg:5-clique}.

\begin{proof}[Proof of \autoref{lem:5-clique-free}]
    By the induction hypothesis, each bin uses at most $\frac{r-2}{A} \cdot n_{r-1}$ colors. Since there are at most $n_r / n_{r-1} + J$ bins, the number of colors used across all bins is at most
    \begin{equation*}
        \frac{r-2}{A} \cdot n_{r-1} \cdot \left(\frac{n_r}{n_{r-1}} + J\right) = \frac{r-2}{A} \cdot n_r + \frac{r-2}{A} \cdot n_{r-1} \cdot \frac{n_r}{3r \cdot n_{r-1}} \leq \frac{r-2}{A} \cdot n_r + \frac{1}{3A} \cdot n_r.
    \end{equation*}
    For online object coloring, by \autoref{lem:5-object-coloring} (with $h = 3A$), the number of colors used is at most
    \begin{equation*}
        \frac{n_r}{3A} + (2\mu)^{3A} = \frac{n_r}{3A} + \left\{6r \cdot (n_{r-1})^{r-1}\right\}^{3A} = \frac{1}{3A} \cdot n_r + \frac{1}{3A} \cdot n_r.
    \end{equation*}
    Therefore, we use at most $\frac{r-2}{A} \cdot n_r + (\frac{1}{3A} + \frac{1}{3A} + \frac{1}{3A}) \cdot n_r = \frac{r-1}{A} \cdot n_r$ colors in total.
\end{proof}

\begin{lemma} \label{lem:5-clique-free-upper}
    For every $r \geq 2$, we have $n_r \leq (7A)^{(3A)^{r-2} (r-1)!}$.
\end{lemma}

\begin{proof}
    By definition of $n_r$, we have $\log n_r = 3A(r-1) \log n_{r-1} + \{3A \log (6r) + \log (3A)\}$. Therefore, when we define $M_r := (3A)^{r-2} (r-1)!$, we have
    \begin{equation*}
        \frac{\log n_r}{M_r} = \frac{\log n_{r-1}}{M_{r-1}} + \frac{3A \log (6r) + \log (3A)}{M_r}.
    \end{equation*}
    Therefore, given that $n_2 = A$ and $M_2 = 1$, the following holds:
    \begin{align*}
        \frac{\log n_r}{M_r} & = \log A + \sum_{i=3}^r \frac{3A \log (6i) + \log (3A)}{(3A)^{i-2} (i-1)!} \\
        & \leq \log A + \sum_{i=3}^{\infty} \frac{3 \log(6i) + \log 3}{3^{i-2} (i-1)!} \\
        & = \log A + 2.6602 \dots \ (\leq \log (7A)).
    \end{align*}
    where the first equality is derived by assigning $A = 1$ because this is the worst case. Therefore, $\log n_r = (3A)^{r-2} (r-1)! \log (7A)$, and hence \autoref{lem:5-clique-free-upper} follows.
\end{proof}

\begin{corollary} \label{cor:5-clique-free}
    Assuming that $N$ is large enough, there exists a deterministic online algorithm for $\textsf{CliqueFree}(M+1)$ on $L$ vertices that uses at most $L/M$ colors.
\end{corollary}

\begin{proof}
    Apply \autoref{lem:5-clique-free} with parameter $A = 2M^2$; we can process $n_{M+1}$ vertices with \autoref{alg:5-clique} for $\textsf{CliqueFree}(M+1)$ using $\frac{n_{M+1}}{2M^2} \cdot M$ colors. Hence, if $L \geq n_{M+1}$, then we can color all vertices using
    \begin{equation*}
        \left(\frac{n_{M+1}}{2M^2} \cdot M\right) \cdot \left\lceil \frac{L}{n_{M+1}} \right\rceil = \frac{n_{M+1}}{2M} \cdot \left\lceil \frac{L}{n_{M+1}} \right\rceil \leq \frac{L}{M}
    \end{equation*}
    colors. Now, we prove $L \geq n_{M+1}$. By \autoref{lem:5-clique-free-upper}, we have $n_{M+1} \leq (14M^2)^{(6M^2)^{M-1} M!} = 2^{2^{\Theta(M \log M)}}$. With $M = \lfloor (\log \log N)^{1-\varepsilon/2} \rfloor$, we have $\log \log n_{M+1} = \Theta(M \log M) = o(\log \log N)$, which is smaller than $\log \log L = (1-o(1)) \log \log N$. Therefore, $n_{M+1} \leq L$, and \autoref{cor:5-clique-free} follows.
\end{proof}

\begin{remark} \label{rem:5-clique-free}
    Brada\v{c} et al. \cite{BFS+26} presented, for every \textbf{constant} $r \geq 3$, a deterministic online algorithm that colors $n$-vertex $r$-clique-free graphs with $O(n\log^{(r-1)}n / \log^{(r-2)} n)$ colors, where $\log^{(j)} n$ denotes the $j$-fold iterated logarithm. In contrast, our deterministic algorithm uses only $O(n (\log \log n / \log n)^{1/(r-2)})$ colors by setting the parameter to $A = \Theta((\log n / \log \log n)^{1/(r-2)})$ with a sufficiently small constant factor, due to \autoref{lem:5-clique-free-upper}. This bound is asymptotically better than \cite{BFS+26} for every constant $r \geq 4$.
\end{remark}

\subsection{Key Idea: Shortening the KYY27 Recursion}\label{sec:5.4}

From this subsection onward, we present an algorithm that colors $M^2$-colorable graphs using $O(N/M)$ colors. The core idea is to incorporate our novel algorithm for $(M+1)$-clique-free graphs (\autoref{alg:5-clique}) into the framework of Kawarabayashi, Yoneda, and Yoneda \cite{KYY27}. They obtained an algorithm that, for every constant $k \geq 2$, colors $k$-colorable graphs using $\widetilde{O}(n^{1-2/(k(k-1))})$ colors and achieves a competitive ratio of $O(n / \log \log n)$.

\paragraph{Kierstead's method.}

We first describe the idea behind Kierstead's algorithm \cite{Kie98b}, which colors 3-colorable graphs with $O(n^{1-c})$ colors (for sufficiently small $c > 0$) and is foundational to \cite{KYY27}'s framework. We color each vertex $v$ using First-Fit if it can be colored with one of the first $n^{1-c}$ colors. If First-Fit cannot color $v$, then $v$ has degree at least $n^{1-c}$, so we exploit this fact to color these vertices as follows.

\begin{description}
    \item[Step 1.] Let $S$ be the set of vertices colored by First-Fit. For a fixed $x \in V \setminus S$, let $L_1 := N_S(x)$. Then, in some 3-coloring of $G$, at most 2 colors appear in $L_1$ (\autoref{fig:204}, left).
    \item[Step 2.] Additionally, for a fixed edge $\{y, z\}$ where $y, z \in V \setminus S$, let $L_{2,y} := N_{L_1}(y)$ and $L_{2,z} := N_{L_1}(z)$. Then, in some 3-coloring of $G$, \emph{either} $L_{2,y}$ or $L_{2,z}$ is ``monochromatic'' (\autoref{fig:204}, right).\footnote{This is because for any 3-coloring $\varphi : V \to \{1, 2, 3\}$, either $\varphi(x) \neq \varphi(y)$ or $\varphi(x) \neq \varphi(z)$ holds. If $\varphi(x) \neq \varphi(y)$, then $L_{2,y}$ consists of 1 color; if $\varphi(x) \neq \varphi(z)$, then $L_{2,z}$ consists of 1 color.}
    \item[Step 3.] This means \emph{either} $G[N(L_{2,y})]$ or $G[N(L_{2,z})]$ is 2-colorable (i.e., bipartite) and can be colored online using only $O(\log n)$ colors by the algorithm of Lov\'{a}sz, Saks, and Trotter \cite{LST89}.
\end{description}

Moreover, in favorable cases, both $L_{2,y}$ and $L_{2,z}$ are large (due to the large degree of each vertex in $V \setminus S$), and hence many future vertices belong to $N(L_{2,y})$ or $N(L_{2,z})$. This provides the intuition behind achieving a bound of $O(n^{1-c})$ colors (although \cite{Kie98b} used some additional ideas).

\begin{figure}[!b]
  \centering
  \includegraphics[width=\linewidth]{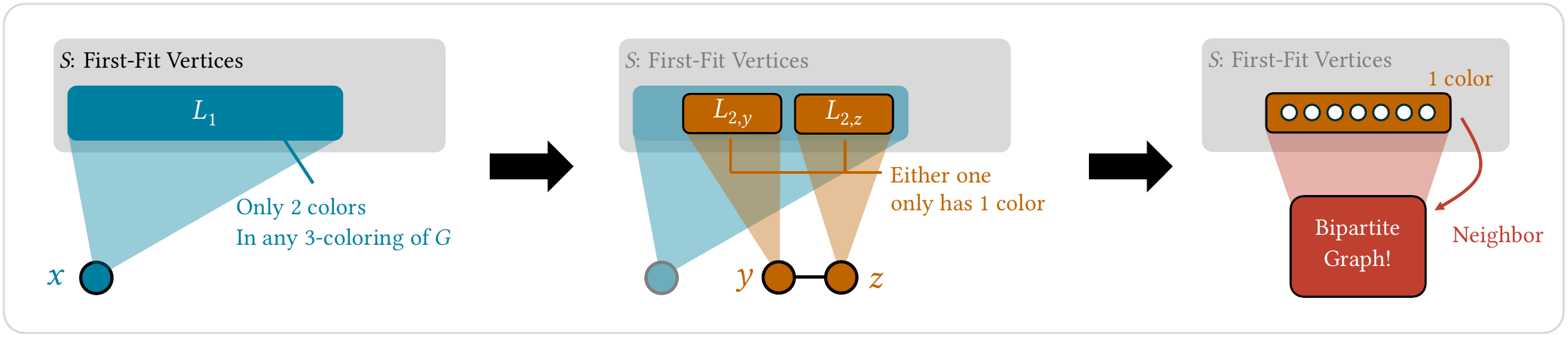}
  \caption{A sketch of Kierstead's algorithm that colors 3-colorable graphs with $O(n^{1-c})$ colors.}
  \label{fig:503}
\end{figure}

\paragraph{The KYY27 framework.}

To generalize to $k$-colorable graphs for any $k \geq 3$, small subgraphs with large chromatic numbers play a crucial role. In the preceding example for 3-colorable graphs, we picked a vertex $x$ (chromatic number $1$) in Step 1 and an edge $\{y, z\}$ (chromatic number $2$) in Step 2. For 4-colorable graphs, we also need to consider subgraphs of chromatic number $3$, such as odd cycles. However, the subgraphs must be small, because a large subgraph creates too many recursive branches. Indeed, in the 3-colorable case, we need only two branches, $L_{2,y}$ and $L_{2,z}$, because $\{y, z\}$ has size two. We create two branches because the algorithm does not know which of $L_{2,y}$ and $L_{2,z}$ is ``the correct one'' (that is, monochromatic in some 3-coloring).

To overcome this difficulty, \cite{KYY27} introduced an online coloring procedure called $\textsf{LOC}(k)$. This procedure must (i) color incoming vertices using a small number of colors, or (ii) terminate by returning a small subgraph $X \subseteq V(G)$ satisfying $\chi(G[X]) > k$, called a \emph{certificate}. The algorithm for $\textsf{LOC}(k)$ in \cite{KYY27} is constructed recursively using $\textsf{LOC}(0), \dots, \textsf{LOC}(k-1)$, and the algorithmic framework is similar to that of the 3-colorable case (First-Fit + Steps 1--3).

\paragraph{Limitation.} The reason why the competitive ratio is limited to $O(n / \log \log n)$ exactly lies in this recursion structure. Let $f_k$ be an upper bound on the size of the certificate returned by $\textsf{LOC}(k)$, e.g., $f_0 = 1$ ($x$ in Step 1) and $f_1 = 2$ ($\{y, z\}$ in Step 2). In \cite{KYY27}, the bound is roughly $f_k \approx f_0 \times f_1 \times \cdots \times f_{k-1}$, due to the recursion toward $\textsf{LOC}(0), \dots, \textsf{LOC}(k-1)$. Since $f_k$ grows double-exponentially (i.e., exceeds $n$ when $k = \Theta(\log \log n)$), the coefficient in the bound on the number of colors used also grows double-exponentially, limiting the competitive ratio.

\paragraph{Our Improvement.} Our key modification is to restrict the recursive calls for $\textsf{LOC}(k)$ to $\textsf{LOC}(0), \dots,$ $\textsf{LOC}(k-M)$. This is made possible by our algorithm for $(M+1)$-clique-free graphs. Consequently, the certificate size is roughly $f_k \approx f_0 \times \cdots \times f_{k-M}$, and $f_k = n^{o(1)}$ at $k = \Theta((\log \log n)^{2-\varepsilon})$.

\begin{figure}[!b]
  \centering
  \includegraphics[width=\linewidth]{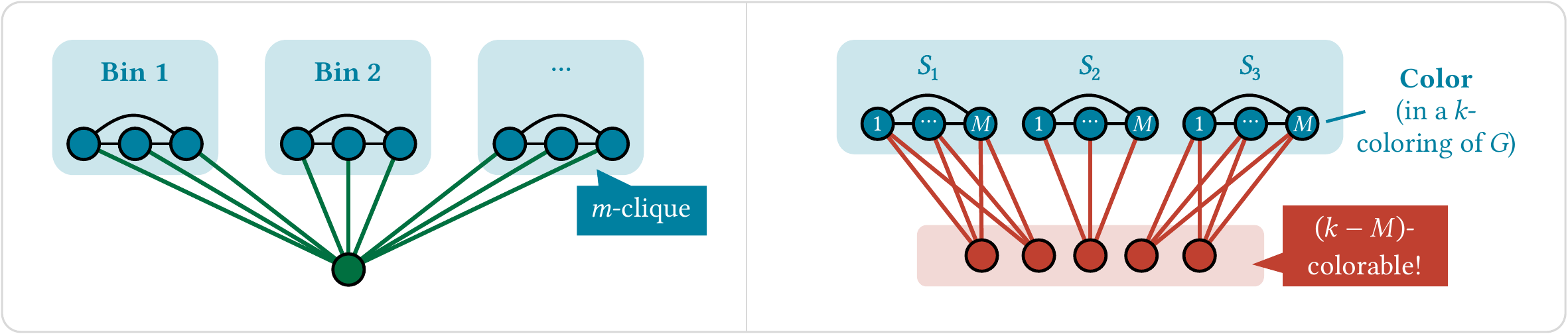}
  \caption{Left: the situation in which $v$ cannot be put in any bin. Right: the intuitive reason why the recursion stops at \textsc{LOC}$(k-M)$.}
  \label{fig:504}
\end{figure}

In our algorithm, we begin with a technique similar to the clique-free algorithm in \autoref{alg:5-clique}, rather than First-Fit. We maintain many bins, and we try to place each arriving vertex into a bin while ensuring that each bin remains $(M+1)$-clique-free (note that each bin can be colored efficiently by \autoref{cor:5-clique-free}). If no bin is available, then $v$ has a \emph{large clique degree}, that is, $v$ is adjacent to many $M$-cliques (\autoref{fig:504} left). For these vertices, we take a recursive strategy analogous to Steps 1--3, which utilizes small subgraphs of large chromatic numbers to iteratively restrict the colors allowed across a collection of cliques (note that, in Steps 1--3, we iteratively restrict the number of colors allowed for a set of vertices using the large degree, but this time we consider cliques).

Indeed, this strategy is stronger than \cite{KYY27}. Let us define a vertex $v$ to be \emph{completely adjacent} to a clique $S$ if $v$ is adjacent to all vertices in $S$. Suppose that we have succeeded in restricting the allowed colors in a collection of $M$-cliques $S_1, \dots, S_t$ to $M$ colors in some $k$-coloring of $G$; in other words, all these cliques use the same $M$ colors. Then, the set of future vertices that are completely adjacent to at least one of $S_1, \dots, S_t$ is $(k-M)$-colorable (\autoref{fig:504} right). This is the intuition for why our algorithm for $\textsf{LOC}(k)$ only recurses to $\textsf{LOC}(0), \dots, \textsf{LOC}(k-M)$.

\subsection{Notation and Definitions}\label{sec:5.5}

We begin a formal discussion of a deterministic online algorithm that colors $M^2$-colorable graphs using $O(N/M)$ colors. First, we present notation and definitions, followed by the ``subproblems'' that serve as subroutines for solving the main problem.

\paragraph{Notation and definitions.}

Let $G = (V, E)$ be a graph. For $v \in V$ and a clique $S \subseteq V$, we say that $v$ is \emph{``completely adjacent''} to $S$ if the vertex $v$ is adjacent to all vertices in $S$. A set of $M$-cliques $\mathcal{S} = \{S_1, S_2, \dots\}$ $(S_i \subseteq V)$ is called a \emph{``clique block''}. For a clique block $\mathcal{S} = \{S_1, S_2, \dots\}$, we define $\mathrm{vert}(\mathcal{S}) := \bigcup_i S_i$ for the set of vertices that belong to $\mathcal{S}$.

\paragraph{Locally $k$-colorable graph.}

Kawarabayashi, Yoneda, and Yoneda \cite{KYY27} introduced a notion of \emph{``locally $k$-colorable graph''}: a graph that has no small non-$k$-colorable subgraph. In this paper, a graph $G = (V, E)$ is locally $k$-colorable if $\chi(G[X]) \leq k$ for all $X \subseteq V$ with $|X| \leq f_k$, where the parameter $f_k$ is a \textcolor{purple}{constant} discussed in \autoref{sec:5.9}. They considered the following problem, $\textsf{LOC}(k)$, which stands for online coloring for locally $k$-colorable graphs.

\begin{definition}[Problem LOC]
    \label{def:5-loc}
    Let $k \geq 0$ be an integer. We define the problem $\textsf{LOC}(k)$ as follows: the online coloring problem in which, for each arrival of a vertex $v$, we must (i) color a vertex, and (ii) may then terminate by outputting a \emph{``certificate''} that $G$ is not locally $k$-colorable, that is, a set $X \subseteq V$ such that $|X| \leq f_k$ and $\chi(G[X])>k$. Importantly, \textcolor{purple}{even if we terminate, we must color the last received vertex}. We also define the problem $\textsf{LOC}'(k)$ to be the version of $\textsf{LOC}(k)$ where the final number of vertices $n$ is known in advance.
\end{definition}

\paragraph{The clique block subproblem.}

We define a subproblem that characterizes the recursive subroutine that, as mentioned in \autoref{sec:5.4}, is used \emph{``to iteratively restrict the colors allowed across a collection of cliques''}. (Note that this can be viewed as an extension of ``subproblem $\textsf{SP}(k, \ell, \nu)$'' defined in \cite{KYY27} to the clique block setting, although we will not mention that here.)

\begin{figure}[t]
  \centering
  \includegraphics[width=\linewidth]{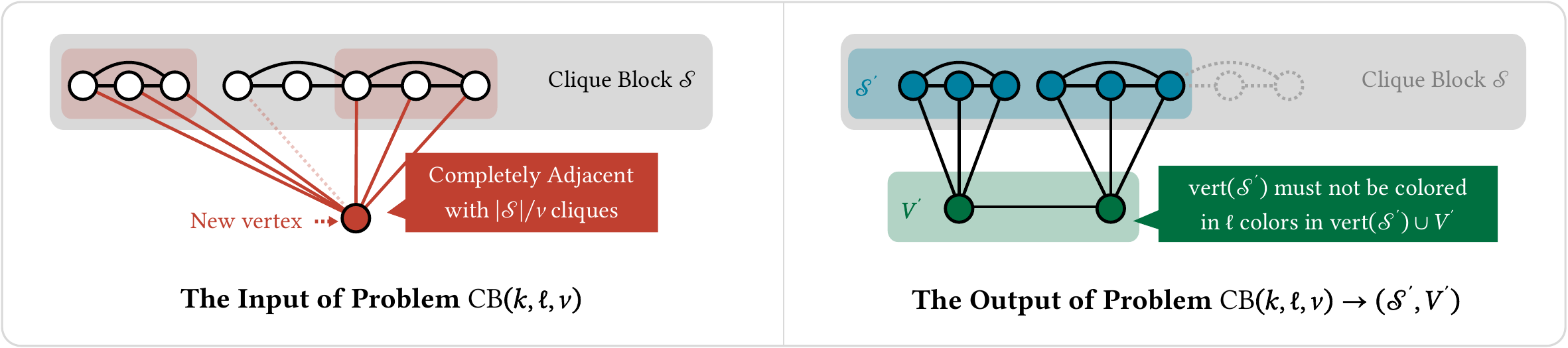}
  \caption{The input and output of the problem $\textsf{CB}(k, \ell, \nu)$.}
  \label{fig:506}
\end{figure}

\begin{definition}[Problem CB] \label{def:5-cb}
    Let $k, \ell$ be integers such that $k \geq \ell \geq M$, and let $\nu \geq 1$ be a real parameter. We define problem $\textsf{CB}(k, \ell, \nu)$ as follows. There is a fixed clique block $\mathcal{S} \neq \emptyset$. Then, vertices are given one by one. For each vertex $v$, a set of cliques that are completely adjacent to $v$, denoted by $\mathcal{C}(v) \ (\subseteq \mathcal{S})$, is also given. It is guaranteed that $|\mathcal{C}(v)| \geq |\mathcal{S}|/\nu$. (\autoref{fig:506}, left). For each arrival of a vertex $v$, we must (i) color a vertex $v$, and (ii) may then terminate by outputting a pair $(\mathcal{S}', V')$ ($\mathcal{S}' \subseteq \mathcal{S}, V' \subseteq V$) that satisfies the following:
    \begin{quote}
        \textbf{Condition:} There exists no $k$-coloring of $G[\mathrm{vert}(\mathcal{S}') \cup V']$, $\varphi: (\mathrm{vert}(\mathcal{S}') \cup V') \to [k]$, such that $|\{\varphi(s) : s \in \mathrm{vert}(\mathcal{S}')\}| \leq \ell$ (\autoref{fig:506}, right). In addition, $|\mathrm{vert}(\mathcal{S}') \cup V'| \leq g_{k,\ell}$ must hold, where the parameter $g_{k,\ell}$ is a \textcolor{purple}{constant} discussed in \autoref{sec:5.9}. 
    \end{quote}
    Importantly, \textcolor{purple}{even if we terminate, we must color the last vertex}. For convenience, we assume that the clique block $\mathcal{S}$ is \textcolor{purple}{not known in advance}, even though each $\mathcal{C}(v)$ is given; see \autoref{rem:5-cb} for clarification.
\end{definition}

\begin{remark} \label{rem:5-cb}
    The assumption that the clique block $\mathcal{S}$, including its size $|\mathcal{S}|$, is unknown is similar to that of online object coloring (\autoref{def:5-object-coloring} and \autoref{rem:5-object-coloring}). This means that the algorithm for $\textsf{CB}(k, \ell, \nu)$ can only ``see'' the elements in $\mathcal{S}$ that appear in $\mathcal{C}(v)$ for some vertex $v$ that has already arrived. Consequently, for a ``parent'' algorithm that uses $\textsf{CB}(k, \ell, \nu)$ as a subroutine, \textcolor{purple}{it is not required to specify $\mathcal{S}$ from the beginning}; it can specify the set of completely adjacent cliques for each arriving vertex, and finally fix $\mathcal{S}$ to be the union of these sets only for the purpose of the analysis. See also \autoref{rem:5-object-coloring} for related details.

    However, when we solve $\textsf{CB}(k, \ell, \nu)$ for $\ell \in \{M+1, \dots, k\}$ (in \autoref{sec:5.8}), we initiate instances $I$ for $\textsf{CB}(k, \ell-1, 2\nu)$, and here, we do fix a clique block $\mathcal{S}$ upon initiating each such instance $I$ (although we assume that this child instance does not know $\mathcal{S}$). We denote the clique block for this instance by $\mathcal{S}(I)$.
\end{remark}

\paragraph{Proof roadmap.}

\begin{figure}[t]
  \centering
  \includegraphics[width=\linewidth]{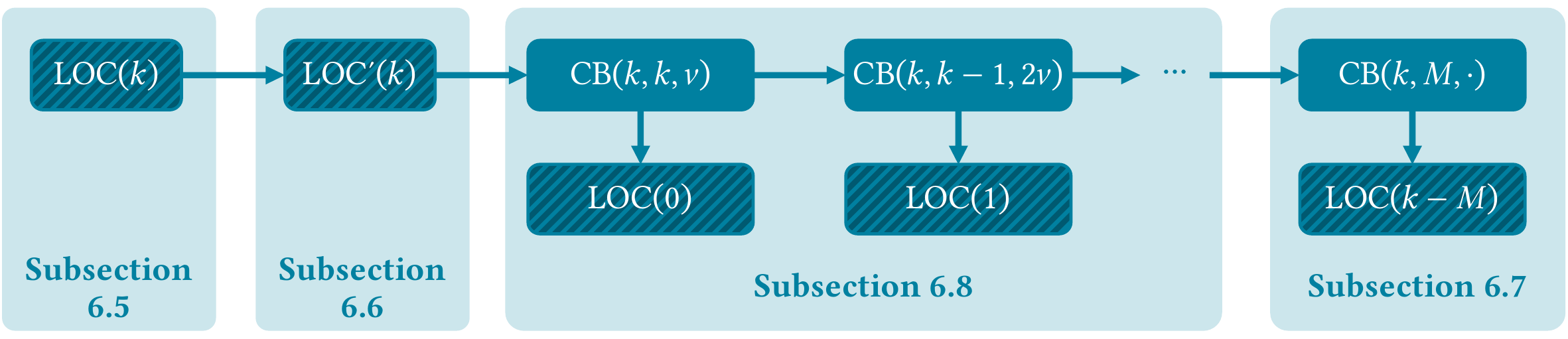}
  \caption{The overall roadmap for solving $\textsf{LOC}(k)$. The recursive algorithm invokes $\textsf{LOC}(0)$, $\dots$, $\textsf{LOC}(k-M)$.}
  \label{fig:505}
\end{figure}

We develop an algorithm to solve $\textsf{LOC}(M^2)$ using at most $3N/M$ colors. This suffices because $\textsf{LOC}(M^2)$ applies to $M^2$-colorable graphs (as every $M^2$-colorable graph is locally $M^2$-colorable). Thus, we develop the algorithm by induction on $k$ and $\ell$ as follows.
\begin{itemize}
    \item \autoref{sec:5.6-pre}: an algorithm to solve $\textsf{LOC}(k)$ using the algorithm for $\textsf{LOC}'(k)$.
    \item \autoref{sec:5.6}: an algorithm to solve $\textsf{LOC}'(k)$ using the algorithm for $\textsf{CB}(k, k, \nu)$ (for some $\nu$).
    \item \autoref{sec:5.7}: an algorithm to solve $\textsf{CB}(k, M, \nu)$ using the algorithm for $\textsf{LOC}(k-M)$. This serves as a kind of ``base case'' for the induction.
    \item \autoref{sec:5.8}: an algorithm to solve $\textsf{CB}(k, \ell, \nu)$ ($\ell \in \{M+1, \dots, k\}$) using the algorithms for $\textsf{CB}(k, \ell-1, 2\nu)$ and $\textsf{LOC}(k-\ell)$.
\end{itemize}
These four parts together constitute the overall algorithm. Its analysis is given in \autoref{sec:5.10}--\autoref{sec:5.11}. This roadmap is illustrated in \autoref{fig:505}.

\subsection{Algorithm for LOC\texorpdfstring{$(k)$}{(k)}}\label{sec:5.6-pre}

First, we describe an algorithm to solve $\textsf{LOC}(k)$ where the final number of vertices $n$ is unknown, using the algorithm for $\textsf{LOC}'(k)$ where $n$ is known in advance.

\paragraph{Algorithm.}

For $t = 0, 1, \dots$, we process the next $2^t$ vertices (or until all vertices are used up) using the algorithm for $\textsf{LOC}'(k)$ with $2^t$ vertices. If $\textsf{LOC}'(k)$ returns a certificate $X$, then $\textsf{LOC}(k)$ immediately returns $X$. Note that we use disjoint colors for different instances of $\textsf{LOC}'(k)$.

\paragraph{Observation.}

If the final number of vertices is $n$, then the final value of $t$ equals $\lfloor \log_2 n \rfloor$. This implies that at most $\lfloor \log_2 n \rfloor + 1$ instances of $\textsf{LOC}'(k)$ are created, and the specified number of vertices in each $\textsf{LOC}'(k)$ is at most $2^t \leq n$.

\subsection{Algorithm for LOC'\texorpdfstring{$(k)$}{(k)}}\label{sec:5.6}

In this section, we describe an algorithm for $\textsf{LOC}'(k)$, where the final number of vertices $n$ is known in advance. The algorithm we describe below bears a substantial similarity to the clique-free algorithm (\autoref{alg:5-clique}).

\paragraph{$k \leq M$ case.} We maintain bins $B_1, B_2, \dots$, and we put the $i$-th arriving vertex $(i \geq 1)$ into bin $B_b$ where $b = \lceil \frac{i}{L} \rceil$. For each bin $B_b$, we use the algorithm for $\textsf{CliqueFree}(M+1)$ given in \autoref{cor:5-clique-free}. When a $(M+1)$-clique appears in $B_b$, we output such a clique as a certificate and terminate. Note that we use one extra color to color the last received vertex.

\paragraph{$k > M$ case.} First, we prepare bins $B_1, B_2, \dots$, initially empty. It is assumed that each $B_i$ is $(M+1)$-clique-free and contains at most $L$ vertices. For each bin, we use the algorithm for $\textsf{CliqueFree}(M+1)$ given in \autoref{cor:5-clique-free}, using disjoint colors for different bins. In addition, we initiate a single instance of $\textsf{CB}(k, k, \nu)$, which uses colors disjoint from those used in the bins, where
\begin{equation*}
    J := \max\left\{1, \left\lfloor \frac{n}{\max(L, n^{1/k^5})} \right\rfloor\right\}, \quad \nu = \frac{2n \cdot L^{M-1}}{J}
\end{equation*}
and its clique block $\mathcal{S}$ is the ``final'' set of all $M$-cliques whose vertices all belong to the same bin. (Some readers may wonder if this is valid in the online setting, but this is justified in \autoref{rem:5-cb}).

\paragraph{Processing each vertex.} The pseudocode for the algorithm is given in \autoref{alg:5-loc}. When a new vertex $v$ arrives, we try to put $v$ into a bin. Specifically, we compute the smallest $j$ such that $|B_j| < L$ and $G[B_j \cup \{v\}]$ is $(M+1)$-clique-free (line 6). Then, we put $v$ into bin $B_j$, except when $B_j$ is empty and there are exactly $J$ ``incomplete'' bins (i.e., bins of sizes between $1$ and $L-1$).

In this exceptional case, we color $v$ using $\textsf{CB}(k, k, \nu)$ instead (lines 7--8). Importantly, $v$ meets the input condition of $\textsf{CB}(k, k, \nu)$ defined in \autoref{def:5-cb}. This is because:
\begin{enumerate}
    \item Since $v$ is completely adjacent to at least one clique in each incomplete bin, $v$ is completely adjacent to at least $J$ cliques in $\mathcal{S}$ (i.e., $|\mathcal{C}(v)| \geq J$).
    \item For each vertex $v$, the number of $M$-cliques that contain $v$ and whose vertices all belong to the same bin is at most $L^{M-1}$, so we have $|\mathcal{S}| \leq n \cdot L^{M-1}$.
\end{enumerate}
Finally, once $\textsf{CB}(k, k, \nu)$ terminates by outputting $(\mathcal{S}', V')$, we output the certificate $\mathrm{vert}(\mathcal{S}') \cup V'$ (lines 9--10). The validity of the output is proven as follows (\autoref{lem:5-loc-valid}), and its size is bounded in the discussion in \autoref{sec:5.9}.

\begin{algorithm}[t]
    \caption{An algorithm for $\textsf{LOC}'(k)$ when $k > M$}
    \label{alg:5-loc}
    \begin{algorithmic}[1]
        \State $B_1, B_2, \dots \leftarrow \emptyset$
        \State $J \leftarrow \max\{1, \lfloor n / \max(L, n^{1/k^5}) \rfloor\}$
        \State $\nu \leftarrow (2n \cdot L^{M-1}) / J$
        \State Initiate an instance of $\textsf{CB}(k, k, \nu)$, where the clique block $\mathcal{S}$ represents the set of all $M$-cliques whose vertices all belong to the same bin
        \For {each arrival of vertex $v$}
            \State $j \leftarrow$ minimum $j$ such that $|B_j| < L$ and $G[B_j \cup \{v\}]$ is $(M+1)$-clique-free
            \If {$B_j = \emptyset$ and $|\{\ell : 1 \leq |B_\ell| < L\}| = J$}
                \State Color $v$ using the instance of $\textsf{CB}(k, k, \nu)$
                \If {the instance of $\textsf{CB}(k, k, \nu)$ terminated by outputting $(\mathcal{S}', V')$}
                    \State \textbf{return} $\text{vert}(\mathcal{S}') \cup V'$
                \EndIf
            \Else
                \State $B_j \leftarrow B_j \cup \{v\}$
                \State Color $v$ with an instance of $\textsf{CliqueFree}(M+1)$ created for $B_j$
            \EndIf
        \EndFor
    \end{algorithmic}
\end{algorithm}

\begin{lemma}\label{lem:5-loc-valid}
    The outputted $\mathrm{vert}(\mathcal{S}') \cup V'$ is not $k$-colorable, assuming the output of $\textsf{CB}(k, k, \nu)$ is valid.
\end{lemma}

\begin{proof}
    Since the output of $\textsf{CB}(k, k, \nu)$ is valid, there is no $k$-coloring $\varphi: (\mathrm{vert}(\mathcal{S}') \cup V') \to [k]$ such that $\mathrm{vert}(\mathcal{S}')$ has at most $k$ colors. This directly implies that $\mathrm{vert}(\mathcal{S}') \cup V'$ is not $k$-colorable.
\end{proof}

\subsection{Algorithm for CB\texorpdfstring{$(k, M, \nu)$}{(k, M, nu)}}\label{sec:5.7}

We describe an algorithm for the base case $\textsf{CB}(k, M, \nu)$.

\paragraph{Algorithm.} Our algorithm simply colors all incoming vertices using $\textsf{LOC}(k-M)$. Let $\mathrm{adj}(v) \in \mathcal{C}(v)$ be an arbitrarily chosen clique in $\mathcal{C}(v)$. If $\textsf{LOC}(k-M)$ returns a certificate set $X$ and terminates, then we return a pair $(\{\mathrm{adj}(v) : v \in X\}, X)$. The validity of the output is shown as follows (\autoref{lem:5-cb-valid-1}), and its size is bounded in \autoref{sec:5.9}.

\begin{lemma}
    \label{lem:5-cb-valid-1}
    Let $(\mathcal{S}', V') := (\{\mathrm{adj}(v) : v \in X\}, X)$ be the outputted pair of the algorithm above, and assume that the output of $\textsf{LOC}(k-M)$ is valid. Then, there is no $k$-coloring of $G[\mathrm{vert}(\mathcal{S}') \cup V']$, $\varphi: (\mathrm{vert}(\mathcal{S}') \cup V') \to [k]$, such that $|\{\varphi(v) : v \in \mathrm{vert}(\mathcal{S}')\}| \leq M$.
\end{lemma}

\begin{proof}
    Suppose there exists a $k$-coloring $\varphi$ with $|\{\varphi(v) : v \in \mathrm{vert}(\mathcal{S}')\}| \leq M$. Without loss of generality, we can assume that $\{\varphi(v) : v \in \mathrm{vert}(\mathcal{S}')\} = \{1, \dots, M\}$. Then, $\{\varphi(v) : v \in S\} = \{1, \dots, M\}$ holds for each clique $S \in \mathcal{S}'$. For each $v \in V'$, $v$ is completely adjacent to $\mathrm{adj}(v) \in \mathcal{S}'$, and thus $\varphi(v)$ must be one of $M+1, \dots, k$. However, this implies that the certificate $V'$ returned by $\textsf{LOC}(k-M)$ is $(k-M)$-colorable, which contradicts the assumption. Therefore, \autoref{lem:5-cb-valid-1} holds.
\end{proof}

\subsection{Algorithm for CB\texorpdfstring{$(k, \ell, \nu)$}{(k, l, nu)} when \texorpdfstring{$\ell > M$}{l > M}}\label{sec:5.8}

Next, we describe an algorithm for the inductive case $\textsf{CB}(k, \ell, \nu)$, where $\ell \in \{M+1, \dots, k\}$.

\begin{figure}[t]
  \centering
  \includegraphics[width=\linewidth]{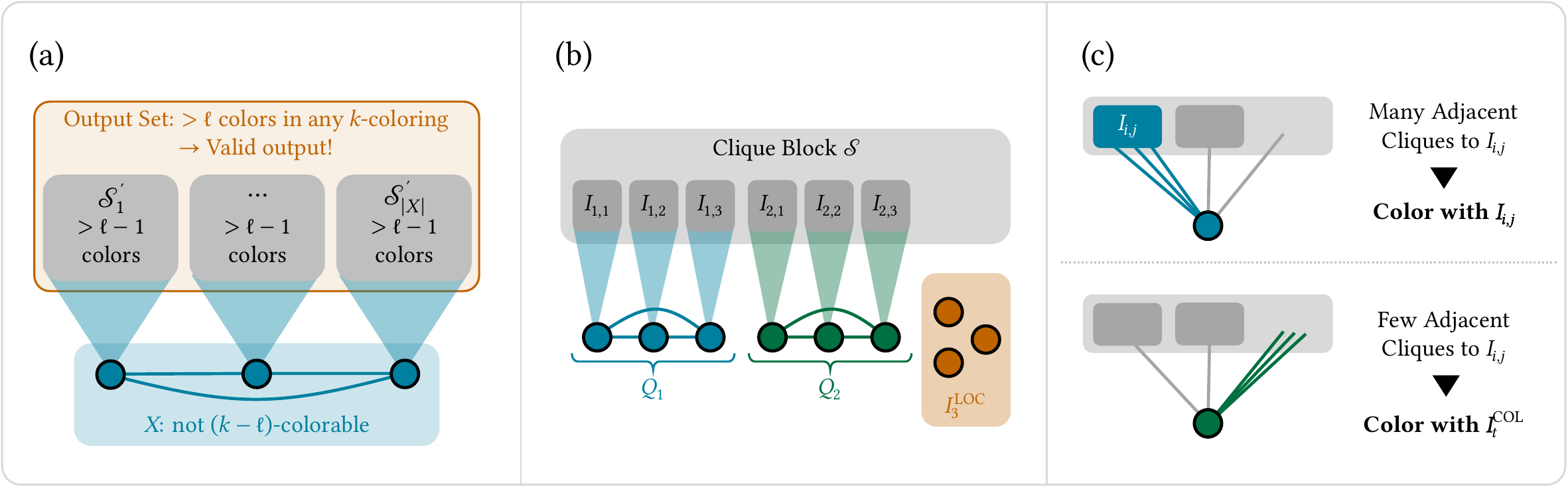}
  \caption{(a) A sketch illustrating the main intuition. Let $(\mathcal{S}', V')$ be the outputted set. For any $k$-coloring $\phi$, $|\{\phi(c) : c \in \text{vert}(\mathcal{S}')\}| > \ell$ must hold. (b) The relationships between instances. (c) A sketch of the procedures when a new vertex arrives.}
  \label{fig:507}
\end{figure}

\paragraph{Intuition.} The key idea is that, by using a small set $X \subseteq V$ such that $\chi(G[X]) > k-\ell$, we can reduce $\textsf{CB}(k, \ell, \nu)$ to $|X|$ subproblems of $\textsf{CB}(k, \ell-1, 2\nu)$. Indeed, let $X = \{x_1, \dots, x_{|X|}\}$, and for each $x_j$, we initiate an instance $I_j$ of $\textsf{CB}(k, \ell - 1, 2\nu)$ with $\mathcal{S}(I_j) := \mathcal{C}(x_j)$. If all instances $I_1, \dots, I_{|X|}$ have terminated and each $I_j$ has returned a pair $(\mathcal{S}_j', V_j')$, then the pair
\begin{equation*}
    \left(\mathcal{S}_1' \cup \cdots \cup \mathcal{S}_{|X|}', V_1' \cup \cdots \cup V_{|X|}' \cup X\right)
\end{equation*}
\noindent
is a valid output for $\textsf{CB}(k, \ell, \nu)$; see \autoref{fig:507} (a). The validity will be proven in \autoref{lem:5-cb-valid-2}. Therefore, by running $\textsf{LOC}(k-\ell)$ to find an appropriate set $X$, we can proceed by induction on $\ell$. Now, we explain how to implement this idea.

\paragraph{Algorithm.} The pseudocode for the following algorithm is given in \autoref{alg:5-cb-induction}. We maintain $I_1^{\mathrm{LOC}}, \dots, I_t^{\mathrm{LOC}}$, the instances of $\textsf{LOC}(k-\ell)$, listed in chronological order. Initially, only $I_1^{\mathrm{LOC}}$ is initiated, that is, $t = 1$ (line 1).

When $I_t^{\mathrm{LOC}}$ terminates with a certificate $X = \{x_1, \dots, x_{|X|}\}$, we initiate new instances $I_{t,1}, \dots, I_{t,|X|}$ of $\textsf{CB}(k, \ell-1, 2\nu)$ (corresponding to $I_1, \dots, I_{|X|}$ in the ``Intuition'' paragraph), where $\mathcal{S}(I_{t,j})$ is fixed to $\mathcal{C}(x_j)$ (lines 12--15). We also record the output set as $Q_t \leftarrow X$. After that, we initiate $I_{t+1}^{\mathrm{LOC}}$ and let $t \leftarrow t+1$ (line 16). The relationship between $I_t^{LOC}$ and $I_{t,j}$ is illustrated in \autoref{fig:507} (b). Note that we use pairwise disjoint colors for all of these instances. Now, when a new vertex $v$ arrives, we proceed as follows:

\begin{enumerate}
    \item If there exists a non-terminated instance $I_{i,j}$ such that $|\mathcal{S}(I_{i,j}) \cap \mathcal{C}(v)| \geq |\mathcal{S}(I_{i,j})| / (2\nu)$, i.e., $v$ is completely adjacent to at least a $1/(2\nu)$ fraction of the cliques in $\mathcal{S}(I_{i,j})$, we color $v$ using $I_{i,j}$ (line 6). This is a valid input to $\textsf{CB}(k, \ell-1, 2\nu)$ (note the parameter $2\nu$). If all of $I_{i,1}, \dots, I_{i,|Q_i|}$ are terminated, we return a pair as written in the ``Intuition'' paragraph and terminate (lines 7--9).
    \item Otherwise, we color $v$ with $I_t^{\mathrm{LOC}}$. When $I_t^{\mathrm{LOC}}$ terminates, we initiate new instances $I_{t,1}, \dots, I_{t,|X|}$, as discussed above (lines 12--16).
\end{enumerate}

\noindent
A sketch of the procedure when $v$ arrives is illustrated in \autoref{fig:507} (c). Importantly, this procedure guarantees $t \leq 3\nu$; we prove this in \autoref{lem:5-t-is-low}.

\begin{lemma}[Validity of Output]
    \label{lem:5-cb-valid-2}
    Let $(\mathcal{S}', V')$ be the outputted pair of $\textsf{CB}(k, \ell, \nu)$ (in line 9), and assume that the outputs of all instances of $\textsf{CB}(k, \ell-1, 2\nu)$ and $\textsf{LOC}(k-\ell)$ are valid. Then, there is no $k$-coloring of $G[\mathrm{vert}(\mathcal{S}') \cup V']$, $\varphi: (\mathrm{vert}(\mathcal{S}') \cup V') \to [k]$, such that $|\{\varphi(v) : v \in \mathrm{vert}(\mathcal{S}')\}| \leq \ell$.
\end{lemma}

\begin{proof}
    Suppose that there exists a $k$-coloring $\varphi$ with $|\{\varphi(v) : v \in \mathrm{vert}(\mathcal{S}')\}| \leq \ell$. Let $C = \{\varphi(v) : v \in \mathrm{vert}(\mathcal{S}')\}$, and $C_i = \{\varphi(v) : v \in \mathrm{vert}(\mathcal{S}'_i)\}$ for each $i$. Since the output for all instances of $\textsf{CB}(k, \ell-1, 2\nu)$ are valid, $|C_i| > \ell-1$ holds. By $C_i \subseteq C$, $|C| \leq \ell$, and $|C_i| > \ell-1$, we obtain $|C_i| = \ell$. This means that
    \begin{equation*}
        C_1 = C_2 = \dots = C_{|X|}.
    \end{equation*}
    Without loss of generality, we may assume $C_i = \{1, 2, \dots, \ell\}$ for each $i$. Let $Q_i = \{q_{i, 1}, \dots, q_{i, |Q_i|}\}$ (at line 9 of \autoref{alg:5-cb-induction}). Since $q_{i, j}$ is completely adjacent to all cliques in $\mathcal{S}(I_{i, j})$, the color $\varphi(q_{i, j})$ must be one of $\ell+1, \dots, k$. However, this means that $Q_i$ is $(k-\ell)$-colorable, which contradicts the assumption that $Q_i$ is a valid output of $\textsf{LOC}(k-\ell)$. Therefore, \autoref{lem:5-cb-valid-2} holds.
\end{proof}

\begin{algorithm}[t]
    \caption{An algorithm for $\textsf{CB}(k, \ell, \nu)$ when $\ell \in \{M+1, \dots, k\}$.}
    \label{alg:5-cb-induction}
    \begin{algorithmic}[1]
        \State $I_1^{\mathrm{LOC}} \leftarrow$ a new instance of \textsc{LOC}$(k-\ell)$
        \State $Q_1, Q_2, \dots \leftarrow \emptyset$
        \State $t \leftarrow 1$
        \For {each arrival of vertex $v$}
            \If {there exists a non-terminated instance $I_{i, j}$ such that $|\mathcal{S}(I_{i,j}) \cap \mathcal{C}(v)| \geq |\mathcal{S}(I_{i,j})| / (2\nu)$}
                \State color $v$ using $I_{i, j}$ (note that ``$\mathcal{C}(v)$'' in the instance $I_{i,j}$ is set to $\mathcal{S}(I_{i,j}) \cap \mathcal{C}(v)$)
                \If {$I_{i, 1}, \dots, I_{i, |Q_i|}$ are all terminated}
                    \State $(\mathcal{S}_j', V_j') \leftarrow$ the returned output of $I_{i, j}$ for all $1 \leq j \leq |Q_i|$
                    \State \textbf{return} $(\mathcal{S}_1' \cup \cdots \cup \mathcal{S}_{|Q_i|}', V_1' \cup \cdots \cup V_{|Q_i|}' \cup Q_i)$
                \EndIf
            \Else
                \State color $v$ using $I_t^{\mathrm{LOC}}$
                \If {$I_t^{LOC}$ terminated by returning $X = \{x_1, \dots, x_{|X|}\}$}
                    \State $Q_t \leftarrow X$
                    \For {$j$ in $1, 2, \dots, |X|$}
                        \State $I_{t, j} \leftarrow$ a new instance of $\textsf{CB}(k, \ell-1, 2\nu)$ with $\mathcal{S}(I_{t, j}) := \mathcal{C}(x_j)$
                    \EndFor
                    \State $I_{t+1}^{\mathrm{LOC}} \leftarrow$ a new instance of $\textsf{LOC}(k-\ell)$ and $t \leftarrow t+1$
                \EndIf
            \EndIf
        \EndFor
    \end{algorithmic}
\end{algorithm}

\subsection{Number of Rounds in CB\texorpdfstring{$(k, \ell, \nu)$}{(k, l, nu)}}\label{sec:5.10}

We show that the event of creating new instances only occurs at most $3\nu$ times in $\textsf{CB}(k, \ell, \nu)$, i.e., $t \leq 3\nu$ in \autoref{alg:5-cb-induction}.

\begin{lemma}[\cite{KYY27}] \label{lem:5-set-common}
    Let $\nu \geq 1$ be a real number. Let $S$ be a non-empty set, and let $A_1, \dots, A_t \subseteq S$ be sets such that $|A_i| \geq \frac{|S|}{\nu}$. If $|A_i \cap A_j| \leq \frac{|A_i|}{2\nu}$ for each $i < j$, then $t < 2\nu$ holds.
\end{lemma}

\begin{proof}
    Suppose that $t \geq 2\nu$, and let $s = \lceil 2\nu \rceil$. Let $B_j = A_j \setminus (A_1 \cup A_2 \cup \dots \cup A_{j-1})$. By the assumptions $|A_i \cap A_j| \leq \frac{|A_i|}{2\nu}$ and $|A_i| \geq \frac{|S|}{\nu}$,
    \begin{align*}
        \sum_{j=1}^{s} |B_j|
        & \geq \sum_{j=1}^{s} \left(|A_j| - \sum_{i=1}^{j-1} |A_i \cap A_j|\right) \\
        & \geq \sum_{j=1}^{s} \left(|A_j| - \sum_{i=1}^{j-1} \frac{1}{2\nu} \cdot |A_i| \right) \\
        & \geq \sum_{j=1}^{s} |A_j| - \sum_{i=1}^{s} \frac{s-i}{2\nu} \cdot |A_i| \geq \sum_{i=1}^{s} \frac{2\nu-s+i}{2\nu} \cdot |A_i| \geq \sum_{i=1}^{s} \frac{2\nu-s+i}{2\nu^2} \cdot |S|.
    \end{align*}
    However, we have $\sum_{i=1}^s |B_i| \leq |S|$ because $B_1, \dots, B_s$ are disjoint. This contradicts the inequality above at $s = \lceil 2\nu \rceil$. Therefore, $t < 2\nu$ holds.
\end{proof}

\begin{lemma}\label{lem:5-t-is-low}
    In each run of \autoref{alg:5-cb-induction} for $\textsf{CB}(k, \ell, \nu)$, we have $t \leq 3\nu$.
\end{lemma}

\begin{proof}
    For each $i \in \{1, \dots, t-1\}$, let $p_i$ be an index such that $I_{i, p_i}$ has not yet been terminated (such an index must exist while \textsc{CB}$(k, \ell, \nu)$ is not terminated). Then, for each $i < j$,
    \begin{equation*}
        \left|\mathcal{S}(I_{i, p_i}) \cap \mathcal{S}(I_{j, p_j})\right| \leq \frac{|\mathcal{S}(I_{i, p_i})|}{2\nu}.
    \end{equation*}
    This is because, otherwise, the vertex $q_{j, p_j}$ must be colored by instance $I_{i,p_i}$ (in lines 5--6 of \autoref{alg:5-cb-induction}) and thus $I_{j, p_j}$ would not be created, a contradiction. In addition, $|\mathcal{S}(I_{i, p_i})| \geq |\mathcal{S}|/\nu$ holds by the input assumption of $\textsf{CB}(k, \ell, \nu)$. By \autoref{lem:5-set-common} (with $A_i := \mathcal{S}(I_{i, p_i})$ and $S := \mathcal{S}$), $t < 2\nu + 1 \leq 3\nu$ holds.
\end{proof}

\subsection{Analysis Part I: Certificate Size \& Correctness}\label{sec:5.9}

Now, we start the analysis of the algorithm. First, we analyze the sizes of the returned certificates. We let $f_k$ be the upper bound on the size $|X|$ of a certificate $X$ returned by $\textsf{LOC}(k)$. We also let $g_{k, \ell}$ be the upper bound on the size $|\mathrm{vert}(\mathcal{S}') \cup V'|$ of a certificate $(\mathcal{S}', V')$ returned by $\textsf{CB}(k, \ell, \nu)$. Note that these bounds can also be used in the definition of each subproblem; see \autoref{def:5-loc} and \autoref{def:5-cb}. Then, the following inequalities hold:

\begin{lemma}\label{lem:5-certificate-recurrence-0}
    For $k \leq M$, we have $f_k \leq M+1$.
\end{lemma}

\begin{proof}
    By \autoref{sec:5.6}, $\textsf{LOC}(k)$ returns an $(M+1)$-clique, so \autoref{lem:5-certificate-recurrence-0} holds.
\end{proof}

\begin{lemma}\label{lem:5-certificate-recurrence-1}
    For $k > M$, we have $f_k \leq g_{k, k}$.
\end{lemma}

\begin{proof}
    By \autoref{sec:5.6}, $\textsf{LOC}(k)$ returns $\mathrm{vert}(\mathcal{S}') \cup V'$ when $\textsf{CB}(k, k, \nu)$ terminates by returning a pair $(\mathcal{S}', V')$. Therefore, $f_k \leq g_{k, k}$.
\end{proof}

\begin{lemma}\label{lem:5-certificate-recurrence-2}
    For $k \geq M$, we have $g_{k, M} \leq (M+1) \cdot f_{k-M}$.
\end{lemma}

\begin{proof}
    We denote the returned pair in $\textsf{CB}(k, M, \nu)$ as $(\mathcal{S}', V')$. By \autoref{sec:5.7}, $\textsf{CB}(k, M, \nu)$ returns a pair $(\{\mathrm{adj}(v) : v \in X\}, X)$ when an instance of $\textsf{LOC}(k-M)$ terminates by returning a set $X$. Since $\mathrm{adj}(v)$ is an $M$-clique (see \autoref{sec:5.7}), $|\mathrm{vert}(\mathcal{S}')| \leq M |X|$, and therefore $|\text{vert}(\mathcal{S}') \cup V'| \leq (M+1)|X| \leq (M+1)f_{k-M}$, proving \autoref{lem:5-certificate-recurrence-2}.
\end{proof}

\begin{lemma}\label{lem:5-certificate-recurrence-3}
    For $k > M$ and $\ell \in \{M+1, \dots, k\}$, we have $g_{k, \ell} \leq f_{k-\ell} \cdot (g_{k, \ell-1} + 1)$.
\end{lemma}

\begin{proof}
    We denote the pair returned in $\textsf{CB}(k, \ell, \nu)$ as $(\mathcal{S}', V')$. By \autoref{sec:5.8}, as shown in lines 8--9 of \autoref{alg:5-cb-induction}, $\textsf{CB}(k, \ell, \nu)$ returns
    \begin{equation*}
        (\mathcal{S}', V') := \left(\mathcal{S}_1' \cup \cdots \cup \mathcal{S}_{|Q_i|}', V_1' \cup \cdots \cup V_{|Q_i|}' \cup Q_i\right)
    \end{equation*}
    if the instances $I_{i, 1}, \dots, I_{i, |Q_i|}$ returned $(\mathcal{S}_1', V_1'), \dots, (\mathcal{S}_{|Q_i|}', V_{|Q_i|}')$, respectively. Since $|\text{vert}(\mathcal{S}'_i) \cup V'_i| \leq g_{k, \ell-1}$ and $|Q_i| \leq f_{k-\ell}$, $|\text{vert}(\mathcal{S}') \cup V'| \leq f_{k-\ell} \cdot (g_{k, \ell-1} + 1)$ holds, proving \autoref{lem:5-certificate-recurrence-3}.
\end{proof}

\begin{lemma}[Correctness]
    \label{lem:5-overall-correctness}
    When we set $f_k$ and $g_{k, \ell}$ according to \autoref{lem:5-certificate-recurrence-0}--\autoref{lem:5-certificate-recurrence-3}, the outputs of the algorithm for $\textsf{LOC}(k)$ and $\textsf{CB}(k, \ell, \nu)$ are valid.
\end{lemma}

\begin{proof}
    We show by induction on both $k$ and $\ell$. For $\textsf{LOC}(k)$ (see \autoref{def:5-loc} for the definition), the output is valid for $k \leq M$ because the output is an $(M+1)$-clique and by \autoref{lem:5-certificate-recurrence-0}, and the output is also valid for $k > M$ by \autoref{lem:5-loc-valid} and \autoref{lem:5-certificate-recurrence-1}. For $\textsf{CB}(k, \ell, \nu)$ (see \autoref{def:5-cb} for the definition), the output is valid for $\ell = M$ by \autoref{lem:5-cb-valid-1} and \autoref{lem:5-certificate-recurrence-2}, and the output is also valid for $\ell > M$ by \autoref{lem:5-cb-valid-2} and \autoref{lem:5-certificate-recurrence-3}. Therefore, \autoref{lem:5-overall-correctness} holds.
\end{proof}

Furthermore, we analyze the growth of $f_k$ in the following two lemmas.

\begin{lemma}\label{lem:5-certificate-recurrence-4}
    For $k > M$, we have $f_k + 1 \leq (M + 1) \times (f_0 + 1) \times \cdots \times (f_{k-M} + 1)$.
\end{lemma}

\begin{proof}
    First, we prove by induction on $\ell$ that $g_{k, \ell} + 1 \leq (M + 1) \times (f_{k-\ell} + 1) \times \cdots \times (f_{k-M} + 1)$ holds for all $\ell \geq M$. When $\ell = M$, the inequality holds due to \autoref{lem:5-certificate-recurrence-2}. Assume that the inequality holds at $\ell = \ell'-1$. Then, this also holds at $\ell = \ell'$ as follows:
    \begin{align*}
        g_{k, \ell'} + 1 & \leq f_{k-\ell'} \cdot (g_{k, \ell'-1} + 1) + 1 \\
        & \leq f_{k-\ell'} \cdot ((M + 1) \times (f_{k-\ell'+1} + 1) \times \cdots \times (f_{k-M} + 1)) + 1\\
        & \leq (M + 1) \times (f_{k-\ell'} + 1) \times \cdots \times (f_{k-M} + 1)
    \end{align*}
    where the first inequality follows from \autoref{lem:5-certificate-recurrence-3} and the second inequality is given by induction on $\ell$. Therefore, $g_{k, k} + 1 \leq (M + 1) \times (f_0 + 1) \times \cdots \times (f_{k-M} + 1)$. By \autoref{lem:5-certificate-recurrence-1}, \autoref{lem:5-certificate-recurrence-4} holds.
\end{proof}

\begin{lemma}\label{lem:5-certificate-recurrence-5}
    Assuming $N$ is large enough, $f_k \leq L$ holds for all $k \leq M^2$.
\end{lemma}

\begin{proof}
    Let $F_k = \log_{M+1}(f_k + 1)$. Then, by \autoref{lem:5-certificate-recurrence-4}, $F_k \leq M$ for $k \leq M$ and $F_k \leq F_0 + \cdots + F_{k-M} + 1$ for $k > M$. Thus, we have $F_k \leq M^{2\lceil k/M \rceil + 1}$ for $k \leq M^2$, since this obviously holds for $k \leq M$ and
    \begin{equation*}
        F_k \leq 1 + \sum_{i=0}^{k-M} F_i \leq 1 + \sum_{i=0}^{k-M} M^{2 \lceil k/M \rceil - 1} \leq M^2 \cdot M^{2 \lceil k/M \rceil - 1} = M^{2 \lceil k/M \rceil + 1}
    \end{equation*}
    for $M < k \leq M^2$, and $F_{k} \leq M^{2M+1}$ holds ($k \leq M^2$). By $M = \lfloor (\log \log N)^{1-\epsilon/2} \rfloor$ and $L = \lfloor N^{1/M^{11}} \rfloor$,
    \begin{align*}
        \log \log (f_k + 1) & = \log (F_k \log (M + 1)) \leq \log (M^{2M+1} \log (M+1)) \\
        & \leq \log (M^{3M} \log (M+1)) = 3M \log M + \log \log (M + 1) \\ 
        & = 3 (\log \log N)^{1-\epsilon/2} \cdot \log (\log \log N)^{1-\epsilon/2} + o(\log \log N) = o(\log \log N), \\
        \log \log L & \geq (\log \log N^{1/M^{11}})/2 = \log ((\log N) / M^{11})/2 \\
        & = \log ((\log N) / (\log \log N)^{11-5.5\epsilon})/2 \\
        & = (\log \log N - \log (\log \log N)^{11-5.5\epsilon})/2 = \Theta(\log \log N).
    \end{align*}
    Therefore, $f_k + 1 \leq L$ for $k \leq M^2$, and \autoref{lem:5-certificate-recurrence-5} follows.
\end{proof}

\subsection{Analysis Part II: Number of Colors Used}\label{sec:5.11}

Finally, we analyze the number of colors used in the algorithm to solve $\textsf{LOC}(M^2)$ on $N$ vertices; we eventually prove that we use at most $3N/M$ colors. In the following proof, we may omit parameters by using notation such as $\textsf{LOC}(\cdot)$, which refers to $\textsf{LOC}(k)$ for any $k$.

\begin{lemma}\label{lem:5-nu}
    Assuming $N$ is sufficiently large, each call to $\textsf{CB}(k, \ell, \nu)$ satisfies $\nu \leq N^{3/M^5}$.
\end{lemma}

\begin{proof}
    Consider a call to $\textsf{CB}(k, \ell, \nu)$. The algorithm for $\textsf{LOC}'(k)$ (for the same $k$) initiates an instance of $\textsf{CB}(k, k, \nu)$ where
    \begin{align*}
        \nu = \frac{2n \cdot L^{M-1}}{J}
        & \leq (2n \cdot L^{M-1}) \cdot \frac{2\max(L, n^{1/k^5})}{n} \\
        & \leq 4 \cdot L^{M-1} \cdot \max(L, n^{1/k^5}) \\
        & \leq 4 \cdot N^{1/M^{10}} \cdot \max(N^{1/M^{11}},N^{1/M^5})
        \leq 4N^{1/M^{10} + 1/M^5} \leq N^{2/M^5}.
    \end{align*}
    This follows from the definitions of $J$ and $\nu$ (in \autoref{sec:5.6}) and the facts that $L = N^{1/M^{11}}$ and $k > M$. Denote this value of $\nu$ by $\nu_0$. Then, for $\ell = k-1, \dots, M$, the algorithm recursively calls $\textsf{CB}(k, \ell, \nu)$ with $\nu = 2^{k-\ell} \nu_0$. Since $k \leq M^2$, we have $\nu \leq 2^{M^2} \cdot N^{2/M^5} \leq N^{3/M^5}$. Therefore, \autoref{lem:5-nu} holds.
\end{proof}

\begin{lemma}\label{lem:5-initiated-instances-1}
    Assume that $N$ is sufficiently large. Let $k \in \{M+1, \dots, M^2\}$. The algorithm for $\textsf{LOC}(k)$ invokes at most $N^{4/M^3}$ instances of $\textsf{CB}(k, \ell, \cdot)$ for each $\ell \in \{M, \dots, k\}$, including indirectly invoked instances.
\end{lemma}

\begin{proof}
    Let $h_\ell$ be the number of $\textsf{CB}(k, \ell, \cdot)$ instances initiated during the run of $\textsf{LOC}(k)$. By \autoref{def:5-cb} and \autoref{lem:5-t-is-low}, one $\textsf{CB}(k, \ell, \cdot)$ instance initiates at most $3\nu \cdot f_{k-\ell}$ instances of $\textsf{CB}(k, \ell - 1, \cdot)$. Thus, $h_{\ell - 1} \leq (h_\ell \times 3\nu) \times f_{k-\ell}$. In addition, since $\textsf{LOC}(k)$ invokes $\textsf{LOC}'(k)$ at most $\log_2 N + 1$ times, we have $h_k \leq \log_2 N + 1$. Therefore, for $\ell \geq M$,
    \begin{align*}
        h_\ell & \leq (3\nu)^{k-\ell} \times (f_0 \times \cdots \times f_{k-\ell-1}) \times (\log_2 N + 1) \\
        & \leq (3\nu)^{(M^2)} \times L^{k-\ell} \times N^{1/M^{11}} \\
        & \leq (3N^{3/M^5})^{(M^2)} \times N^{1/M^9} \times N^{1/M^{11}} \leq N^{4/M^3}.
    \end{align*}
    The first inequality is due to $h_{\ell-1} \leq (h_\ell \times 3\nu) \times f_{k-\ell}$, the second inequality is due to $k \leq M^2$ and \autoref{lem:5-certificate-recurrence-5}, and the third inequality is due to \autoref{lem:5-nu}, $L = N^{1/M^{11}}$, and $k - \ell \leq M^2$. This completes the proof.
\end{proof}

\begin{lemma}\label{lem:5-initiated-instances-2}
    Assume that $N$ is sufficiently large. An instance of $\textsf{LOC}(M^2)$ invokes a total of at most $N^{5/M}$ instances of $\textsf{LOC}(\cdot)$, including indirectly invoked instances.
\end{lemma}

\begin{proof}
    Let $r_k$ be the number of $\textsf{LOC}(\cdot)$ instances invoked by $\textsf{LOC}(k)$, including $\textsf{LOC}(k)$ itself. For $k \leq M$, obviously $r_k = 1$ holds. For $k > M$, due to \autoref{lem:5-initiated-instances-1}, $\textsf{LOC}(k)$ initiates at most $N^{4/M^3}$ instances of $\textsf{CB}(k, \ell, \cdot)$ for each $M \leq \ell \leq k$. By \autoref{lem:5-t-is-low}, each \textsc{CB}$(k, \ell, \cdot)$ invokes at most $3\nu$ instances of $\textsf{LOC}(k-\ell)$. Therefore,
    \begin{align*}
        r_k \leq (3\nu \cdot N^{4/M^3}) \cdot (r_0 + \cdots + r_{k-M}) + 1
        & \leq 3N^{4/M^3 + 3/M^5} \cdot (r_0 + \cdots + r_{k-M}) + 1 \\
        & \leq N^{5/M^3} \cdot (r_0 + \cdots + r_{k-M})
    \end{align*}
    \noindent
    and the inequality above holds with $r_k = N^{5k/M^3}$ $(k > M)$. Therefore, $\textsf{LOC}(M^2)$ invokes at most $N^{5/M}$ instances of $\textsf{LOC}(\cdot)$.
\end{proof}

\begin{lemma}\label{lem:5-bins}
    Consider an instance $\textsf{LOC}(k)$ $(k \leq M^2)$ received $n$ vertices and $b$ of them are (directly) put into bins (note that $n$ and $b$ are unknown). Assuming $N$ is large enough, at most $\frac{b}{L} + 4 \cdot n^{1-1/M^{10}}$ bins are created.
\end{lemma}

\begin{proof}
    Consider the case $k > M$, because the case $k \leq M$ is trivial. Suppose that $\textsf{LOC}(k)$ initiated instances for $\textsf{LOC}'(k)$ on $n_0, n_1, \dots, n_{m-1}$ vertices where $n_i = 2^i$, and the $n_i$-vertex instance (directly) put $b_i$ vertices to bins. Then, since $J \leq (n_i)^{1-1/M^{10}}$, the $n_i$-vertex instance creates at most $(n_i)^{1-1/M^{10}} + \frac{b_i}{L}$ bins. Therefore, at most
    \begin{equation*}
        \sum_{i=0}^{m-1} (n_i)^{1-1/M^{10}} + \frac{b_i}{L} \leq \sum_{i=0}^{\lfloor \log_2 n \rfloor} (2^i)^{1-1/M^{10}} + \frac{b}{L} \leq 4 \cdot n^{1-1/M^{10}} + \frac{b}{L}
    \end{equation*}
    \noindent
    bins are created in $\textsf{LOC}(k)$, which proves the lemma.
\end{proof}

\begin{theorem}\label{thm:5-colors}
    Assuming $N$ is sufficiently large, $\textsf{LOC}(M^2)$ uses at most $\frac{3N}{M}$ colors.
\end{theorem}

\begin{proof}
    Assume that $\textsf{LOC}(M^2)$ invokes $m$ instances of $\textsf{LOC}(\cdot)$, the $i$-th instance receives $n_i$ vertices, and $b_i$ of them are put into bins. First, by \autoref{lem:5-initiated-instances-2}, $\textsf{LOC}(M^2)$ invokes at most $N^{5/M}$ instances of $\textsf{LOC}(\cdot)$, and therefore $m \leq N^{5/M}$. Second, $\sum_i b_i \leq N$ holds because each vertex $v$ that arrives during the run of $\textsf{LOC}(k)$ is eventually placed in one of the bins for some $\textsf{LOC}(\cdot)$ instance. Third, $\sum_i n_i \leq NM$ holds, because for each arrival of vertex $v$, at most $M$ calls to $\textsf{LOC}(\cdot)$ instances happen (the worst case is $\textsf{LOC}(M^2) \to \textsf{LOC}(M^2-M) \to \dots \to \textsf{LOC}(M)$).
    
    Therefore, by \autoref{lem:5-bins}, we can bound the total number of bins created as follows. For readability, we highlight the terms in different colors.
    \begin{align*}
        \sum_{i=1}^{m} \left(4 \cdot n_i^{1-1/M^{10}} + \frac{b_i}{L}\right)
        & \leq \textcolor{darkgreen}{4 \cdot m \cdot \left(\frac{\sum_i n_i}{m}\right)^{1-1/M^{10}}} + \textcolor{darkred}{\frac{\sum_i b_i}{L}} \\
        & \leq \textcolor{darkgreen}{4 \cdot N^{5/M} \cdot \left(\frac{NM}{N^{5/M}}\right)^{1-1/M^{10}}} + \textcolor{darkred}{\frac{N}{L}} \\
        & \leq \textcolor{darkgreen}{4 \cdot N^{5/M} \cdot N^{(1-5/M)(1-1/M^{10})} \cdot M} + \textcolor{darkred}{\frac{N}{L}} \\
        & \leq \textcolor{darkgreen}{4 \cdot N^{5/M + (1-5/M)(1-1/M^{10})} \cdot N^{1/M^{12}}} + \textcolor{darkred}{\frac{N}{L}} \\
        & \leq \textcolor{darkgreen}{4 \cdot N^{1 - 1/M^{10} + 5/M^{11} + 1/M^{12}}} + \textcolor{darkred}{\frac{N}{L}} \leq \textcolor{darkgreen}{\frac{N}{L}} + \textcolor{darkred}{\frac{N}{L}} = \frac{2N}{L}.
    \end{align*}
    \noindent
    In addition, each bin uses at most $\frac{L}{M}+1$ colors (the $+1$ term comes from the constraint that we must color the last received vertex). Therefore, $\textsf{LOC}(M^2)$ uses at most $(\frac{L}{M}+1) \cdot \frac{2N}{L} \leq \frac{3N}{M}$ colors.
\end{proof}

\subsection{The Grand Finale} \label{sec:5.12}

We are now ready to prove our main theorem, stated as follows.

\begingroup
\def\thetheorem{\ref{thm:main-3}}
\begin{theorem}
    For every fixed $\varepsilon > 0$, there exists a polynomial-time deterministic online coloring algorithm that attains a competitive ratio of $O(n / (\log \log n)^{2-\varepsilon})$. 
\end{theorem}
\addtocounter{theorem}{-1}
\endgroup

We prove that our algorithm runs in polynomial time (\autoref{lem:5-polytime}), and then evaluate the performance of Kawarabayashi, Yoneda, and Yoneda's algorithm \cite{KYY27} for $M$-colorable graphs (\autoref{thm:5-kyy}). Finally, we prove the main theorem.

\begin{lemma} \label{lem:5-polytime}
    The proposed algorithm for the problem $\textsf{LOC}(M^2)$ runs in polynomial time.
\end{lemma}

\begin{proof}
    We assume that $N$ is sufficiently large. First, we evaluate the time complexity of \autoref{alg:5-clique} for solving $\textsf{CliqueFree}(r)$ $(r \leq M+1)$. In the algorithm for object coloring (\autoref{lem:5-object-coloring}), each set can be processed in $O(|U|)$ time. In \autoref{alg:5-clique}, each bin has size $n_r \leq L$, so the number of $(r-2)$-cliques in each bin is at most $L^{r-2} \leq L^{M-1} \leq N$, totaling at most $N^2$ across all bins. Hence, we can assume $|U| \leq N^2$, so \autoref{alg:5-clique} runs in $O(N^2 M^2)$ time per vertex, excluding child instances; note that we find all these cliques in $O(N^2 M^2)$ time by brute force. Since the recursion depth is at most $M$, the total time for solving $\textsf{CliqueFree}(r)$, including recursion, is $O(N^2 M^3)$ time per vertex.
    
    Next, we evaluate the time complexity for solving $\textsf{LOC}(M^2)$.
    \begin{itemize}
        \item The running time of \autoref{alg:5-loc} excluding child instances is $O(N^2 M^3)$ time per vertex. This follows from an argument similar to the one above, where the bottleneck is solving $\textsf{CliqueFree}(M+1)$.
        \item The running time of \autoref{alg:5-cb-induction} excluding child instances is $O(N^3)$ time per vertex, due to $|\mathcal{S}| \leq N^2$ (by the same discussion as $|U| \leq N^2$ above) and at most $N$ child instances are created due to $|Q_1| + \dots + |Q_t| \leq N$.
        \item The recursion depth for solving $\textsf{LOC}(M^2)$, excluding $\textsf{CliqueFree}(\cdot)$ (but including child instances of $\textsf{LOC}(\cdot), \textsf{CB}(\cdot, \cdot)$), is at most $M \cdot M^2 = M^3$.
    \end{itemize}
    Therefore, the algorithm uses $O(\max\{N^2 M^3, N^3\}) \cdot M^3 = O(N^3 M^3)$ time per vertex in total. This sums to $O(N^4 (\log \log N)^{3(1-\varepsilon/2)})$ time for all vertices.
\end{proof}

\begin{theorem}[\cite{KYY27}] \label{thm:5-kyy}
    There exists a polynomial-time deterministic online algorithm that colors $N$-vertex $M$-colorable graphs using $O(N/M^2)$ colors.
\end{theorem}

\begin{proof}
    By \cite[Lemma 3.18]{KYY27}, for every $k \geq 0$, there is a polynomial-time deterministic algorithm to color $N$-vertex $k$-colorable graphs online using at most $f(k) := 2^{2^{k+3}} \cdot N^{1-\frac{1}{k(k-1)/2+1}}$ colors. Here,
    \begin{align*}
        \log_2 f(M)
        & = 2^{M+3} + \left(1 - \frac{1}{M(M-1)/2+1}\right) \log_2 N \\
        & = (\log N)^{o(1)} + \log_2 N - \Theta\left(\frac{\log N}{(\log \log N)^{2-\varepsilon}}\right) = \log_2 N - \omega(\log \log N).
    \end{align*}
    Therefore, $f(M) = O(N / (\log N)^c)$ for every fixed $c > 0$, and \autoref{thm:5-kyy} follows.
\end{proof}

\begin{proof}[Proof of \autoref{thm:main-3}]
    For every integer $k \geq 1$, there exists a polynomial-time deterministic algorithm that colors $N$-vertex $k$-colorable graphs online in $k \cdot O(N/M^2)$ colors: we can use \autoref{thm:5-kyy} for $k \leq M$, we can use our algorithm for $\textsf{LOC}(M^2)$ for $M < k \leq M^2$, and we can use distinct colors for all vertices for $k > M^2$ (which uses $N$ colors).
    
    We note that $\textsf{LOC}(M^2)$ applies to $M^2$-colorable graphs, in which case the problem does not terminate. The algorithm for $\textsf{LOC}(M^2)$ runs in polynomial time by \autoref{lem:5-polytime}. Therefore, by \autoref{lem:2-competitive-ratio}, we have a polynomial-time deterministic online coloring algorithm that attains a competitive ratio of $O(N/M^2)$, which is $O(N / (\log \log N)^{2-\varepsilon})$.
\end{proof}

\section{Conclusion}\label{sec:6}

In this paper, we studied the competitive ratio of the online coloring problem. The best previous upper and lower bounds are shown in \autoref{tab:conclusion}; large poly-logarithmic gaps remained for both randomized and deterministic algorithms. However, we have improved three of these results. In \autoref{sec:3}, we established the $\Omega(n / \log n)$ lower bound for deterministic algorithms. In \autoref{sec:4}, we established the $O(n (\log \log n)^2 / \log^2 n)$ upper bound for randomized algorithms. Finally, in \autoref{sec:5}, we established the $O(n / (\log \log n)^{2-\varepsilon})$ upper bound for deterministic algorithms.

\begin{table}[!b]
    \centering
    \footnotesize
    \renewcommand{\arraystretch}{1.4}

    \caption{The improvements in upper bounds and lower bounds
    of competitive ratio for the online graph coloring problem.}

    \begin{tabular}{>{\centering\arraybackslash}p{2.3cm}lccYY}
        \toprule
        \multicolumn{2}{c}{\multirow{2}{*}{Type}} &
        \multicolumn{2}{c}{Previous Results} &
        \multicolumn{2}{>{\columncolor{resultyellow}}c}{Our Results}
        \\

        \multicolumn{2}{c}{} & Result & Gap & Result & Gap
        \\
        \midrule

        \multirow{2}{*}{\shortstack{Randomized\\Algorithms}}
        & Upper Bound
        & $O(n/\log n)$
        & \multirow{2}{*}{$O(\log n)$}
        & $O(n (\log \log n)^2 / \log^2 n)$
        &
        \\

        & Lower Bound
        & $\Omega(n/\log^2 n)$
        &
        & $-$
        & \multirow{-2}{*}{$O\left((\log\log n)^2\right)$}
        \\

        \multirow{2}{*}{\shortstack{Deterministic\\Algorithms}}
        & Upper Bound
        & $O(n/\log\log n)$
        & \multirow{2}{*}{$\displaystyle
            O\left(\frac{\log^2 n}{\log\log n}\right)$}
        & $\displaystyle O(n/(\log\log n)^{2-\varepsilon})$
        &
        \\

        & Lower Bound
        & $\Omega(n/\log^2 n)$
        &
        & $\Omega(n/\log n)$
        & \multirow{-2}{*}{$\displaystyle
            O\left(
            \frac{\log n}{(\log\log n)^{2-\varepsilon}}
            \right)$}
        \\

        \bottomrule
    \end{tabular}\label{tab:conclusion}
\end{table}

As a result, we have nearly closed the gap for randomized algorithms, and the gap for deterministic algorithms has become significantly smaller. Most importantly, we finally affirmatively solved the 40-year-old open question of whether randomized algorithms are asymptotically better than deterministic algorithms. Since there exists an $O(n (\log \log n)^2 / \log^2 n)$-competitive randomized algorithm while any deterministic algorithm is $\Omega(n / \log n)$-competitive, the answer to this question is yes. However, for the deterministic side, the gap is still poly-logarithmic. In addition, for the randomized side, the gap is almost closed but not completely closed. Therefore, it is natural to present the following conjectures:

\begin{conjecture}
    Every deterministic online coloring algorithm is $\Omega(n / \log^c n)$-competitive for every fixed $c > 0$.
\end{conjecture}

\begin{conjecture}
    There exists a randomized online coloring algorithm, against an oblivious adversary, that attains a competitive ratio of $O(n / \log^2 n)$.
\end{conjecture}

Especially, for the gap in deterministic algorithms, we are currently leaning toward the lower-bound side; we believe no deterministic algorithm achieves a competitive ratio of $O(n / \log n)$. This is because our adversary framework in \autoref{sec:3}, which decides the base color after two vertices share the same online color, is still loose and restricted.

\section*{AI Usage Disclosure}

We extensively used OpenAI's large language model, GPT-6 Astra, during the proof exploration phase, in collaboration with the authors. For \autoref{thm:main-1} and \autoref{thm:main-2}, the AI initially tried to prove the theorem using the low-chromatic approach, which considers improvements on $k$-colorable graphs for constant $k$ to achieve a better competitive ratio. This research direction is also historically popular, as seen in \cite{Vis92, Kie98b, KYY27}. However, it failed.

Therefore, the human authors introduced a new research direction that considers the large-chromatic case, especially from the viewpoint of \autoref{prop:1-lv2}. In particular, we posed a question: \emph{``Can we construct an algorithm where almost all colors are used on exactly three vertices?''} With this new approach, the AI managed to generate the proof. However, we consider the essence of our contribution to be the new high-chromatic approach; for example, in \autoref{sec:3}, the new adversary model that replaces the transparent model is natural given the question of whether the adversary can force ``at most two vertices per color''. For \autoref{thm:main-2} and \autoref{thm:main-3}, the AI generated complicated and unnatural proofs, and we substantially simplified them.

This paper was written entirely by the human authors, based on discussions of the intuition behind the algorithm. The authors used OpenAI's GPT models to polish the paper's presentation. The authors take full responsibility for the correctness, content, and readability of the paper.

\printbibliography

@article{ACF+02,
  author       = {Auer, Peter and Cesa-Bianchi, Nicol{\`o} and Freund, Yoav and Schapire, Robert E.},
  title        = {The Nonstochastic Multiarmed Bandit Problem},
  journal      = {SIAM Journal on Computing},
  volume       = {32},
  number       = {1},
  pages        = {48--77},
  year         = {2002},
  doi          = {10.1137/S0097539701398375}
}

@inproceedings{AEL+23,
  author       = {Akbari, Amirreza and Eslami, Navid and Lievonen, Henrik and Melnyk, Darya and S{\"a}rkij{\"a}rvi, Joona and Suomela, Jukka},
  title        = {Locality in Online, Dynamic, Sequential, and Distributed Graph Algorithms},
  booktitle    = {50th International Colloquium on Automata, Languages, and Programming ({ICALP} 2023)},
  series       = {Leibniz International Proceedings in Informatics ({LIPIcs})},
  volume       = {261},
  pages        = {10:1--10:20},
  year         = {2023},
  publisher    = {Schloss Dagstuhl -- Leibniz-Zentrum f{\"u}r Informatik},
  doi          = {10.4230/LIPIcs.ICALP.2023.10}
}

@article{Alb03,
  author       = {Albers, Susanne},
  title        = {Online algorithms: a survey},
  journal      = {Mathematical Programming},
  volume       = {97},
  number       = {1-2},
  pages        = {3--26},
  year         = {2003},
  doi          = {10.1007/s10107-003-0436-0}
}

@article{AS21,
  author       = {Albers, Susanne and Schraink, Sebastian},
  title        = {Tight Bounds for Online Coloring of Basic Graph Classes},
  journal      = {Algorithmica},
  volume       = {83},
  number       = {1},
  pages        = {337--360},
  year         = {2021},
  doi          = {10.1007/s00453-020-00759-7}
}

@article{BBH+14,
  author       = {Bianchi, Maria Paola and B{\"o}ckenhauer, Hans-Joachim and Hromkovi{\v{c}}, Juraj and Keller, Lucia},
  title        = {Online Coloring of Bipartite Graphs with and without Advice},
  journal      = {Algorithmica},
  volume       = {70},
  number       = {1},
  pages        = {92--111},
  year         = {2014},
  doi          = {10.1007/s00453-013-9819-7}
}

@inproceedings{BCL+18,
  author       = {Bubeck, S{\'e}bastien and Cohen, Michael B. and Lee, Yin Tat and Lee, James R. and M{\k{a}}dry, Aleksander},
  title        = {{$k$}-server via multiscale entropic regularization},
  booktitle    = {Proceedings of the 50th Annual {ACM} Symposium on Theory of Computing ({STOC} 2018)},
  pages        = {3--16},
  year         = {2018},
  publisher    = {Association for Computing Machinery},
  doi          = {10.1145/3188745.3188798}
}

@article{Bea76,
  author       = {Bean, Dwight R.},
  title        = {Effective coloration},
  journal      = {Journal of Symbolic Logic},
  volume       = {41},
  number       = {2},
  pages        = {469--480},
  year         = {1976},
  doi          = {10.2307/2272247}
}

@misc{BFS+26,
  author        = {Brada{\v{c}}, Domagoj and Fox, Jacob and Steiner, Raphael and Sudakov, Benny and Zhang, Shengtong},
  title         = {Coloring small locally sparse degenerate graphs and related problems},
  year          = {2026},
  eprint        = {2601.15245},
  archiveprefix = {arXiv},
  primaryclass  = {math.CO}
}

@inproceedings{BSV+24,
  author       = {Blikstad, Joakim and Svensson, Ola and Vintan, Radu and Wajc, David},
  title        = {Online edge coloring is (nearly) as easy as offline},
  booktitle    = {Proceedings of the 56th Annual {ACM} Symposium on Theory of Computing ({STOC} 2024)},
  pages        = {36--46},
  year         = {2024},
  publisher    = {Association for Computing Machinery},
  doi          = {10.1145/3618260.3649741}
}

@inproceedings{BSV+25b,
  author       = {Blikstad, Joakim and Svensson, Ola and Vintan, Radu and Wajc, David},
  title        = {Online Edge Coloring: Sharp Thresholds},
  booktitle    = {2025 {IEEE} 66th Annual Symposium on Foundations of Computer Science ({FOCS})},
  pages        = {1787--1812},
  year         = {2025},
  publisher    = {IEEE},
  doi          = {10.1109/FOCS63196.2025.00094}
}

@inproceedings{Erg25,
  author       = {Ergen, Ekin},
  title        = {Online Makespan Scheduling Under Scenarios},
  booktitle    = {33rd Annual European Symposium on Algorithms ({ESA} 2025)},
  series       = {Leibniz International Proceedings in Informatics ({LIPIcs})},
  volume       = {351},
  pages        = {27:1--27:16},
  year         = {2025},
  publisher    = {Schloss Dagstuhl -- Leibniz-Zentrum f{\"u}r Informatik},
  doi          = {10.4230/LIPIcs.ESA.2025.27}
}

@article{FKL+91,
  author       = {Fiat, Amos and Karp, Richard M. and Luby, Michael and McGeoch, Lyle A. and Sleator, Daniel D. and Young, Neal E.},
  title        = {Competitive paging algorithms},
  journal      = {Journal of Algorithms},
  volume       = {12},
  number       = {4},
  pages        = {685--699},
  year         = {1991},
  doi          = {10.1016/0196-6774(91)90041-V}
}

@inproceedings{GKM+14,
  author       = {Gutowski, Grzegorz and Kozik, Jakub and Micek, Piotr and Zhu, Xuding},
  title        = {Lower bounds for on-line graph colorings},
  booktitle    = {Algorithms and Computation ({ISAAC} 2014)},
  series       = {Lecture Notes in Computer Science},
  volume       = {8889},
  pages        = {507--515},
  year         = {2014},
  publisher    = {Springer},
  doi          = {10.1007/978-3-319-13075-0_40}
}

@article{GL88,
  author       = {Gy{\'a}rf{\'a}s, Andr{\'a}s and Lehel, Jen{\H{o}}},
  title        = {On-line and first fit colorings of graphs},
  journal      = {Journal of Graph Theory},
  volume       = {12},
  number       = {2},
  pages        = {217--227},
  year         = {1988},
  doi          = {10.1002/jgt.3190120212}
}

@article{Hal97,
  author       = {Halld{\'o}rsson, Magn{\'u}s M.},
  title        = {Parallel and On-Line Graph Coloring},
  journal      = {Journal of Algorithms},
  volume       = {23},
  number       = {2},
  pages        = {265--280},
  year         = {1997},
  doi          = {10.1006/jagm.1996.0836}
}

@article{Hal00,
  author       = {Halld{\'o}rsson, Magn{\'u}s M.},
  title        = {Online coloring known graphs},
  journal      = {The Electronic Journal of Combinatorics},
  volume       = {7},
  number       = {1},
  pages        = {R7},
  year         = {2000},
  doi          = {10.37236/1485}
}

@article{HS94,
  author       = {Halld{\'o}rsson, Magn{\'u}s M. and Szegedy, Mario},
  title        = {Lower bounds for on-line graph coloring},
  journal      = {Theoretical Computer Science},
  volume       = {130},
  number       = {1},
  pages        = {163--174},
  year         = {1994},
  doi          = {10.1016/0304-3975(94)90157-0}
}

@article{Ira94,
  author       = {Irani, Sandy},
  title        = {Coloring inductive graphs on-line},
  journal      = {Algorithmica},
  volume       = {11},
  number       = {1},
  pages        = {53--72},
  year         = {1994},
  doi          = {10.1007/BF01294263}
}

@article{Kie91,
  author       = {Kierstead, Hal A.},
  title        = {A polynomial time approximation algorithm for dynamic storage allocation},
  journal      = {Discrete Mathematics},
  volume       = {88},
  number       = {2-3},
  pages        = {231--237},
  year         = {1991},
  doi          = {10.1016/0012-365X(91)90011-P}
}

@incollection{Kie98a,
  author       = {Kierstead, Hal A.},
  title        = {Coloring graphs on-line},
  editor       = {Fiat, Amos and Woeginger, Gerhard J.},
  booktitle    = {Online Algorithms: The State of the Art},
  series       = {Lecture Notes in Computer Science},
  volume       = {1442},
  pages        = {281--305},
  year         = {1998},
  publisher    = {Springer},
  doi          = {10.1007/BFb0029574}
}

@article{Kie98b,
  author       = {Kierstead, Hal A.},
  title        = {On-line coloring {$k$}-colorable graphs},
  journal      = {Israel Journal of Mathematics},
  volume       = {105},
  pages        = {93--104},
  year         = {1998},
  doi          = {10.1007/BF02780324}
}

@article{KT81,
  author       = {Kierstead, Hal A. and Trotter, William T.},
  title        = {An extremal problem in recursive combinatorics},
  journal      = {Congressus Numerantium},
  volume       = {33},
  pages        = {143--153},
  year         = {1981}
}

@inproceedings{KYY27,
  author        = {Kawarabayashi, Ken-ichi and Yoneda, Hirotaka and Yoneda, Masataka},
  title         = {Online Graph Coloring for {$k$}-Colorable Graphs},
  booktitle     = {Proceedings of the 2027 Annual {ACM}-{SIAM} Symposium on Discrete Algorithms ({SODA})},
  year          = {2027},
  publisher     = {Society for Industrial and Applied Mathematics},
  eprint        = {2511.16100},
  archiveprefix = {arXiv},
  primaryclass  = {cs.DS},
  note          = {To appear}
}

@misc{Li26,
  author        = {Li, Yaqiao},
  title         = {New bounds for online coloring three-colorable graphs},
  year          = {2026},
  eprint        = {2610.41516},
  archiveprefix = {arXiv},
  primaryclass  = {cs.DS}
}

@article{LST89,
  author       = {Lov{\'a}sz, L{\'a}szl{\'o} and Saks, Michael and Trotter, William T.},
  title        = {An on-line graph coloring algorithm with sublinear performance ratio},
  journal      = {Discrete Mathematics},
  volume       = {75},
  number       = {1-3},
  pages        = {319--325},
  year         = {1989},
  doi          = {10.1016/0012-365X(89)90096-4}
}

@inproceedings{MMS88,
  author       = {Manasse, Mark S. and McGeoch, Lyle A. and Sleator, Daniel D.},
  title        = {Competitive algorithms for on-line problems},
  booktitle    = {Proceedings of the 20th Annual {ACM} Symposium on Theory of Computing ({STOC} 1988)},
  pages        = {322--333},
  year         = {1988},
  publisher    = {Association for Computing Machinery},
  doi          = {10.1145/62212.62243}
}

@misc{Ope26,
  author       = {{OpenAI}},
  title        = {Squared-logarithmic randomized {$k$}-server on arbitrary metrics},
  year         = {2026},
  url          = {https://github.com/openai/math/blob/main/preprints/Squared-logarithmic-randomized-k-server-on-arbitrary-metrics-September-24-2026/Squared-logarithmic-randomized-k-server-on-arbitrary-metrics-September-24-2026.pdf},
  note         = {OpenAI Math Release preprint}
}

@inproceedings{RW24,
  author       = {Rajaraman, Rajmohan and Wasim, Omer},
  title        = {Competitive Capacitated Online Recoloring},
  booktitle    = {32nd Annual European Symposium on Algorithms ({ESA} 2024)},
  series       = {Leibniz International Proceedings in Informatics ({LIPIcs})},
  volume       = {308},
  pages        = {95:1--95:17},
  year         = {2024},
  publisher    = {Schloss Dagstuhl -- Leibniz-Zentrum f{\"u}r Informatik},
  doi          = {10.4230/LIPIcs.ESA.2024.95}
}

@article{Vis92,
  author       = {Vishwanathan, Sundar},
  title        = {Randomized Online Graph Coloring},
  journal      = {Journal of Algorithms},
  volume       = {13},
  number       = {4},
  pages        = {657--669},
  year         = {1992},
  doi          = {10.1016/0196-6774(92)90061-G}
}

@inproceedings{YY26,
  author       = {Yoneda, Hirotaka and Yoneda, Masataka},
  title        = {Online Coloring for Graphs of Large Odd Girth},
  booktitle    = {34th Annual European Symposium on Algorithms ({ESA} 2026)},
  series       = {Leibniz International Proceedings in Informatics ({LIPIcs})},
  volume       = {388},
  pages        = {127:1--127:15},
  year         = {2026},
  publisher    = {Schloss Dagstuhl -- Leibniz-Zentrum f{\"u}r Informatik},
  doi          = {10.4230/LIPIcs.ESA.2026.127}
}

\end{document}